\documentclass[10pt,letterpaper]{article}
\usepackage[top=1in,bottom=1in,left=1in,right=1in,footskip=0.75in]{geometry}

\usepackage{amsmath,amssymb}

\usepackage{changepage}

\usepackage{textcomp,marvosym}

\usepackage{cite}

\usepackage{xr}
\usepackage{nameref,hyperref}

\usepackage[inline]{enumitem}

\usepackage[right]{lineno}

\usepackage[nopatch=eqnum]{microtype}
\DisableLigatures[f]{encoding = *, family = * }

\usepackage[table,dvipsnames]{xcolor}

\usepackage{array}

\newcolumntype{+}{!{\vrule width 2pt}}

\newlength\savedwidth

\usepackage{setspace} 

\raggedright
\usepackage[aboveskip=1pt,labelfont=bf,labelsep=period,justification=raggedright,singlelinecheck=off]{caption}

\makeatletter
\renewcommand{\@biblabel}[1]{\quad#1.}
\makeatother

\usepackage{lastpage,fancyhdr,graphicx}
\usepackage{epstopdf}
\fancyheadoffset[L]{2.25in}
\fancyfootoffset[L]{2.25in}
\usepackage{mathtools}
\DeclarePairedDelimiter\abs{\lvert}{\rvert}%

\DeclarePairedDelimiter\norm{\lVert}{\rVert}%

\newcommand{\rl}{\textbf{r}} %
\newcommand{\vl}{\textbf{v}} %
\newcommand{\rloc}[1]{\textbf{r}_{loc #1}} %
\newcommand{\anytime}{(\cdot)}

\newcommand{\xr}{x_r} %
\newcommand{\yr}{y_r} %
\newcommand{\epar}{\textbf{e}_{\parallel}} %
\newcommand{\eperp}{\textbf{e}_{\perp}}    %

\newcommand{\vpar}{v_{||}}
\newcommand{\vrelpar}{\tilde{v}_{||}}

\newcommand{\amountdyads}{$6\,M$} %
\newcommand{\amounttrajectories}{$21\,M$} %
\newcommand{\avgTravlers}{$24\,k$} %

\newcommand{\distance}{d_{ij}}
\newcommand{\distanceThreshold}{d_{min}}

\newcommand{\dyadangle}{\phi}
\newcommand{\dyadproxangle}{\alpha} %
\newcommand{\dyadbins}{B}
\newcommand{\binstanding}{\dyadbins_{\text{standing}}}
\newcommand{\binfree}{\dyadbins_{\text{free}}}
\newcommand{\binwith}{\dyadbins_{\text{coflow}}}
\newcommand{\binagainst}{\dyadbins_{\text{counterflow}}}
\newcommand{\binacross}{\dyadbins_{\text{crossflow}}}

\newcommand{\vcom}{\textbf{v}_{com}}
\newcommand{\vprox}{\textbf{v}_{prox}} %
\newcommand{\comspeed}{\text{v}_{\text{com}}}
\newcommand{\cond}{\mathbb{E}}
\newcommand{\condcomspeed}{\cond[\comspeed \mid (\xr, \yr)]}
\newcommand{\rcom}{\boldsymbol{r}_{\text{com}}}
\newcommand{\dotrcom}{\dot{\boldsymbol{r}}_{\text{com}}}
\newcommand{\conddensity}{\cond[\density \mid (\xr, \yr)]}

\newcommand{\vcomlow}{\text{v}_{_L}} %
\newcommand{\vcomhigh}{\text{v}_{_H}}

\newcommand{\statesud}{{\Gamma^{\,\updownarrow\,}}}
\newcommand{\stateslr}{{\Gamma^{\leftrightarrow}}}
\newcommand{\param}{\boldsymbol{\theta}} %
\newcommand{\cb}{\,|\,} %

\newcommand{\lgr}{\Pi_\text{OLO}} %
\newcommand{\hatlgr}{\hat\Pi_\text{OLO}} %
\newcommand{\Prob}{\mathbb{P}} %

\newcommand{\closecount}{N_{\closeradius}} %
\newcommand{\proxcount}{N_{prox}} %

\newcommand{\closeradius}{R} %
\newcommand{\density}{\rho} %
\newcommand{\densityunit}{\mathrm{p/m}^2}
\newcommand{\densityzerocross}{\rho_{c}} %
\newcommand{\densityfreeflow}{\rho_{\text{free}}} %

\newcommand{\trajstart}{t_{i,0}} %
\newcommand{\trajend}{t_{i,f}}   %
\newcommand{\trajinterval}{[\trajstart, \trajend]}
\newcommand{\traj}{\gamma}
\newcommand{\trajtime}{\mathcal{T}}

\newcommand{\scalefactor}{A}
\newcommand{\addfactor}{E}

\newcommand{\modalletter}{g}
\newcommand{\modaldistance}{D}
\newcommand{\modalbandlr}{\modalletter^\leftrightarrow}
\newcommand{\modalbandud}{\modalletter^\updownarrow}
\newcommand{\Gausfit}{G}

\newcommand{\vstandingthresh}{v_s}
\newcommand{\deltaband}{\Delta b}

\newcommand{\DfitWithDens}{5.10}
\newcommand{\vcomlowfitWithDens}{0.54}
\newcommand{\vcomlowslopefitWithDens}{0.68}
\newcommand{\vcomhighfitWithDens}{1.58}
\newcommand{\vcomhighslopefitWithDens}{-0.44}
\newcommand{\fitScalefitWithDens}{18.24}
\newcommand{\fitScaleslopefitWithDens}{-5.49}
\newcommand{\dBandMeter}{0.15}

\usepackage[normalem]{ulem} %

\usepackage{etoolbox}
\newtoggle{showcaptionslabels}
\togglefalse{showcaptionslabels} %

\newcommand{\reffig}[1]{Fig~\ref{#1}}
\newcommand{\refeqq}[1]{Eq~\ref{#1}}
\newcommand{\reftab}[1]{Table~\ref{#1}}
\newcommand{\refsec}[1]{Section~\ref{#1}}
\newcommand{\SIAppendix}{S1~Appendix}
\newcommand{\refSIsec}[1]{Section~\ref{#1} in~\SIAppendix}

\usepackage[percent]{overpic}
\usepackage[toc,page]{appendix}
\usepackage{enumitem}
\usepackage{tikz}
\newcommand{\figinnerlab}[4]{%
\begin{tikzpicture}
\draw (0, 0) node[anchor=south west,inner sep=0, yshift=#3,xshift=#4] {#1 }; %
\draw (0, 0) node[fill=none] {\footnotesize{(#2)}};
\end{tikzpicture}%
}
\newcommand{\figinnerlabB}[7]{%
\begin{tikzpicture}
\draw (0, 0) node[anchor=south west,inner sep=0, yshift=#3,xshift=#4] {#1 }; %
\draw (0, 0) node[fill=none] {\footnotesize{(#2)}};
\draw (#6, #7) node[fill=none] {\footnotesize{(#5)}};
\end{tikzpicture}%
}

\newcommand{\figinnerlabLoopB}[2]{%
\begin{tikzpicture}
\draw (0,0) node[anchor=south west,inner sep=0] {#1};

\foreach \lab/\x/\y in {#2} {
  \draw (\x,\y) node[fill=none] {\footnotesize{(\lab)}};
}
\end{tikzpicture}
}

\newcommand{\figinnerlabLoop}[2]{%
\begin{tikzpicture}
\draw (0,0) node[anchor=south west,inner sep=0] {#1};

\foreach \lab/\x/\y in {#2} {
  \draw (\x,\y) node[draw,fill=white,inner sep=1pt,font=\tiny] {\footnotesize{(\lab)}};
}
\end{tikzpicture}
}

\begin{document}
\vspace*{0.2in}

\begin{flushleft}
	{\Large
		\textbf{Probabilistic dynamics of small groups in crowd flows} %
	}
	\newline
	\\
	Chiel van der Laan\textsuperscript{1,2},
	Alessandro Corbetta\textsuperscript{1,2,*},
	\\
	\bigskip
	\textbf{1} Department of Applied Physics and Science Education, Eindhoven University of Technology, Eindhoven, The Netherlands
	\\
	\textbf{2} Eindhoven Artificial Intelligence Systems Institute, Eindhoven, The Netherlands
	\\
	\bigskip

	* \nolinkurl{a.corbetta@tue.nl}

\end{flushleft}

\section*{Abstract}

Pedestrians in crowds frequently move as part of small groups, which can constitute up to 70\% of individuals in public spaces. Dyads (groups of two) are most frequent.
Understanding quantitatively the dynamics of dyads walking in crowds is therefore an essential building block towards a fundamental comprehension of the crowd behavior as a whole, and is mandatory for accurate crowd dynamics models. Unavoidably, due to the non-deterministic behavior of pedestrians, characterizations of the dynamics must be probabilistic.

In this work, we analyse the dynamics of over \amountdyads~dyads: a statistical ensemble of unprecedented  resolution within a multi-year real-life pedestrian trajectory measurement campaign  (about \amounttrajectories~trajectories, collected at Eindhoven Central Station, The Netherlands).
We provide phenomenological models for dyad behavior depending on the surrounding crowd state.
We present a thorough collection of fundamental diagrams that probabilistically relate both dyad velocity and dyad formation to the state of the surrounding crowd (density, relative velocity).
Depending on the surrounding crowd, dyads adjust their interpersonal distance and may shift in formation, possibly moving from abreast states (known to favor social interaction) to in-file (which favors navigation through dense crowds).
To quantitatively investigate formation changes,  we introduce a scalar indicator, which we dub Orientation Log-Odds (OLO), that  quantifies the relative log-likelihood of abreast versus in-file formations.
Conceptually, for any given crowd state, the OLO quantifies the energy difference between the abreast and the in-file configuration under a Boltzmann-like assumption. We model how OLO depends on the crowd state, showcasing that its derivative is a product of two velocity-density fundamental diagrams.

Together, these results provide a statistically robust, data-driven description of dyad configuration dynamics in real-world crowds, establishing a foundation towards new predictive, group-aware crowd models.

\section*{Introduction}
Understanding and quantitatively modeling pedestrian dynamics in crowds is essential for designing safe and comfortable
civil infrastructures~\cite{still2021applied,feliciani2023trends}, and is an outstanding scientific challenge, deeply connected with the statistical physics of active matter~\cite{corbetta-annurev-2023}.
Prevalent in crowds are groups of individuals, i.e., \textit{sets of people  intentionally walking together}~\cite{adrian_glossary_2025}, which can account for even $70\%$ of the whole mass~\cite{DyadDistanceMoussaid2010} and influence the collective dynamics~\cite{feliciani_social_2023,hu_social_2021,ye_traffic_2021,DyadDistanceMoussaid2010}.  Among groups, dyads, i.e., \textit{two-person groups}, are the most frequent~\cite{DyadDistanceMoussaid2010},
with larger groups often fragmenting into stable two- or three-person subunits during movement~\cite{zanlungo_walking_2013}. Ultimately, this makes dyads one of the fundamental building blocks of collective pedestrian behavior.
The characterization of group dynamics -- and of dyads specifically -- is therefore among the necessary ingredients towards predictive models for the crowd, and a must to surpass current modeling approaches where crowds are often only collections of independent agents~\cite{nicolas_social_2023}.
Notably, due to stochasticity in pedestrian behavior, physical characterizations of groups must be probabilistic, encompassing average trends, typical fluctuations, and unlikely states.
Such a probabilistic characterization has recently been enabled by advances in automated computer vision (e.g.,~\cite{seer2014kinects,pouw2024high,gu2025emergence,parisi2021pedestrian}), which have enabled the collection of vast, anonymous, real-life pedestrian trajectory datasets.
Large trajectory datasets, even in the  millions, have already unlocked detailed probabilistic characterizations of some pedestrian dynamics (e.g., in low-density cases~\cite{corbetta-pre-2018,vleuten-pre-2024}, or effective/macroscopic descriptions at large spatial scales~\cite{gabbana-pnasn-2022,pouw2024high,minartz2025discovering, gu2025emergence,garcimartin2018redefining,parisi2021pedestrian,bonnemain2023pedestrians}).

\noindent The aim of this paper is to provide a thorough probabilistic characterization and phenomenological model of the behavior of dyads as macroscopic parameters of the surrounding crowd change.

A necessary aspect in the analysis of dyads and groups is their correct identification within the crowd. So far, group behavior has been extensively studied in controlled laboratory settings -- where groups are defined \emph{a priori}~\cite{hu_social_2021,hu_experimental_2023,ye_traffic_2021,crociani_micro_2019,gorrini_crossing_2019,gorrini_granulometric_2016,crociani_shape_2018} -- and through moderate-scale, real-life datasets, with  groups identified through exhaustive human annotation.
However, manual identification of groups, while obviously highly accurate, is very labor-intensive to scale beyond thousands of %
cases~\cite{DyadDistanceMoussaid2010,zanlungo_intrinsic_2017,zanlungo_intrinsic_2019,zanlungo_social_2020}.

Laboratory experiments, for example, in uni- and multidirectional flows and bottleneck scenarios (cf. review in~\cite{nicolas_social_2023} for an overview) have provided first approximations to macroscopic descriptors, including
the relation between densities and group velocities (so-called fundamental diagrams~\cite{crociani_micro_2019,ye_traffic_2021,hu_experimental_2023,hu_social_2021}).
Yet, limited participant numbers, typically fewer than one hundred units, have bottlenecked finer-scale investigation, e.g.,  of  fluctuations around ensemble averages.
Conversely, research based on real-life datasets has enabled more detailed microscopic analyses, e.g., in the case of a dyad walking  in complex environments~\cite{zanlungo_spatial-size_2015,DyadDistanceMoussaid2010}. In~\cite{zanlungo_spatial-size_2015}, it has been observed that
the crowd surrounding a dyad can, for instance, affect its  spatial formation: a dyad can transition from abreast configuration (pedestrians side-by-side, efficient for communication~\cite{zanlungo_potential_2014}) at low density, to an in-file configuration that streamlines motion in higher-density regimes.

A quantitative probabilistic characterization of how dyads behave as we change critical parameters -- such as their walking speed, the local density, and the relative speed of the surrounding crowd, ultimately yielding also changes of formation -- has remained, however, beyond the statistical resolution of current investigations.
Inevitably, to analyse group dynamics in datasets with more than $10^3$ %
trajectories, which are routine in recent data collections, the group identification must be automated.
Anonymity in data collection, often mandatory when operating in public spaces, limits measurements to trajectories only and generates a substantial technical challenge. In fact, pedestrians are reduced to positional information only, with no additional kinematic or anatomical features, which forbids tracking and group detection hinging on visual cues (e.g., appearance and body language~\cite{khan_detection_2015,solera_socially_2016,li_social_2023}).
Given the exclusive availability of trajectories of centers of mass, group detection can rely only on proxemics indicators, using heuristics to distinguish genuine groups from coincidental proximity.
Heuristics proposed so far consider the time consistency or spatio-temporal clustering of distances~\cite{yucel_deciphering_2013,brscic_modelling_2017,pouw_monitoring_2020}, sufficient correlation in velocities~\cite{yucel_modeling_2018}, or even in gait~\cite{gregorj_gait_2025}.
A key bottleneck is the computational cost of pair-wise metrics/correlation indicators among all trajectory pairs in space and time. For this purpose, in~\cite{pouw_monitoring_2020}, some authors of this work proposed a graph-based additive algorithm requiring a single pass in time of the dataset, allowing the extraction of groups within datasets of sizes exceeding $10^4$ trajectories.%

\begin{figure*}[h]
				\centering
		\figinnerlab{
		\begin{overpic}[width=1\linewidth]{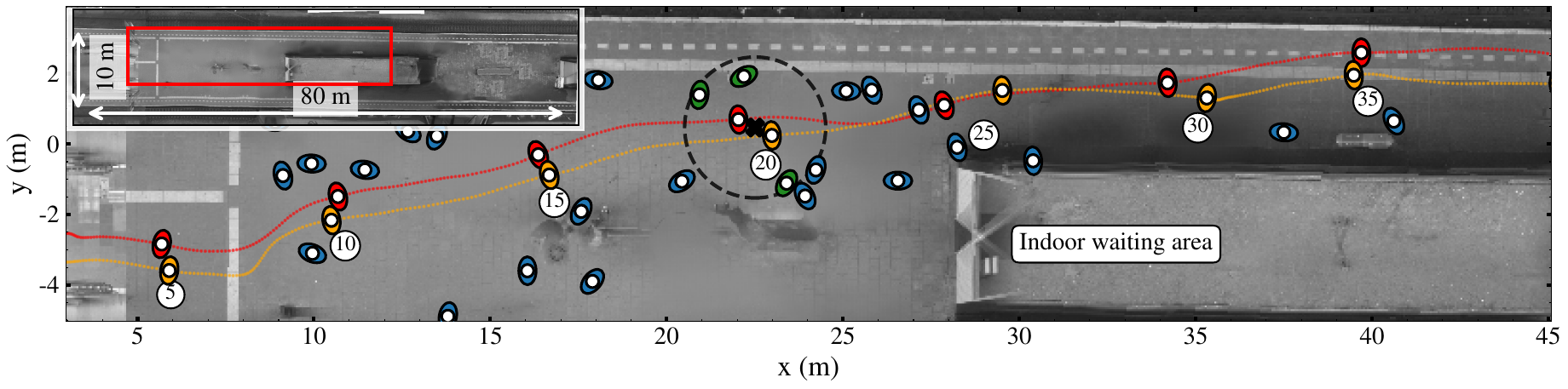}
		\put(73,-4){\includegraphics[width=0.3\linewidth]{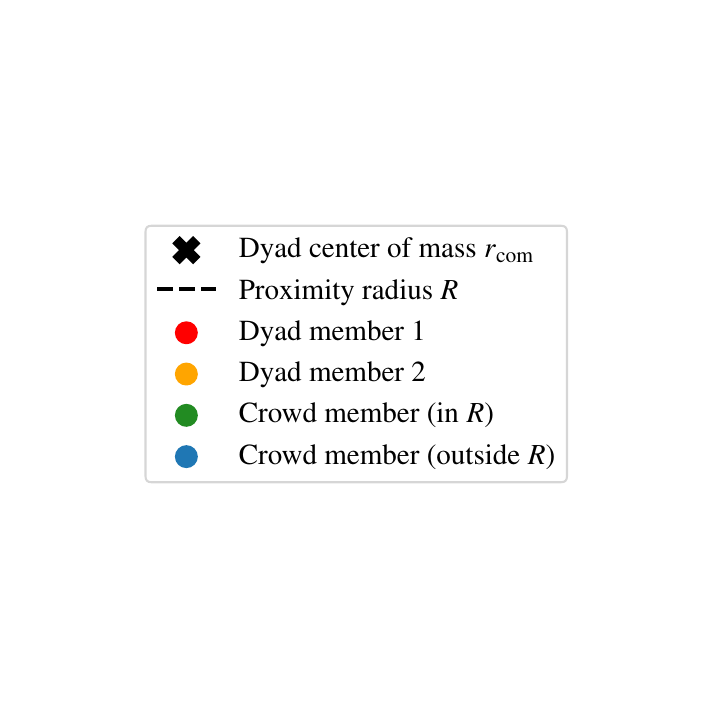}}
		\end{overpic}}{a}{-0.6cm}{-.46cm}
		\figinnerlab{
		\begin{overpic}[width=0.26\linewidth]{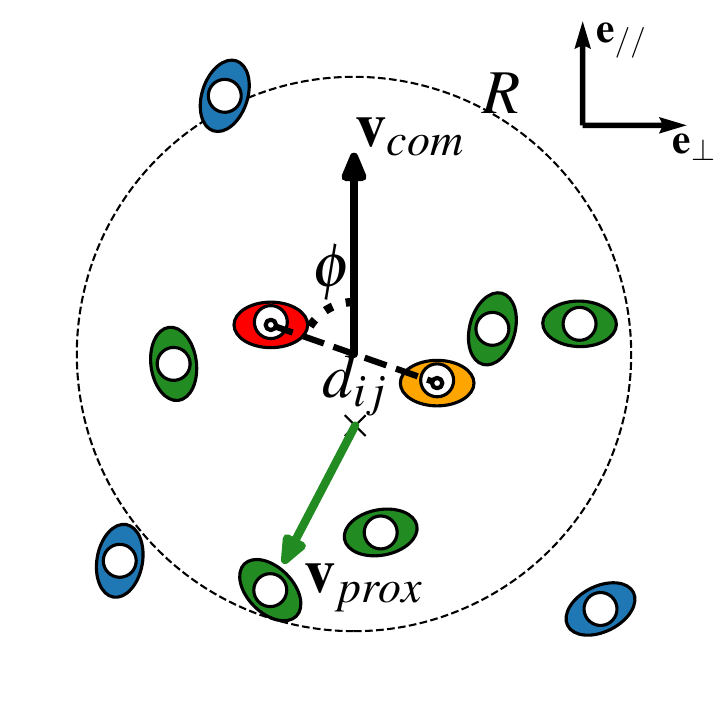}
		\end{overpic}}{b}{-0.4cm}{-.56cm}
		\figinnerlab{
		\includegraphics[width=.26\linewidth]{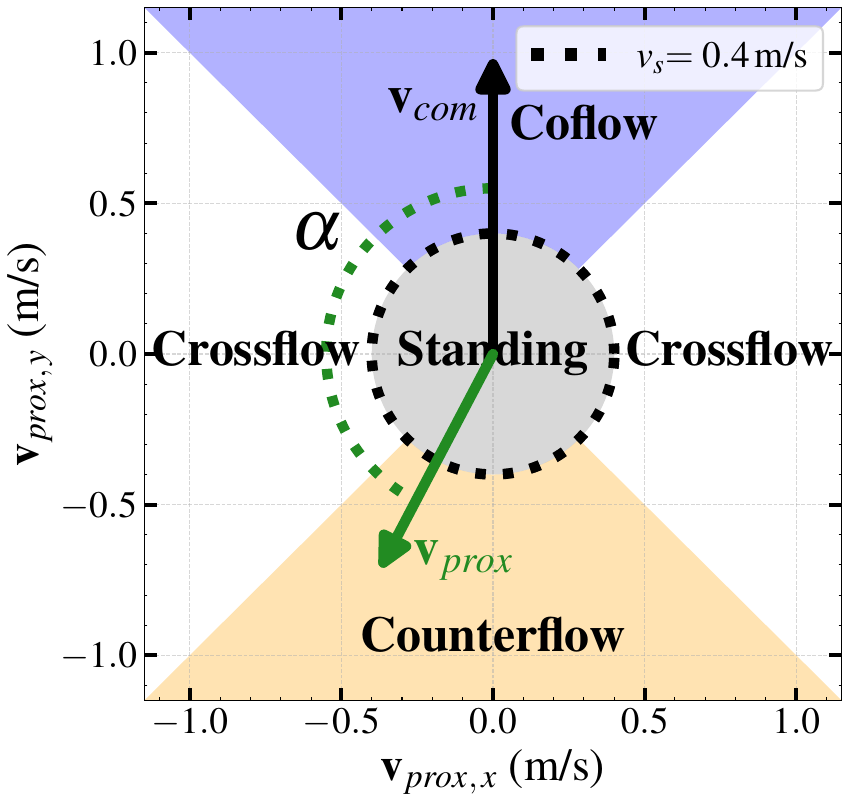}}{c}{-0.4cm}{-0.52cm}
		\figinnerlab{
		\includegraphics[width=0.44\linewidth]{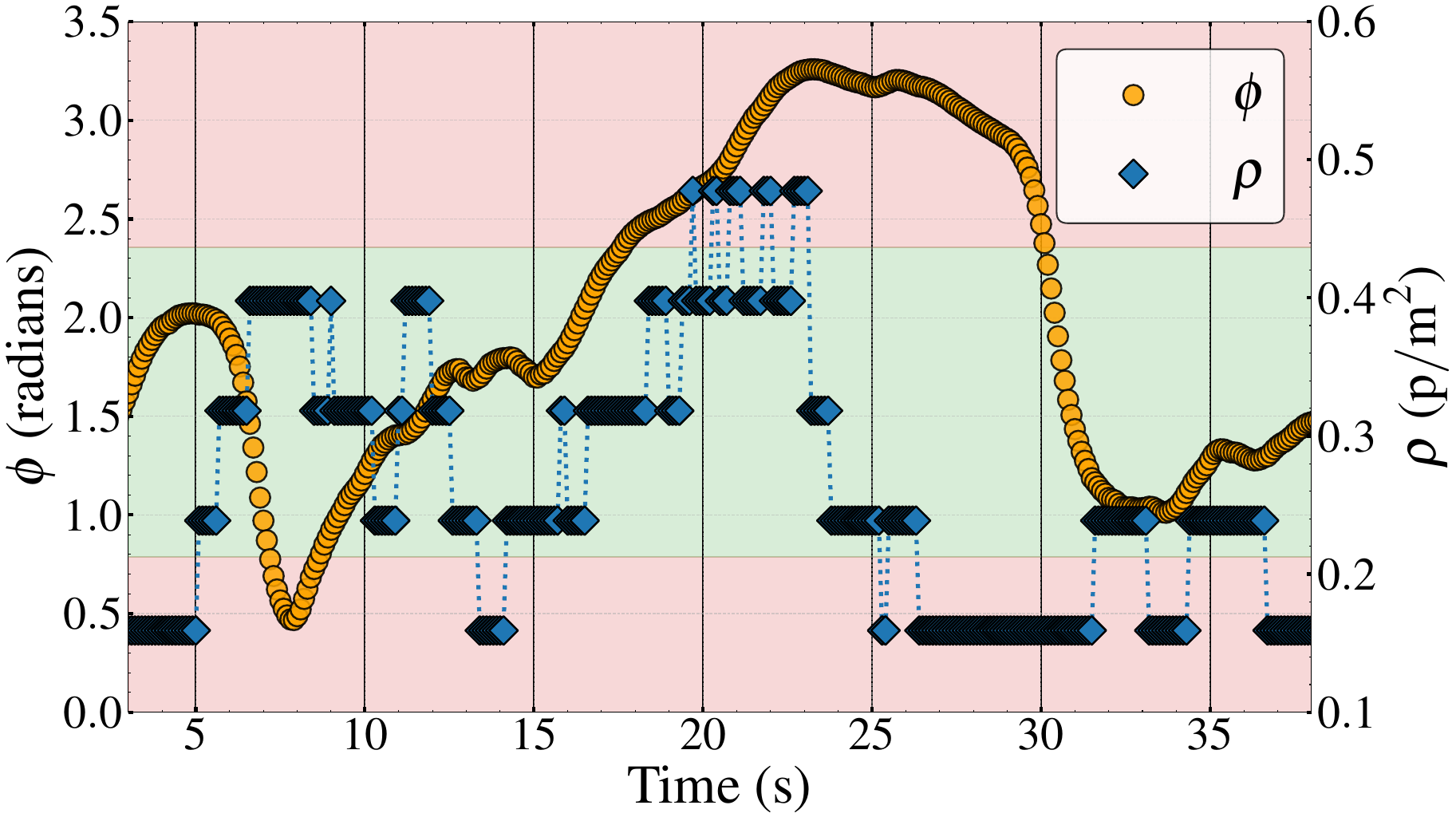}{}}{d}{-0.4cm}{-0.04cm}
		\caption{Example dyad trajectory and definition of dyad observables.
			(a) Subsection of the monitoring area at Eindhoven Central Station platform 3-4 (full monitoring area: $80\,\mathrm{m}\times 10\,\mathrm{m}$ displayed in the top left, with the subsection highlighted by the red box) with an example trajectory (red and orange lines) of a dyad entering from the bottom left and moving toward the top right of this domain.
			The red and orange markers show the positions of the dyad members at $5~\mathrm{s}$ intervals.
			Other pedestrians' positions (blue) are from a snapshot at time $20~\mathrm{s}$; at this time, the dashed circle marks a $2~\mathrm{m}$ proximity radius around the dyad center-of-mass $\rcom$ with pedestrians inside colored in green.
			(b) Observables in the dyad's frame of reference.
			The dyad's frame of reference is
			defined by the basis $(\epar, \eperp)$~(\refeqq{eq:dyadframe}),
			where the center-of-mass velocity $\vcom$~(\refeqq{eq:basicdyadstuff}) is parallel to $\vpar$.
			Besides $\vcom$,
			the interpersonal distance $\distance$, and the dyad angle $\dyadangle=-\operatorname{atan2}(\xr,\yr)$ are illustrated,
			where $\operatorname{atan2}$ is the two-argument arctangent function.
			Considered observables that depend on crowd members within the radius $R$ are their mean velocity $\vprox$ and local density~$\density$~(\refeqq{eq:density}), here  $\density = 0.56\,\densityunit$.
			(c) Dyad-crowd interaction regimes $\dyadbins$ (\refeqq{eq:binningdef}) defined depending on $\vprox$ and the angle $\dyadproxangle$ between $\vcom$ and $\vprox$. 
			We consider dynamics only in free-flow (no crowd), co-flow, counter-flow, and moving through a standing crowd.
			(d) $\dyadangle$ and $\density$ over time as the dyad in (a) traverses the domain, highlighting the abreast (red) and in-file (green) regions.
			The vertical lines correspond to time snapshots in (a).
			Between $t=15\,\mathrm{s}$ and $t=35\,\mathrm{s}$ the dyad enters a region of higher $\density$, temporarily going in in-file formation before going back to an abreast formation.
		}\label{fig1}
\end{figure*}

In this study, we investigate and characterize probabilistically how a crowd surrounding a dyad determines its kinematics, e.g., dyad velocity, relative distance between the dyad components, and, in general, the spatial formation.
We show how the crowd exerts different effects depending on its density and the relative flow with respect to the dyad, considering counter-flowing, standing, and co-flowing crowds.  
Respectively, these are cases in which the crowd velocity is opposed, vanishes, and has the same orientation with respect to the dyad velocity.
For this purpose, we establish an unprecedented dataset  composed of $\approx 6\mathrm{M}$~%
dyads that have been part of the highly variable daily crowd traffic at a train platform in Eindhoven,  The Netherlands (cf. \reffig{fig1}a).
Specifically, we derive a comprehensive set of fundamental diagram-like relationships,
revealing how dyad configuration is modulated by the surrounding crowd.
To move beyond descriptive statistics and enable predictive modeling, we introduce a scalar indicator, which we dub \emph{Orientation Log-Odds} (OLO), quantifying the relative probability that a dyad is in an abreast vs. in-file configuration.
Not only do we use the OLO indicator to provide a probabilistic characterization of how formation is influenced by crowd variables and dyad velocity, but we also show that this indicator can be phenomenologically modeled, with high accuracy, as a product of fundamental diagram-like relations connected with local minima and maxima of the formation probability.

\noindent We constructed our dyad dataset by automatically identifying \amountdyads~dyads throughout the exhaustive tracking data collected at Eindhoven train station between March 2021 and March 2024 (avg. \avgTravlers~trajectories per day). To achieve scalable dyad identification, we build on  the graph-based method in~\cite{pouw_monitoring_2020} and efficiently classify groups based on the joint time-consistency of distance.
Overall, our computational and modeling framework allows us, for the first time, to provide accurate phenomenological models for the dyad behavior  (velocity and configuration) across the full spectrum of crowd densities and flow regimes observed: free flows, co-, counter-flows, and flows through standing crowds.

The remainder of the paper is structured as follows:
in~\refsec{sec:StatsAndDetection}, we describe our dataset and analysis pipeline. We introduce the large-scale pedestrian trajectory dataset from Eindhoven Central Station, define the coordinate system and key observables, and detail the unsupervised dyad detection methodology.
In~\refsec{sec:analyses}, we present a comprehensive empirical characterization of dyad dynamics. This includes an analysis of fundamental diagram relationships across various regimes, and conditional analyses of distance and positions. We also establish our \emph{Orientation Log-Odds} (OLO), a scalar indicator for the probability of the dyad configuration (abreast vs. in-file).
In~\refsec{sec:usingOLO}, we characterize and model OLO behavior across crowd conditions of increasing complexity, from free-flow conditions, progressing to static crowds,  and finally a general scenario including co- and counter-flows. The  discussion in~\refsec{sec:discussion} closes the work.

\section{Dataset, statistical observables, and unsupervised dyad identification}\label{sec:StatsAndDetection}

This study considers the dynamics of dyads walking at Eindhoven Central Station (the fifth-largest station in the country) on platform 3-4.
We leverage an anonymous trajectory dataset of 900 days collected over a three-year period from March 2021 to March 2024 (partially publicly available at~\cite{zenodo}; cf. other usages, e.g., in~\cite{pouw2024data,pouw2021benchmarking,minartz2025discovering}). The measurement setup performed 24/7 anonymous crowd tracking over an $80\,\mathrm{m} \times 10\,\mathrm{m}$ area, with  $10\,\mathrm{Hz}$ time resolution, via a grid of overhead sensors (cf. \reffig{fig1}, see~\cite{pouw2024data,pouw2021benchmarking} for the methodology), outputting positional data on an individual basis.
The platform is mostly unobstructed, except for a central indoor waiting area and a few benches near the right end of the observable area. The indoor waiting area effectively narrows the free walking corridor to approximately $4\,\mathrm{m}$ in the central section of the platform (full overhead image top left of~\reffig{fig1}a).
The dataset includes about \avgTravlers~pedestrian trajectories per weekday across various density levels, typically ranging from $0$ to $1.3~\densityunit$, with occasional outliers reaching up to $2~\densityunit$.
To filter out measurement noise, each trajectory is smoothed using a Savitzky-Golay filter, and the first and last two seconds are excluded to avoid artifacts from partially observed motion at the edges of the sensing area.

To identify, parametrize, and analyse the state of dyads, we focus on four key observables that proxy the dyad's local dynamics and the hydrodynamic/macroscopic state of the surrounding crowd.
Note that the data's anonymous nature (trajectory-only) constrains our parameters to kinematic observables, making our choice possibly the simplest. Postponing formal definitions to the next \refsec{subsec:KeyAndCoord},  for each dyad, at each time instant, we consider the vector state
\begin{equation}\label{eq:masterpar}
	(\rloc{},\,\vcom,\,\density,\,\vprox),
\end{equation}
where
\begin{itemize}
	\item $\rloc{}$ [$m$] is the relative position of the dyad members in the coordinate system of the center-of-mass;
	\item $\vcom$  [$m/s$] is the dyad velocity, in terms of its center-of-mass;
	\item $\density$ [$p/m^2$] is the  local density of the crowd surrounding the dyad;
	\item $\vprox$ [$m/s$] is the (hydrodynamic) velocity of the crowd surrounding the dyad.
\end{itemize}
We will work with derived quantities that are easy to interpret, ranging from the distance between dyad members to their classification within specific flow regimes.
Specifically, our characterization will be probabilistic, and we will work with suitable conditioned (marginal) probability or conditional statistics of the law
\begin{equation}
	\Prob(\rloc{},\,\vcom,\,\density,\,\vprox),
\end{equation}
which we empirically estimate through our dataset. Moreover, we will use the symbol $\param$ to denote some generic conditioning variables (either in general or section-specific).

The observables in \refeqq{eq:masterpar} are also instrumental for the unsupervised identification of dyads in anonymous data, which is the focus of \refsec{subsec:dyad-detection}.

\subsection*{Ethics statement}
This study was approved by the Ethical Review Board of Eindhoven University of Technology (ref. ERB2020AP1; 21 February 2020). The research relies on pedestrian trajectory data collected via overhead sensors designed to ensure anonymity, with part of the dataset already publicly available. Consequently, no personally identifiable information or additional personal attributes were used or stored. 

\subsection{Observables and coordinate system}\label{subsec:KeyAndCoord}

\paragraph{Kinematic state of a dyad.}
Given a dyad of two pedestrians, henceforth indexed with 1 and 2, with instantaneous positions $(\rl_1,\rl_2)$ and velocities $(\vl_1,\vl_2)= (\dot \rl_1,\dot \rl_2)$, we adopt a co-moving coordinate system centered on the dyad's center-of-mass (see \reffig{fig1}b), in the same spirit as~\cite{zanlungo_potential_2014} by Zanlungo et al.
The center-of-mass position $\rcom$ and velocity $\vcom$ are
\begin{equation}
	\rcom = \frac{\rl_1+\rl_2}{2}, \qquad\qquad
	\vcom = \dotrcom = \frac{\vl_1+\vl_2}{2} = \norm{\vcom}\epar,
	\label{eq:basicdyadstuff}
\end{equation}
where the last equality is expressed in terms of the local basis
\begin{equation}
	\epar=\frac{\vcom}{\norm{\vcom}}, \qquad \qquad \eperp = R_{90^\circ}\epar,
	\label{eq:dyadframe}
\end{equation}
in which $\epar$ is parallel to $\vcom$ and $\eperp$ is perpendicular (with $\norm{\cdot}$ being the Euclidean norm).
This basis is ill-defined in case a dyad is standing, since small instantaneous fluctuations (due to body oscillation or measurement noise) map into erratic $\epar$ directions. Since we are interested in walking dyads, we set a ``standing'' velocity threshold at $\vstandingthresh = 0.4\,\mathrm{m/s}$, and require that the walking speed
\begin{equation}
	\comspeed = \norm{\vcom}
\end{equation}
satisfies
\begin{equation}
	\comspeed > \vstandingthresh.
\end{equation}
Note that we choose $\vstandingthresh$ between the standing ($\comspeed = 0\,\mathrm{m/s}$) and slow-walking ($\comspeed \approx 0.6\,\mathrm{m/s}$) peaks in the velocity distribution (see \refSIsec{ap:whystanding}). 
In this basis, which is depicted in~\reffig{fig1}b, the relative positions of the dyad members can be rewritten as
\begin{equation}
	\label{eq:rloc-def}
	\rloc{1} = \xr\epar + \yr\,\eperp,
	\qquad
	\rloc{2} = -\rloc{1} = -\xr\epar - \yr\,\eperp,
\end{equation}
where $(\xr, \yr)$ are the components of the relative position longitudinal and transverse to $\epar$, respectively. %
In this reference, the distance, $d$, between the dyad members can be conveniently written as
\begin{equation}\label{eq:def-distance}
	d = 2 \|\rloc{1}\| = 2 \sqrt{\xr^2 + \yr^2}.
\end{equation}

The $(\xr, \yr)$ coordinates allow us to establish a quantitative definition for abreast and in-file dyad formation, which we regulate through an even partition of the $(\xr, \yr)$ plane into two cones of identical area, $\stateslr$ and $\statesud$, bordered
by the $45^\circ$ bisectors. %
In formulas, we define
\begin{align}
	\text{Abreast} \quad & \Leftrightarrow \quad (x_r, y_r) \in \stateslr = \{(x_r, y_r) \mid x_r^2 - y_r^2 \ge 0\},\label{eq:statelr} \\
	\text{In-file} \quad & \Leftrightarrow \quad (x_r, y_r) \in \statesud = \{(x_r, y_r) \mid x_r^2 - y_r^2 < 0\}.\label{eq:stateud}
\end{align}
In other terms, we name abreast any configuration in which the angle between the relative position and the velocity, $\phi = \arctan(y_r/x_r)$, is larger than $45^\circ$, and in-file otherwise. Note that the abreast and in-file regions of the $(\xr,\yr)$ plane are identical but a $\pm 90^\circ$ rotation.

\paragraph{Hydrodynamic state of the crowd surrounding a dyad.} To parametrize the state of the crowd surrounding a dyad, we consider local estimates of the density $\density$ and of the crowd velocity $\vprox$  within a radius of $\closeradius = 2\,\mathrm{m}$ around the dyad center-of-mass $\rcom$.
We chose \(R = 2\,\mathrm{m}\) such that the individual level-of-service A~\cite{Fruin1971}  (low densities) for each member can be fully resolved within the circle, even under in-file configurations with maximal interpersonal distance \((\approx 1\,\mathrm{m})\).

Specifically, we define the local density as
\begin{equation}
	\density = \frac{\closecount}{\pi \closeradius^2},
	\label{eq:density}
\end{equation}
where $\closecount$ denotes the total number of pedestrians within the proximity radius (including the dyad members).
By extension, the density for a dyad in free flow is $\densityfreeflow = \density(\closecount\,=\,2)  = 0.16\,\densityunit$.
The corresponding proxy crowd velocity is given by the mean velocity of the pedestrians in this region,
\begin{equation}
	\vprox = \frac{1}{\proxcount} \sum_{\substack{i=1\\i\notin \text{dyad}}}^{\proxcount} \mathbf{v}_i,
\end{equation}
where $\proxcount = \closecount - 2$ represents the number of pedestrians within the proximity radius, excluding the dyad members themselves.

The two velocities $\vcom$ and $\vprox$ allow isolating different regimes of relative motion between the dyad and the crowd, including, but not limited to, flow through a standing crowd, co-flow, and counter-flow. For this purpose, let $\dyadproxangle$ be the oriented angle between these two velocities, i.e.
\begin{equation}
	\dyadproxangle = \measuredangle(\vcom, \vprox)  = \operatorname{atan2}\left(\norm{\vcom \times \vprox}, \vcom \cdot \vprox\right), \quad -\pi \leq \dyadproxangle \leq \pi,
\end{equation}
where $\operatorname{atan2}(\cdot)$ is the two-argument arctangent function (positive for counterclockwise angles from $\vcom$ and negative otherwise).
We use the angle $\dyadproxangle$  together with the velocity magnitude  $\|\vprox\|$ to exhaustively classify the dyad-crowd dynamics into five non-overlapping regimes, also depicted in \reffig{fig1}c:
\begin{equation}
	\dyadbins(\dyadproxangle, \|\vprox\|) =
	\begin{cases}
		\binfree     & \text{if } \proxcount=0                                                                      \\
		\binstanding & \text{if } \|\vprox\| < \vstandingthresh = 0.4\,\textrm{m/s}                                 \\
		\binwith     & \text{if } \|\vprox\| \geq \vstandingthresh\, \wedge |\dyadproxangle| < \pi/4                \\
		\binagainst  & \text{if } \|\vprox\| \geq \vstandingthresh\, \wedge |\dyadproxangle| > 3\pi/4               \\
		\binacross   & \text{if } \|\vprox\| \geq \vstandingthresh\, \wedge \pi/4 \leq |\dyadproxangle| \leq 3\pi/4
	\end{cases}\label{eq:binningdef}.
\end{equation}
$\binfree$ identifies the free flow regime, in which the dyad is surrounded by no other pedestrian within radius $R$. 
We detect a crowd standing around the dyad by limiting $\|\vprox\|$ by the previously defined threshold $\vstandingthresh$.
When $\|\vprox\| > \vstandingthresh$, we consider the crowd to be moving, which we further separate, based on $\dyadproxangle$, into co-flowing, counter-flowing, and cross-flowing regimes. Also, in this case, we use the  cones along the $45^\circ$ bisectors as boundaries.

In the following, we opt to focus on free-flow, standing, co- and counter-flow regimes as prototypes of the interaction scenarios, and neglect the many corner cases of the cross-flow regimes in which a  dyad traverses (or is traversed by) a crowd moving sideways relative to its own velocity. Moreover, the long and narrow geometry of the platform ($80\,\mathrm{m}\,\times\,10\,\mathrm{m}$) suppresses cross-flow, as pedestrians tend to move along the platform rather than across it.

Within the considered discrete regimes $\dyadbins$ (\refeqq{eq:binningdef}),  in the cases $\binfree$ and $\binstanding$ -- \refeqq{eq:masterpar} reduces to two variables, $(\rloc{},\,\vcom)$, and three variables, $(\rloc{},\,\vcom,\,\density)$, respectively -- whereas for $\binagainst$ and $\binwith$ \refeqq{eq:masterpar} retains the full four-parameter complexity. We introduce, therefore, a final dimensionless velocity ratio that quantifies the projection of the crowd velocity onto the dyad's direction of motion for non-standing crowds, which approximates both the relative velocity of the dyad and the crowd and proxies the regimes, and which we will use to parametrize continuously between these flow states:
\begin{align}
	\vrelpar = \frac{\vprox \cdot \vcom}{\norm{\vcom}^2}
	= \frac{\norm{\vprox}}{\norm{\vcom}} \cos\dyadproxangle
	\label{eq:velpar} \qquad \text{for } \norm{\vprox} > \vstandingthresh
\end{align}
This scalar quantity has two key properties:
\begin{enumerate}
	\item the sign of $\vrelpar$ discriminates among relative flow regimes of dyad and crowd  and provides a continuous proxy that largely corresponds to the  regimes in Eq~\refeq{eq:binningdef},	specifically,
	      \begin{itemize}
		      \item when $\vrelpar > 0$, the dyad and crowd move in the same direction (co-flow), with  $\vrelpar \in (\vstandingthresh, \infty)$. This primarily captures $\binwith$;
		      \item when  $\vrelpar < 0$, they move in opposite directions (counter-flow) with  $\vrelpar \in (-\infty, -\vstandingthresh)$. This primarily captures $\binagainst$.
	      \end{itemize}
	\item The magnitude of $\vrelpar$ scales as the speed ratio:  when $\vprox$ and $\vcom$ are parallel  ($\cos \dyadproxangle = 1 \Rightarrow  |\dyadproxangle| \in \{0,\pi\}$), $\vrelpar$ directly represents the crowd-to-dyad speed ratio.
	      For example, $|\vrelpar| = 1$ indicates matched speeds (dyad and crowd move at the same speed, either in co- or counter-flow), $\vrelpar = 2$ means the crowd moves twice as fast as the dyad, and so on.
	      Moreover, since $\norm{\vprox} > \vstandingthresh$, $|\vrelpar|$ can be small only if $\norm{\vcom}$ is large (i.e., the dyad is faster than the crowd).
	      Finally, any incidence angle $\dyadproxangle$ such that $|\cos \dyadproxangle| < 1$ reduces the magnitude of $\vrelpar$. 
		  Note that the reduction is at most  $29\%$, as we neglect \(B_{\text{crossflow}}\), which imposes \(|\cos \dyadproxangle| > \sqrt{2}/2 \approx 0.71\), ensuring that \(\vrelpar\) remains a reliable proxy for the speed ratio
\end{enumerate}

\subsection{Unsupervised detection of dyads}\label{subsec:dyad-detection}
Here, we describe the approach for automatically identifying dyads in our dataset.
As we work with anonymous trajectory data,
the approach can rely only on trajectory properties, and heuristically classify dyads based on how mutual proximity is sustained over time.
The technical challenge is distinguishing between genuine dyads and trajectories that came into contact by chance.
We combine two previously proposed approaches~\cite{corbetta-pre-2018, pouw_monitoring_2020} and model pedestrians in a crowd as nodes in a graph. 
Edges carry information about guessed pairwise interactions or lack thereof. Conceptually, dyads will be connected components of size 2 of this graph. %

More formally, let us denote by $\traj_i\anytime$ the trajectory of pedestrian $i$, defined between their entry and exit times, say, the interval $\trajtime_i = \trajinterval$. 
Hence, $\{\traj_i\anytime, i=1,\ldots \}$ is the set of all observed trajectories, i.e., the nodes of our graph.

Let $\mathcal{E}_0$ be the edge set including all pedestrian pairs  appearing together in our measurement area in at least one frame, and thus potentially part of a group, i.e.,
\begin{equation}
    \mathcal{E}_0 = \left\{ (\traj_i\anytime, \traj_j\anytime),\; i > j \;\middle|\;
    \trajtime_i \cap \trajtime_j \neq \emptyset \right\}.
\end{equation}
Since we focus on walking dyads (cf. \refsec{sec:StatsAndDetection}), we introduce the time domain in which both pedestrians in (any) pair in $\mathcal{E}_0$ are walking, i.e., when both move faster than the standing threshold $\vstandingthresh$. This gives the ``walking time domain'' of a pair $(\traj_i\anytime, \traj_j\anytime)$, $\tilde{\trajtime}_{ij}$:
\begin{equation}\label{eq:walkingdomain}
    \tilde{\trajtime}_{ij} = \bigl\{\, t \in \trajtime_i \cap \trajtime_j \;\big|\;
    \|\mathbf{v}_i(t)\| > \vstandingthresh,\
    \|\mathbf{v}_j(t)\| > \vstandingthresh \,\bigr\}.
\end{equation}
In the same time interval, we also compute the average distance of the pair (cf. \refeqq{eq:def-distance}), %
\begin{equation}
    \langle d_{ij} \rangle_t = \langle d_{ij}\anytime \rangle_t = \frac{1}{|\tilde{\trajtime}_{ij}|}\int_{\tilde{\trajtime}_{ij}} d_{ij}(t)\,dt.
\end{equation}
To ensure proximity while walking, generalizing the approach in~\cite{pouw_monitoring_2020}, we retain pairs in $\mathcal{E}0$ who walk for at least a time interval $\tilde{t}{min}$ and have an average distance below a threshold $\distanceThreshold$.
In formulas, this identifies the subset of edges
\begin{equation}
    \mathcal{C} = \left\{ (\traj_i\anytime, \traj_j\anytime) \in \mathcal{E}_0 \;\middle|\;
    \begin{aligned}
        & |\tilde{\trajtime}_{ij}| > \tilde{t}_{min}  \\
        &\langle d_{ij} \rangle_t < \distanceThreshold
    \end{aligned}
    \right\}%
    \quad
    \begin{aligned}
      & \tilde{t}_{min} = 1.5~\mathrm{s}\\
      &\distanceThreshold = 1.5~\mathrm{m}.
    \end{aligned}
    \label{eq:connections}
\end{equation}
In other words, $\mathcal{C}$ is the subset of edges connecting  pedestrians that appear systematically together and maintain proximity. According to our heuristics, such pedestrians belong to the same group.

Conceptually, as in~\cite{corbetta-pre-2018}, we isolate dyads by retaining only connected components in the graph with size two. Practically, to compensate for possible imperfections in our dataset, such as tracking ID changes%
, this dyad selection step entails additional technical procedures that we detail in \refSIsec{app:dyad-detection}.

Finally, we discard exceptionally short trajectories: for all subsequent analyses, we only keep dyads that are in the scene together for $t_{min}$ and are walking for at least $\hat{t}_{min}$, resulting in the final dyad set
\begin{equation}\label{eq:dyadset}
    \mathcal{D} = \left\{ (\traj_i\anytime, \traj_j\anytime) \in \mathcal{C} \;\middle|\;
    \begin{aligned}
        &|\trajtime_{ij}| > t_{\min} \\
        &|\tilde{\trajtime}_{ij}| > \hat{t}_{min}
    \end{aligned}
    \right\}
    \qquad
    \begin{aligned}
      &t_{\min}      = 8.0~\mathrm{s}\\
      &\hat{t}_{min} = 4.0~\mathrm{s},
    \end{aligned}
\end{equation}
with, again, a few technical steps detailed in \refSIsec{app:dyad-detection}.

We validated  the detection via qualitative visual inspection (hundreds of randomly sampled detections confirmed co-walking behavior via animated scene replays). Moreover, in \refSIsec{ap:sensitivity} we report on the robustness of the method through sensitivity analyses with respect to the algorithm parameters.  
It is worth mentioning that requiring dyads to have velocity correlation above a threshold
(conjecturing that a pair walking together should have velocity fluctuations in phase/counter-phase)
makes no practical difference at our scale (while possibly being detrimental in case of smaller sets:
dyads could have velocity fluctuations in counter-phase, or decorrelated at slow center-of-mass
velocity).

\section{Large-scale empirical analysis of dyad configuration dynamics and crowd-dependent formation changes}\label{sec:analyses}
We consider here how dyad configurations depend probabilistically on various crowd conditions, on the basis of our multi-year dataset (\refsec{sec:StatsAndDetection}).
Section \refsec{subsec:FundChar} presents our analysis, where we focus on conditional dependencies in the four-dimensional state vector $(\rloc{},\,\vcom,\,\density,\,\vprox)$ (\refeqq{eq:masterpar}).
Increasing microscopic detail throughout, we consider:
\begin{itemize}[leftmargin=0pt,label={},labelsep=0pt,nosep]
	\item (i)~\textit{Fundamental diagrams} ($\mathbb{E}[\comspeed | \density,\,\rloc{},\, \dyadbins]$): speed-density relationships that characterize dyad behavior across $\dyadbins$ regimes, revealing systematic differences in speed across configurations and densities.
	\item (ii)~\textit{Configuration heatmaps} ($\Prob(\xr, \yr \mid \comspeed,\,\density,\,\dyadbins)$): probability distributions showing how dyad configurations vary with speed, density, and flow regime, revealing a transition from abreast to in-file governed by both density and flow regimes.
	\item (iii)~\textit{Relative-position diagrams} ($\mathbb{E}[\density | \rloc{}]$ and $\mathbb{E}[\comspeed | \rloc{}]$): conditional expectations of density and dyad speed as functions of relative position, revealing the typical crowd conditions and velocities associated with each $\rloc{}$.
	\item (iv)~\textit{Interpersonal distance analysis}: a distance-density relationship quantifying how modal (i.e., most frequent) distances between dyad members depend on density and configuration, revealing distinct trends for abreast and in-file arrangements.
\end{itemize}

Finally, in \refsec{subsec:Pi} we reduce the relative likelihood of each configuration in parameter space to a scalar measure  -- which we dub Orientation Log-Odds (OLO) -- that captures the preference between abreast versus in-file configurations, enabling quantitative modeling of dyad formation dynamics.

\subsection{Analysis of configuration-dependent dyad behavior in varying environment}\label{subsec:FundChar}
\begin{figure}[h]
		\centering
				\figinnerlab{
		\includegraphics[width=0.38\linewidth]{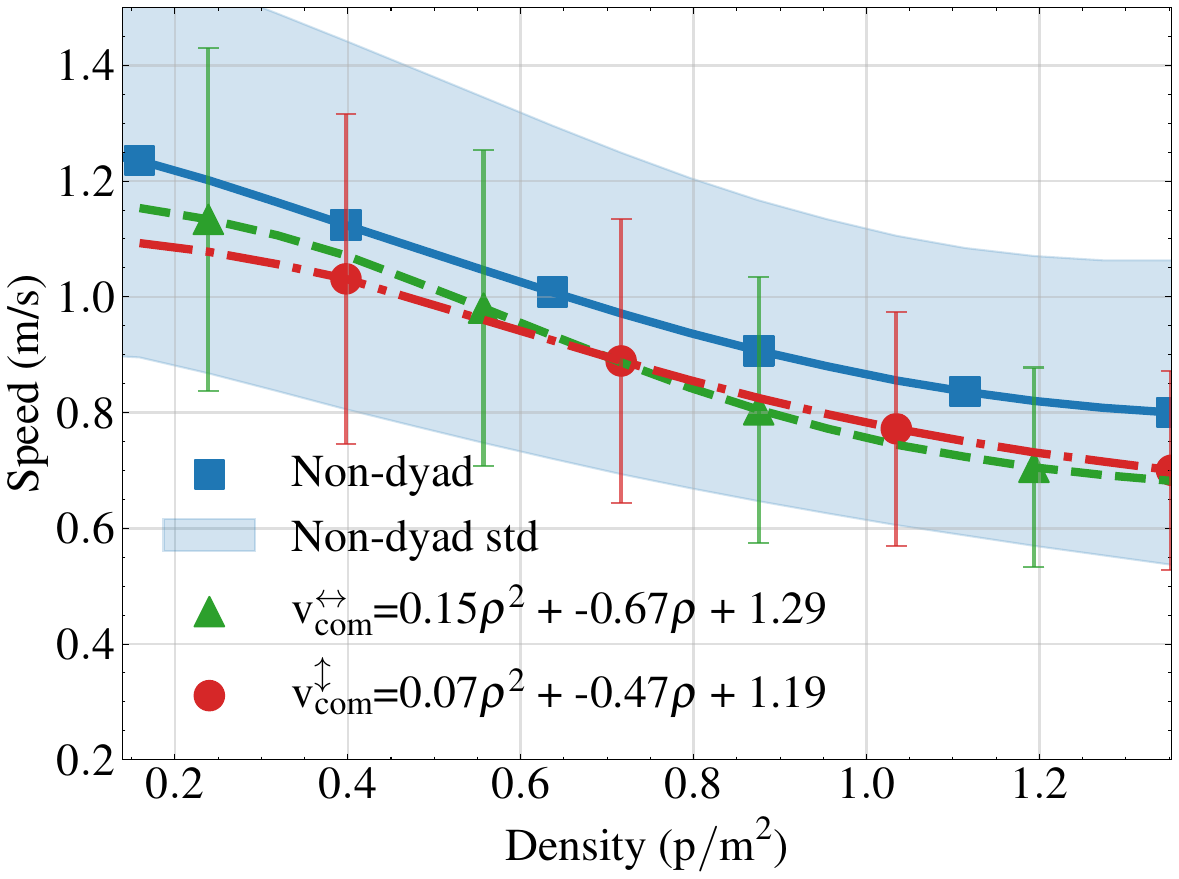}}{a}{-0.3cm}{-0.4cm}
		\figinnerlab{
		\includegraphics[width=0.38\linewidth]{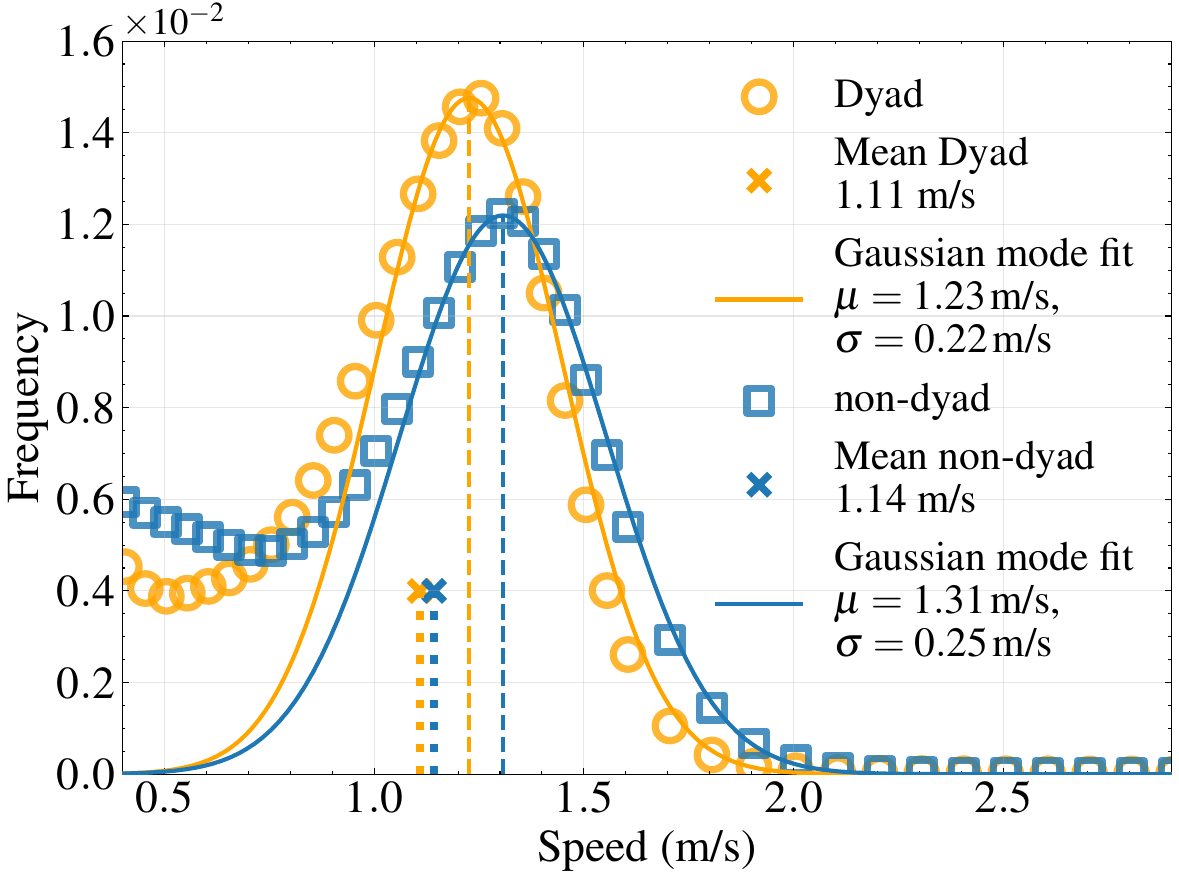}}{b}{-0.3cm}{-0.4cm}
		\figinnerlab{
		\includegraphics[width=0.38\linewidth]{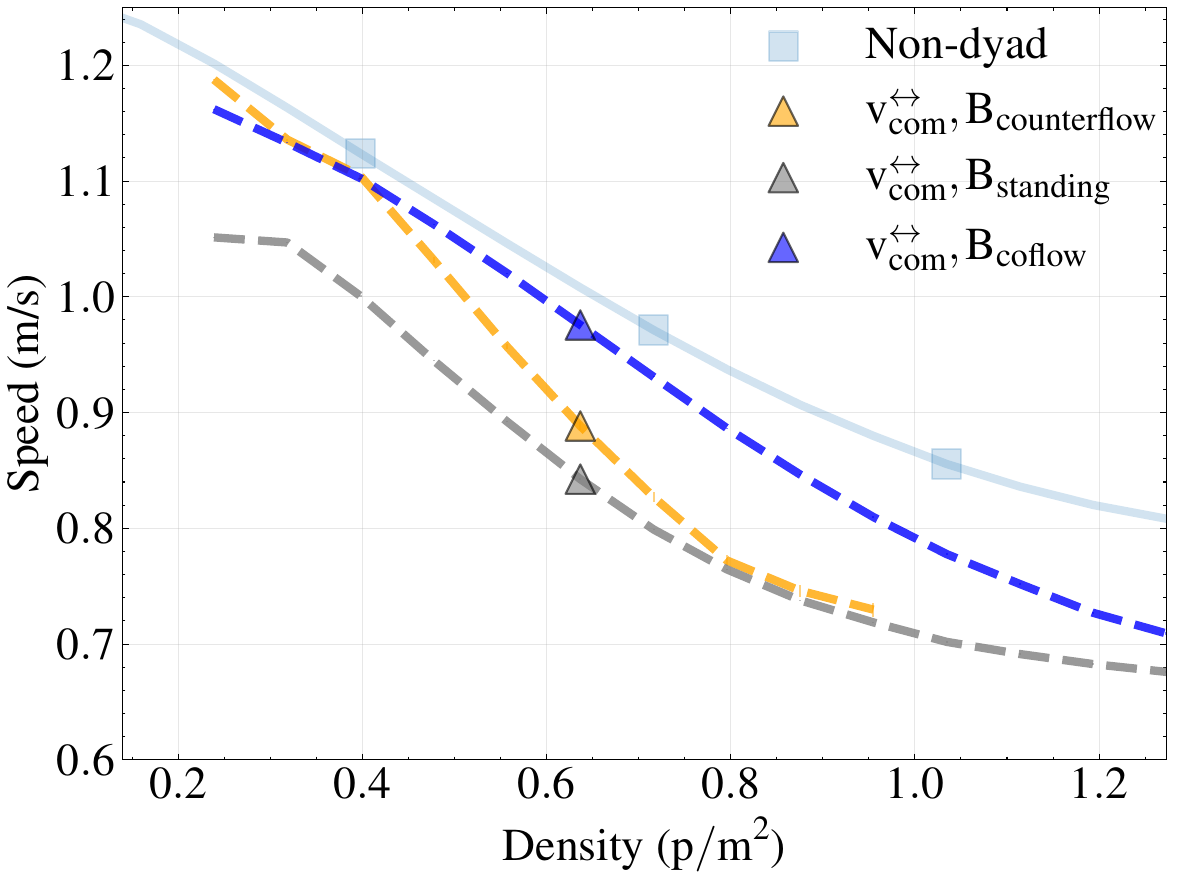}}{c}{-0.3cm}{-0.4cm}
		\figinnerlab{
		\includegraphics[width=0.38\linewidth]{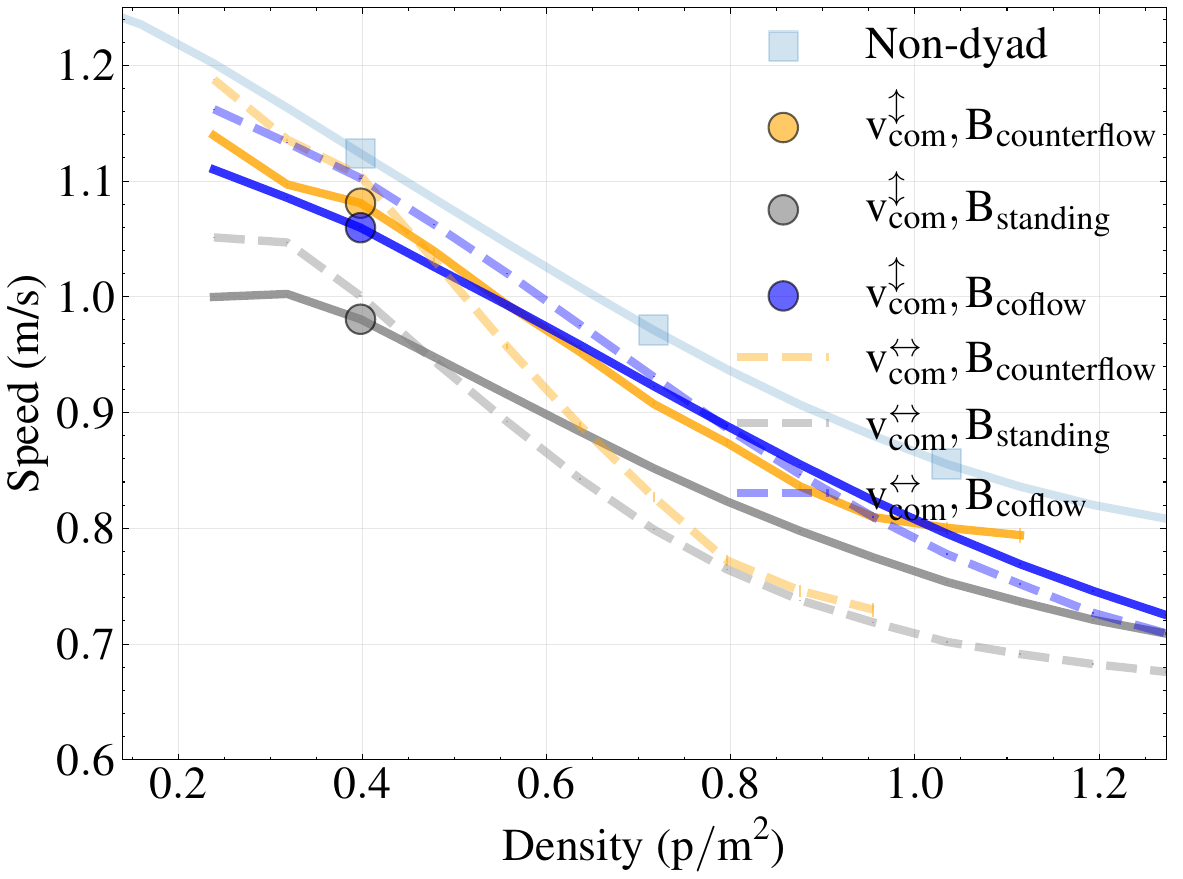}}{d}{-0.3cm}{-0.4cm}
		\caption{%
			Dependency of dyad speed, $\comspeed$, on crowd density, $\density$, for abreast (triangles) and in-file (circles) configurations.
			(a) Fundamental diagram ($\mathbb{E}[\comspeed | \density,\,\rloc{}]$):  mean velocity and standard deviation vs. local density for non-dyad crowd members (blue squares), abreast dyads $\comspeed^{\leftrightarrow}$ (\refeqq{eq:speed_abreast}, green triangles), and in-file dyads $\comspeed^{\updownarrow}$ (\refeqq{eq:speed_infile}, red circles).
			(b) Walking speed probability density in free-flow conditions ($\mathbb{E}[\comspeed | \binfree]$) for dyad members (yellow circles) and non-dyad pedestrians (blue squares).
			Dashed vertical line: modal speed; dotted line with cross: mean speed; solid line: Gaussian fit to modal region. Both modal and mean speeds are slower for dyads vs. non-dyads.
			(c) Abreast dyad speeds conditioned on $\dyadbins$ \refeqq{eq:binningdef} ($\mathbb{E}[\comspeed | \dyadbins,\,\density]$). Note the different y-axis scale from (a).
			Co-moving dyads ($\binwith$, dashed blue) maintain the highest speeds, counter-flow dyads ($\binagainst$, dashed yellow) intermediate speeds, and dyads in stationary crowds ($\binstanding$, dashed gray) the lowest speeds.
			(d) In-file dyad speeds (solid lines) across the same $\dyadbins$ regimes as in (c); abreast cases, as in (c), are included to simplify comparison. The ordering of speeds among the different dyad types is the same as in (c).
			At low densities, in-file dyads are slower than abreast; this reverses at higher densities. Crossover occurs at $\density \approx 0.5\,\mathrm{p/m}^2$ for $\binstanding$ and $\binagainst$, and at $\density \approx 0.8\,\mathrm{p/m}^2$ for $\binwith$.
		}\label{fig2}
\end{figure}

\paragraph{Fundamental diagrams.}
To establish a baseline characterization of dyad configurations, we analyse speed-density relationships (i.e., fundamental diagrams, $\mathbb{E}[\comspeed | \density]$) and compare dyad and non-dyad pedestrian speeds across different interaction regimes.
In \reffig{fig2}a, we show the relationship between dyad speed, $\comspeed$, and density, $\density$, further conditioned on abreast ($\stateslr$, \refeqq{eq:statelr}) and in-file ($\statesud$, \refeqq{eq:stateud}) dyads.
In formulas, we consider the center-of-mass speeds for abreast and in-file configurations as
\begin{align}
	\comspeed^{\leftrightarrow,B}(\rho) & = \mathbb{E}[\comspeed \mid  (\xr, \yr) \in \stateslr,\rho,B ] \label{eq:speed_abreast} \\
	\comspeed^{\updownarrow,B}(\rho)    & = \mathbb{E}[\comspeed \mid (\xr, \yr) \in \statesud,\rho,B] \label{eq:speed_infile}
\end{align}
respectively.
This fundamental diagram demonstrates that dyads consistently move slower than non-dyad pedestrians across all density conditions, corroborating previous findings from controlled laboratory experiments~\cite{hu_social_2021,hu_experimental_2023} and smaller-scale empirical studies~\cite{zanlungo_spatial-size_2015}.
At low densities, abreast dyads move faster than in-file dyads, but this relationship reverses at higher densities ($\density > 0.7\,\densityunit$), indicating a density-driven configuration transition. We hypothesize that in-file movement becomes more efficient for navigating crowded environments at higher densities, whereas at lower densities the abreast configuration is preferred to facilitate communication~\cite{zanlungo_potential_2014}.

To quantify speed differences under uncongested conditions, \reffig{fig2}b compares free-flow speed distributions ($\mathbb{E}[\comspeed | \binfree]$).
Non-dyads exhibit both higher modal and mean speeds than dyads; the modal speed differs by approximately $0.1\,\mathrm{m/s}$, indicating slower dyad movement even in free flow.
While this plot suggests that the modal speed may better reflect intrinsic velocity preferences, we use the mean speed in subsequent analysis, as at higher densities the speed probability distribution becomes bimodal with a secondary peak near $\vstandingthresh$, complicating modal estimation.

We then examine how relative crowd motion modulates the relationship between in-file, abreast, and non-dyads, by analysing mean speeds across the three dyad-crowd interaction regimes (\reffig{fig2}c-d).
Both abreast and in-file configurations exhibit a consistent speed hierarchy across all densities:
dyads co-moving with the crowd ($\binwith$) maintain the highest speeds, those moving against the crowd ($\binagainst$) achieve intermediate speeds, while dyads navigating through stationary crowds ($\binstanding$) move slowest.
We hypothesize that this ordering is likely due to the increased \emph{navigational conflict} (i.e., the extent to which individuals must adjust their trajectories or speeds to avoid collisions) between dyad and crowd, specifically:
\begin{itemize}
	\item For $\binwith$, dyads are likely to benefit from aligned flow direction, reducing navigational conflicts and enabling higher speeds dictated by the general crowd motion.
	\item For $\binagainst$, both dyads and oncoming crowd members are likely aiding in resolving navigational conflict. This could explain the intermediate speeds compared to the other two bins.
	\item For $\binstanding$, we hypothesize that dyads must resolve all navigational conflict by themselves, with little aid from the crowd, resulting in the lowest speeds.
\end{itemize}
This hierarchy demonstrates that relative crowd motion, not density alone, fundamentally shapes dyad behavior.
Moreover, \reffig{fig2}d reveals that the crossover density at which in-file becomes faster than abreast is regime-dependent: it occurs at $\density \approx 0.5~\mathrm{p/m}^2$ for both $\binstanding$ and $\binagainst$, but it occurs at a substantially higher density, $\density \approx 0.8~\mathrm{p/m}^2$, for $\binwith$.
We hypothesize that this regime-dependent transition may be attributed to how in-file configurations reduce navigational conflicts: when dyads traverse stationary crowds or move against the flow, the streamlined in-file arrangement becomes advantageous at lower densities due to the higher frequency of avoidance maneuvers required.
In contrast, co-moving dyads experience fewer conflicts and can maintain the more comfortable abreast formation at higher densities before spatial constraints necessitate the transition to in-file.

\begin{figure}[h]
		\centering
				\figinnerlabLoop{%
		\begin{overpic}[width=.7\linewidth]{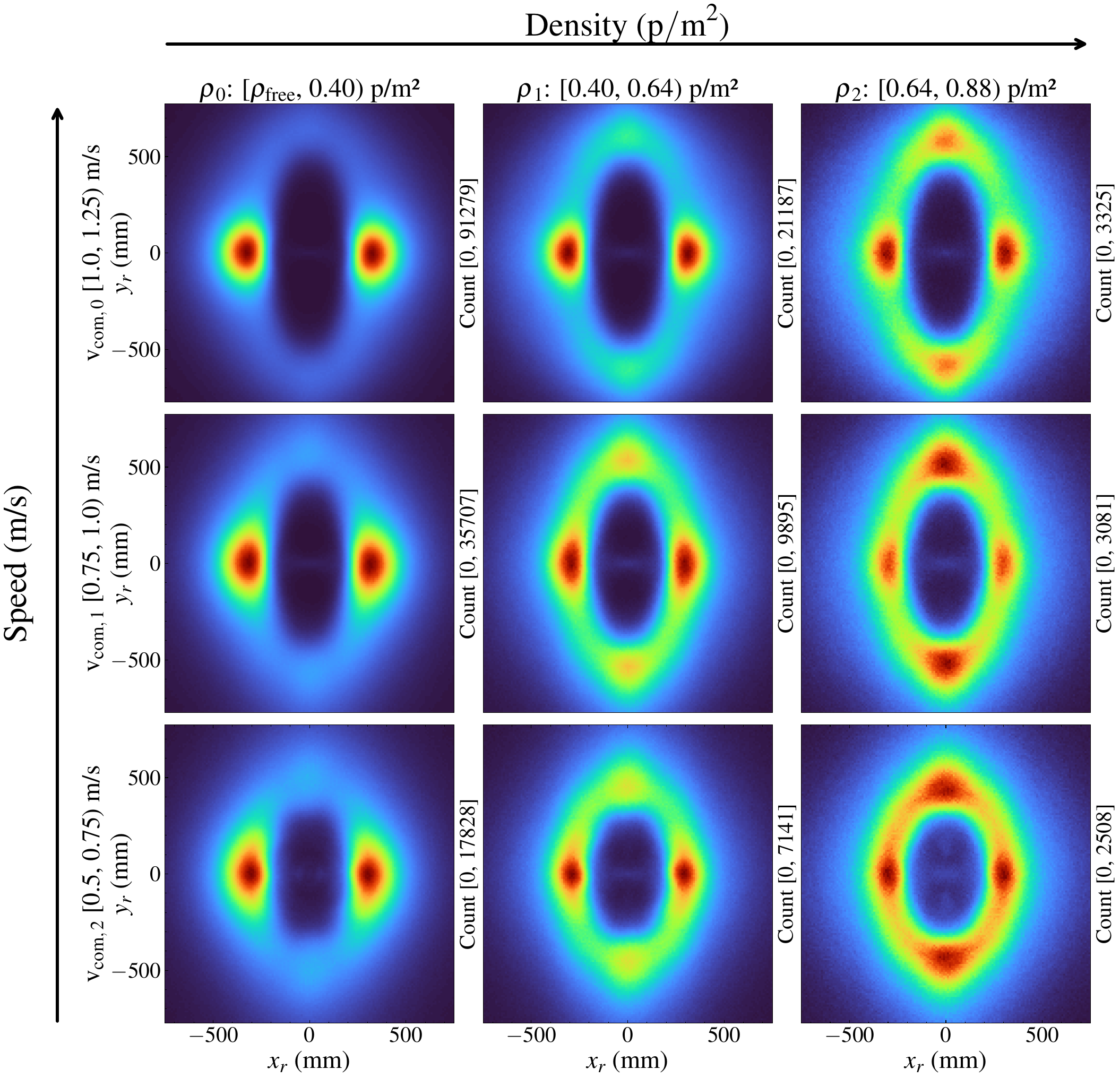}%
		\put(102,23){\includegraphics[width=.07\linewidth]{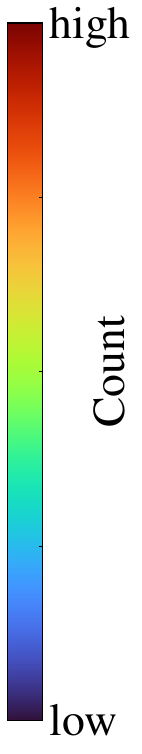}}%
		\end{overpic}}{%
		a/2.0cm/7.41cm,%
		b/5.42cm/7.41cm,%
		c/8.80cm/7.41cm,%
		d/2.0cm/4.1cm,%
		e/5.42cm/4.1cm,%
		f/8.80cm/4.1cm,%
		g/2.0cm/0.8cm,%
		h/5.42cm/0.8cm,%
		i/8.80cm/0.8cm
		}
		\caption{
			Probability density of dyad positions arranged in fundamental diagram format, showing how dyad configurations vary with density and speed.
			The colorization is scaled to the minimum and maximum values of the pdf, as indicated by Count[min, max] in each subplot.
			At low densities (left columns), abreast configurations near the $\xr$ axis dominate across all speed bins (rows).
			As density increases, a secondary peak emerges near the $\yr$ axis, corresponding to in-file configurations.
			This shift indicates that the likelihood of dyad configurations depends systematically on crowd density.
		}\label{fig3}
\end{figure}

\paragraph{Configuration heatmaps.}
To understand how dyads organize themselves during motion and under the influence of the surrounding crowd, we examine their internal spatial configuration using the probability distribution of relative positions $\Prob(\xr, \yr)$.
Specifically, we condition $\Prob(\xr, \yr)$ on speed and density, $\Prob(\xr, \yr \mid \comspeed, \density)$ (\reffig{fig3}).
The plots are arranged in a fundamental diagram format, with $\density$ increasing from left to right and $\comspeed$ increasing from bottom to top.
The heatmaps reveal two distinct most likely configuration states: abreast formations concentrated along the $\xr$ axis and in-file formations along the $\yr$ axis.
At low densities, abreast configurations dominate across all speeds, whereas at higher densities the probability of in-file arrangements increases.
Moreover, this visualization shows the transition from predominantly abreast configurations at low densities ($\density_0$, left column), to a higher chance of mixed configurations as density increases ($\density_2$, right column).
The coexistence of both configurations indicates a need for additional conditioning to isolate more distinct regimes.

\begin{figure}[h]
		\centering
				\figinnerlabLoop{%
		\includegraphics[width=.84\linewidth]{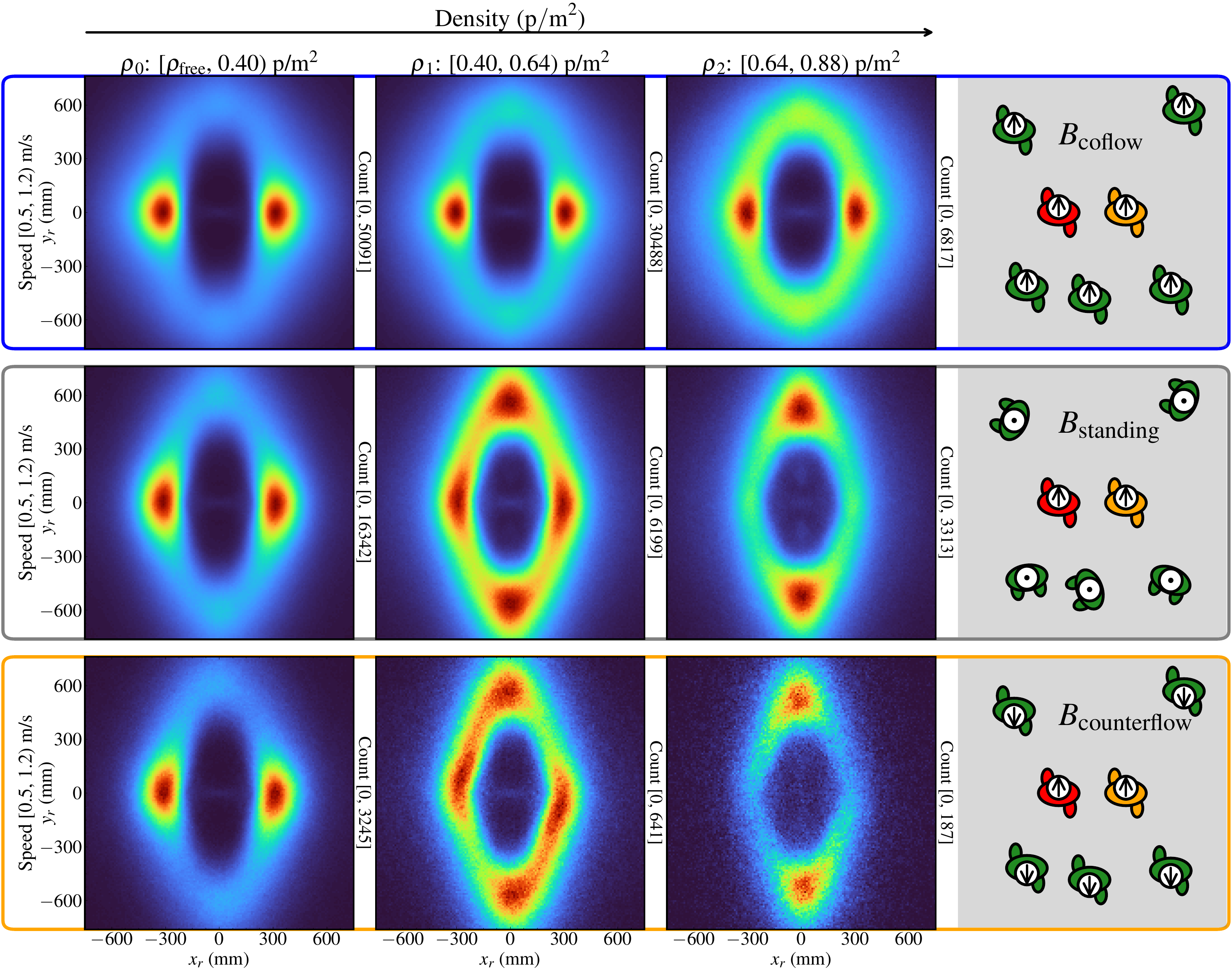}}{%
		a/1.25cm/7.46cm,%
		b/4.65cm/7.46cm,%
		c/8.02cm/7.46cm,%
		d/1.25cm/4.10cm,%
		e/4.65cm/4.10cm,%
		f/8.02cm/4.10cm,%
		g/1.25cm/0.70cm,%
		h/4.65cm/0.70cm,%
		i/8.02cm/0.70cm
		}
		\includegraphics[width=0.42\linewidth]{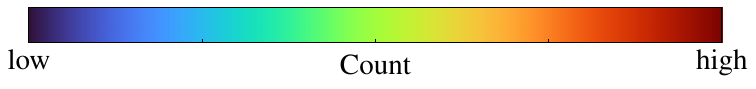}
		\caption{
			Probability density of dyad positions under different dyad-crowd interaction regimes $\dyadbins$ (\refeqq{eq:binningdef}), showing how dyad configurations depend on relative crowd motion and density.
			Minimum and maximum values are indicated by Count[min, max] in each subplot.
			Top row: $\binwith$ (co-moving), middle row: $\binstanding$ (stationary crowd), bottom row: $\binagainst$ (counter-flow).
			In contrast to \reffig{fig3}, this analysis shows that in-file configurations dominate at higher densities (right column) when dyads traverse standing crowds or move against the flow.
			At comparable densities, dyads moving with the crowd remain predominantly abreast, while those in standing or opposing crowds adopt in-file arrangements.
		}\label{fig4}
\end{figure}

To better isolate the factors driving configuration transitions observed in \reffig{fig3}, we refine our analysis by conditioning on both density $\density$ and dyad-crowd interaction regimes $\dyadbins$ (\refeqq{eq:binningdef}), and consider $\Prob(\xr, \yr \mid \dyadbins, \density)$.
This separates the effects of crowd density from those of relative motion direction, and  allows us to determine whether configuration changes are driven by more than just spatial constraints (i.e., $\density$).
In \reffig{fig4}, we present this analysis across three interaction regimes: $\binwith$ (co-moving, top row), $\binstanding$ (stationary crowd, middle row), and $\binagainst$ (counter-flow, bottom row), each plotted across increasing density levels from left to right.
This shows that in-file configurations become relatively more likely than abreast configurations under some conditions. Specifically, this occurs at higher densities when dyads traverse standing crowds (\reffig{fig4}f) or move against the crowd flow (\reffig{fig4}j).
Notably, at comparable density levels, dyads moving with the crowd maintain predominantly abreast configurations (\reffig{fig4}c), clearly distinguishing the configuration preferences that were instead mixed in \reffig{fig3}.
This finding demonstrates that the direction of crowd-dyad interaction, not just density alone, is a key determinant of configuration transitions.

\begin{figure}[h]
		\centering
				\figinnerlabB{
		\includegraphics[width=0.75\linewidth]{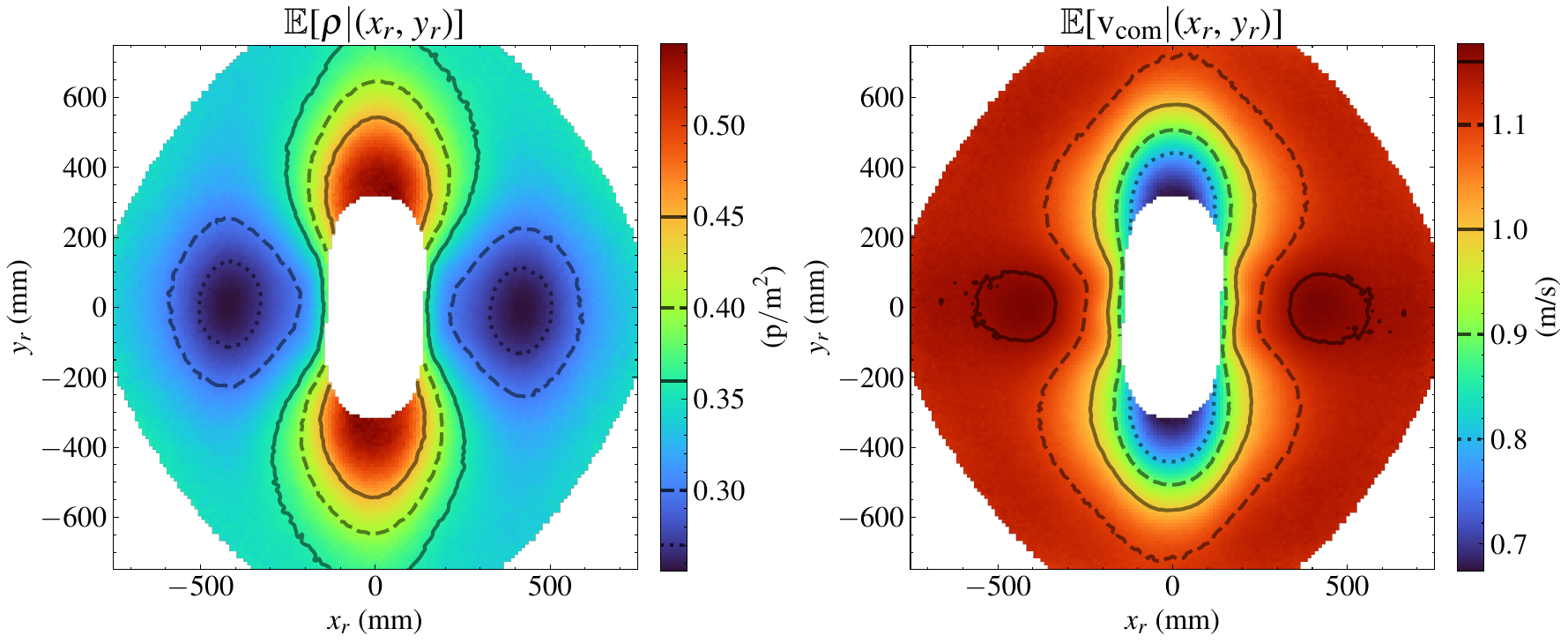}}{a}{-0.3cm}{-0.6cm}{b}{6.5cm}{0cm}
		\caption{
			Dependence of density $\density$ and dyad speed $\comspeed$ on the relative position $\rloc{}$.
			(a) Density-configuration ($\conddensity$) diagram indicating that abreast configurations occur predominantly at lower $\density$ than in-file configurations.
			(b) Speed-configuration ($\condcomspeed$) diagram showing that smaller interpersonal distances correspond to lower speeds in both configurations, with a stronger effect for in-file dyads.
			The highest expected $\comspeed$ is observed in a concentrated region centered along the $\xr$ axis (abreast configurations).
			Note that the point symmetry about the origin follows from the dyad symmetry (\refeqq{eq:rloc-def}): each pair contributes data at both $\rloc{}$ and $-\rloc{}$.
		}\label{fig5}
\end{figure}

\paragraph{Relative position fundamental diagrams.}
To examine how density and speed vary across relative positions $\rloc{}$ within the dyad's local coordinate frame, we build
on~\cite{crociani_micro_2019}, and leverage the scale of our dataset to overcome prior sample size limitations. Specifically, we compute the conditional expectations $\mathbb{E}[\density | \rloc{}]$ (\reffig{fig5}a) and $\mathbb{E}[\comspeed | \rloc{}]$ (\reffig{fig5}b)

The density-configuration relationship (\reffig{fig5}a) indicates that abreast configurations predominantly occur under lower-density conditions,
whereas in-file configurations are associated with higher average densities. This is consistent with the previous results (\reffig{fig3}) and supports that elongated in-file configurations facilitate efficient movement through denser crowd environments. %
The speed-configuration relationship (\reffig{fig5}b) shows a general trend: velocity decreases as interpersonal distance decreases.
Additionally, \reffig{fig5}b reveals that dyads attain maximum average speeds ($1.1\,\mathrm{m/s}$) when positioned in abreast formations and minimum average speeds ($0.7\,\mathrm{m/s}$) in in-file formation. 
This again confirms the known preference~\cite{zanlungo_potential_2014} for abreast walking in uncrowded conditions.

\begin{figure}[h]
		\centering
				\includegraphics[width=0.39\linewidth]{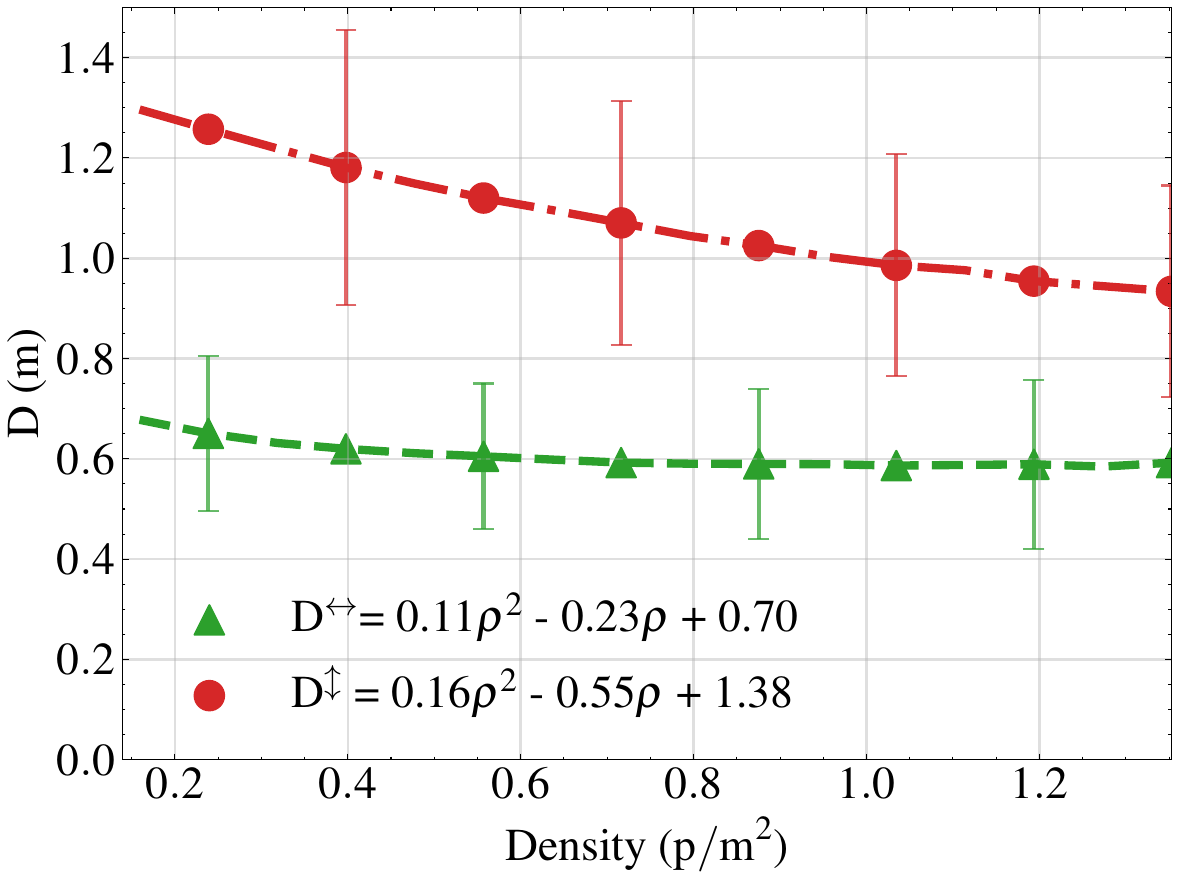}
		\caption{
			Modal interpersonal distances ($\modaldistance$) and their standard deviations as a function of local density $\density$ for abreast (green triangles, $\modaldistance^{\leftrightarrow}$) and in-file (red circles, $\modaldistance^{\updownarrow}$) dyads.
			Abreast configurations show a slight decline that flattens asymptotically at $\modaldistance^{\leftrightarrow} \approx 0.6~\mathrm{m}$, while in-file configurations exhibit a continuous decreasing trend throughout the density range.
		}\label{fig6}
\end{figure}

\paragraph{Interpersonal distance.}
To characterize the most likely dyad formations observed,
we analyse the mode of the interpersonal distance $\distance$ (i.e., the most frequent case)  of both abreast and in-file dyads.
We denote the modal distances for abreast and in-file dyads by $\modaldistance^{\leftrightarrow}$ and $\modaldistance^{\updownarrow}$, respectively.
Operationally, $\modaldistance$ is calculated by extracting the most common distances from the probability distributions $\Prob(\xr, \yr \mid \density)$,
as detailed in \refSIsec{ap:modal}.
Gaussian fits to the resulting marginal distributions yield the modal position $\mu$ and standard deviation $\sigma$, from which we define $\modaldistance$ as:
\begin{equation}
	\modaldistance^{\leftrightarrow} = 2\mu^{\leftrightarrow}, \quad \modaldistance^{\updownarrow} = 2\mu^{\updownarrow}, \quad \sigma_\modaldistance^{\leftrightarrow} = 2\sigma^{\leftrightarrow}, \quad \sigma_\modaldistance^{\updownarrow} = 2\sigma^{\updownarrow}.
	\label{eq:interpersonal_distances}
\end{equation}
In \reffig{fig6}, we present the relationship between these modal interpersonal distances $\modaldistance$ and local density $\density$.

The relationship between interpersonal distance and density reveals an asymmetric response between configurations.
For abreast dyads, $\modaldistance^{\leftrightarrow}$ remains near-constant, decreasing only slightly from $\sim\,0.7\,\mathrm{m}$ at low densities and approaching a plateau at $\sim\,0.6\,\mathrm{m}$ at higher densities, consistent with a lower bound set by body width and personal space. 
In-file dyads, by contrast, compress steadily with density, shrinking from $\sim 1.30\,\mathrm{m}$ at $\density=0.2\,\densityunit$ to $\sim 0.95\,\mathrm{m}$ at $\density=1.0\,\densityunit$.
This asymmetric ``compressibility'' likely reflects the different biomechanical constraints in each configuration: in-file dyads can reduce their longitudinal spacing by adjusting their gait phase and stride length to avoid stepping on each other's heels, while abreast dyads face a hard lower bound of interpersonal distance, set by shoulder width and the physical impossibility of occupying the same space. This demonstrates that spatial constraints imposed by crowd density affect the abreast and in-file configurations differently.

\subsection{Orientation Log-Odds, $\lgr$: scalar reduction of dyad configuration}\label{subsec:Pi}
In \refsec{subsec:FundChar}, we showed how the crowd density and flow regime, together with the  dyad velocity, determine systematic changes in the probability of the dyad spatial configuration. Here, we obtain a compact, interpretable measure of the dyad configuration: we reduce the 2D probability distribution $\Prob(\xr, \yr \mid \param)$, with $\param$ indicating some generic state parameter,  to a single scalar reflecting the relative probability of the dyad being in the abreast versus the in-file state. 
This will enable quantitative analysis and modeling of the probabilistic dependence of dyad formation on different choices of $\param$ (\refsec{sec:usingOLO}).
Specifically, we coarse-grain our observation considering the probability of being within abreast ($\stateslr$, \refeqq{eq:statelr}) or in-file ($\statesud$, \refeqq{eq:stateud}) configurations. This establishes, respectively, the two complementary conditional Bernoulli probabilities
\begin{align}
	 & \Prob(\statesud \mid \param) = \int_{\statesud} d\Prob(\xr, \yr \mid \param) & \text{\textit{in-file probability}}, \\
	 & \Prob(\stateslr \mid \param) = \int_{\stateslr} d\Prob(\xr, \yr \mid \param) & \text{\textit{abreast probability}}.
\end{align}
These two probabilities are linked to each other by the relation
\begin{equation}
	\Prob(\stateslr \mid \param) +  \Prob(\statesud \mid \param) = 1,
\end{equation}
which follows from the fact that $\stateslr$ and $\statesud$ are a disjoint partition of the $(x_r,y_r)$ plane.

\begin{figure}[h]
		\centering
				\figinnerlabLoopB{%
		\begin{overpic}[width=.7\linewidth]{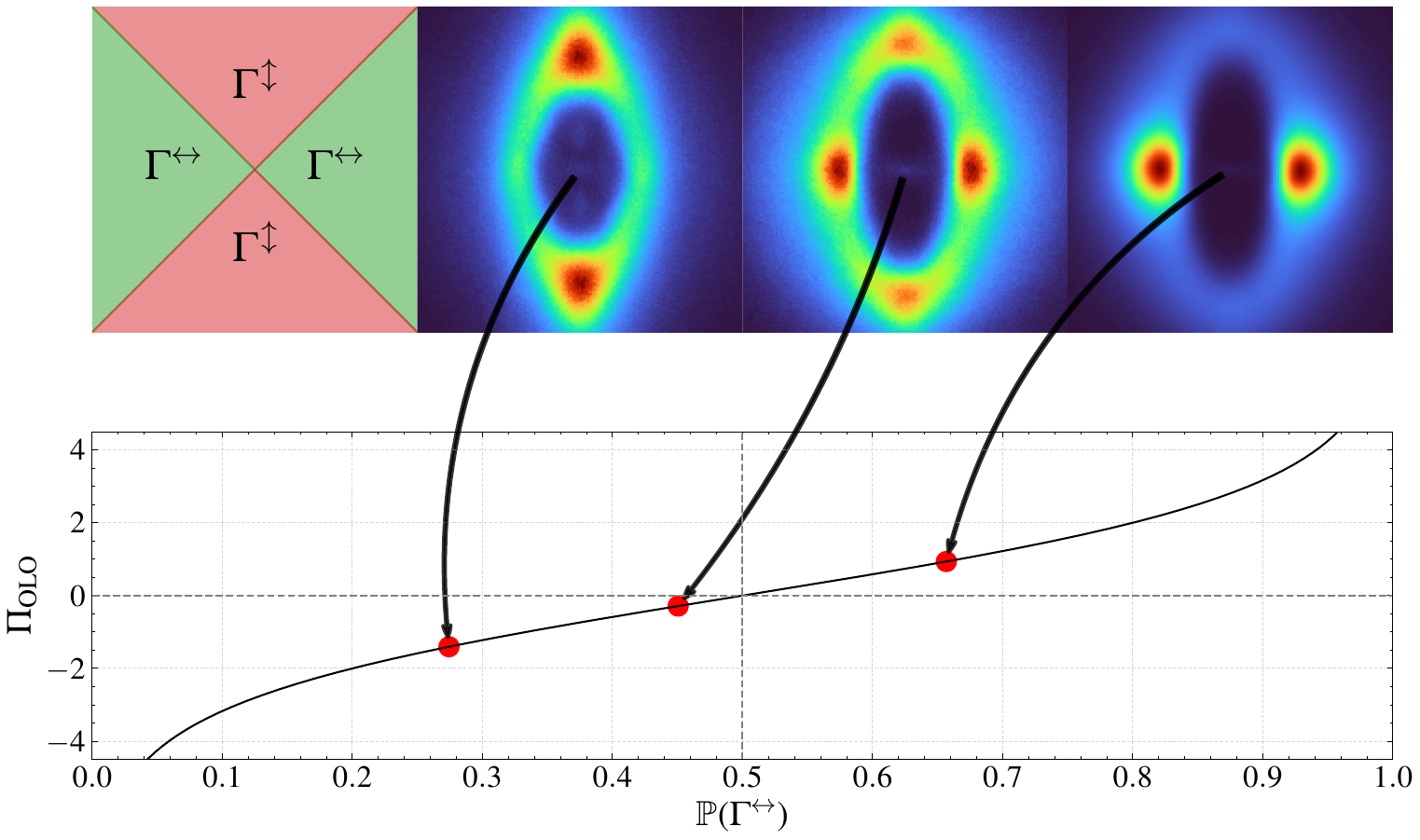}
		\put(99,33.5){\includegraphics[width=.035\linewidth]{figures/quadrant_support_colorbar.pdf}}
		\end{overpic}}{%
		a/0.8cm/4.05cm,%
		b/3.7cm/4.05cm,%
		c/6.4cm/4.05cm,%
		d/9.2cm/4.05cm,%
		e/0.8cm/0.2cm
		}
		\caption{
			Values of $\lgr$ (\refeqq{eq:big_pi}), i.e., relative probability of abreast versus in-file configurations, for different probability densities of dyad positions.
			(a) Division of the $(\xr,\yr)$ plane into abreast states $\stateslr$ (green, \refeqq{eq:statelr}) and in-file states $\statesud$ (red, \refeqq{eq:stateud}).
			(b-d) Example probability density for dyads in different regimes:
			(b) in-file states dominate ($\lgr \approx -1 < 0$),
			(c) in-file and abreast are equally likely ($\lgr \approx 0$), and
			(d) abreast states dominate ($\lgr \approx 1 > 0$).
			(e) $\lgr$ as a function of the probability an abreast state $\Prob(\stateslr)$.}\label{fig7}
\end{figure}

Finally, we introduce a scalar quantity, an invertible reparameterization of $\Prob(\stateslr \mid \param)$  with additional interpretability and geometric features. We dub it Orientation Log-Odds (OLO), $\lgr$, and in formulas it reads:
\begin{equation}
	\lgr(\param) = \log_2 \frac{\Prob(\stateslr \cb \param)}{\Prob(\statesud \cb \param)} = \text{logit}_2(\Prob(\stateslr \cb \param)).
	\label{eq:big_pi}
\end{equation}
$\lgr$ comes with  three useful properties:
\begin{enumerate}
	\item It quantifies the relative likelihood of abreast ($\lgr > 0$) versus in-file ($\lgr < 0$) configurations, with $\lgr = 0$ indicating equal probability for both states (cf. examples  in \reffig{fig7}).

		Any additional OLO unit implies the abreast configuration is twice as likely as the in-file case (due to the $\log_2$-base choice).
	\item It reflects the symmetry of our definition of the regions $\stateslr$ and $\statesud$ that are identical up to a $90^\circ$ rotation of the $(\xr,\yr)$ plane with a sign shift:
	      \begin{align}
		      \stateslr & \xrightarrow{\text{Rot}_{90}} \statesud        \\
		      (\xr,\yr) & \xrightarrow{\text{Rot}_{90}} (\mp\yr,\pm \xr) \\
		      \lgr      & \xrightarrow{\text{Rot}_{90}} -\lgr
	      \end{align}
	\item By analogy with equilibrium statistical mechanics (already used for dyads~\cite{zanlungo_potential_2014}), we associate the observed probability $\Prob$ of a given state with an equilibrium energy $E$ via a Boltzmann-like distribution, $\Prob(x) \propto e^{+E(x)}$. This gives
	      \begin{equation}
		      \lgr = \log_2 \frac{\Prob(\stateslr)}{\Prob(\statesud)} \propto \left( E_{\statesud} - E_{\stateslr} \right),
	      \end{equation}
	      where $E_{\statesud}$ and $E_{\stateslr}$ are respectively the energies of the in-file and abreast states. Such an interpretation aligns with the Langevin-like approaches that have been extensively used to  model pedestrian dynamics~\cite{corbetta-annurev-2023}. %
\end{enumerate}

\section{$\lgr$ and dyad formation depending on crowd conditions}\label{sec:usingOLO}

Here, we examine, in probabilistic terms, how dyad formation changes with the surrounding crowd state.
We leverage the synthetic indicator $\lgr$ (\refeqq{eq:big_pi}), which we characterize through crowd density $\density$, dyad speed, $\comspeed$, flow regime $B$ (\refeqq{eq:binningdef}), and relative speed $\vrelpar$ (\refeqq{eq:velpar}), which we will synthetically refer to with the vector parameter $\param$. Specifically, we model the relation
\begin{equation}
	\lgr = \lgr(\rho,\comspeed,B,\vrelpar) = \lgr(\param).
\end{equation}
We consider three regimes of increasing complexity with the underlying modeling assumption that in the absence of nearby pedestrians, dyads follow some intrinsic ``free-flow'' dynamics that are perturbed by the presence of a crowd. Specifically, these regimes are
\begin{enumerate*}[label=(\arabic*)]  %
	\item free-flow conditions ($\binfree$),
	\item dyads moving through stationary crowds ($\binstanding$), and
	\item the general dynamic conditions in which both dyad and crowd move in co- or counter-flow. %
\end{enumerate*}
In each of these regimes, some components of $\param$ vanish or are set to specific values, which we will clarify by writing a conditioning statement, e.g.,
\begin{equation}
	(\param \mid B = \binfree),
\end{equation}
to mean that only parameters relevant in free-flow are active in the parameter vector $\param$.

\subsection{$\lgr$ in free flow regime, $\binfree$}\label{subsec:OLOfree}
\begin{figure}[h]
		\centering

				\figinnerlab{
		\begin{overpic}[width=0.35\linewidth]{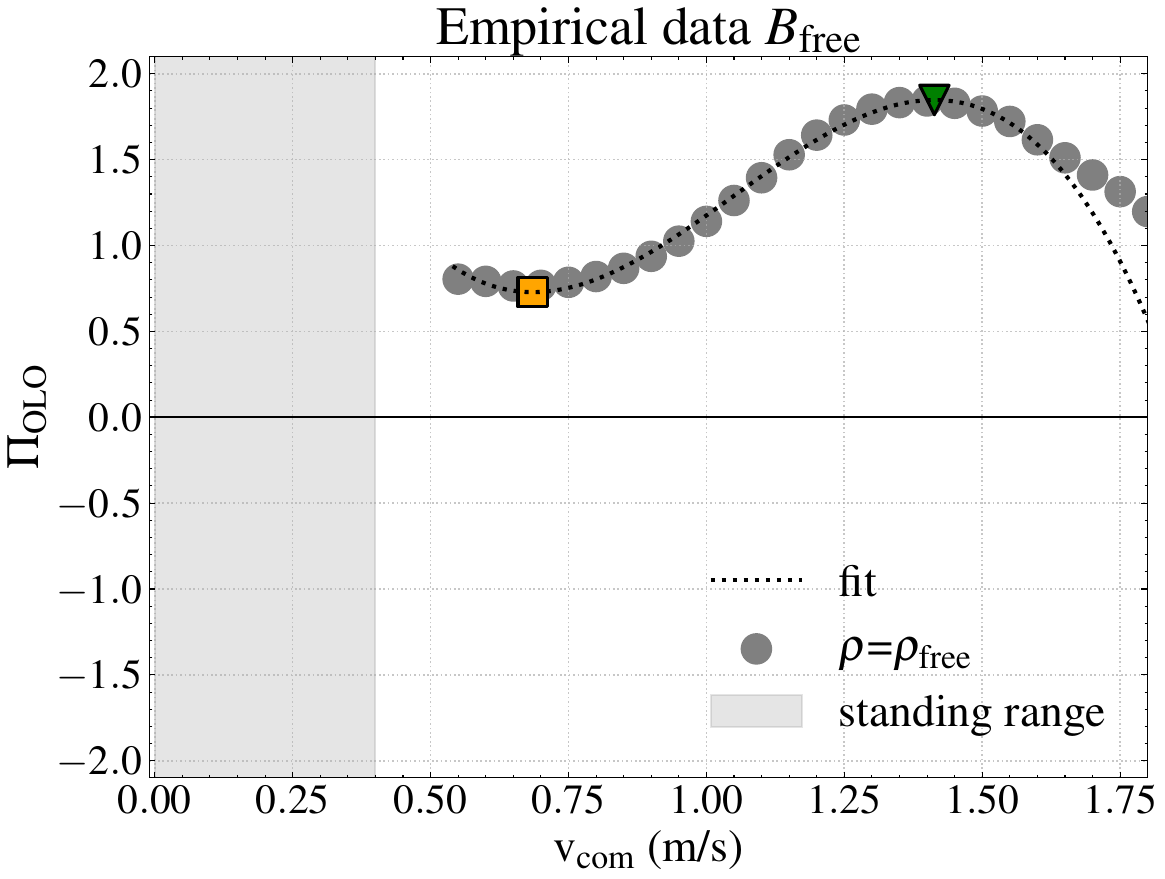}
		\put(15,10){\includegraphics[width=0.2\linewidth]{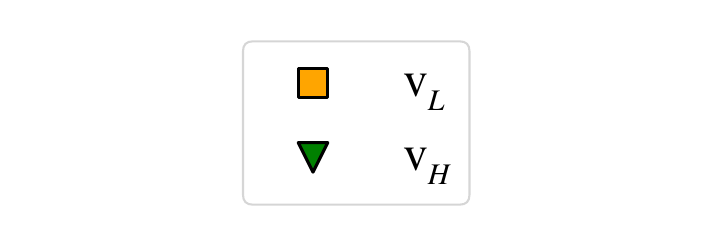}}
		\end{overpic}}{a}{-.2cm}{-.4cm}
		\figinnerlab{
		\includegraphics[width=0.35\linewidth]{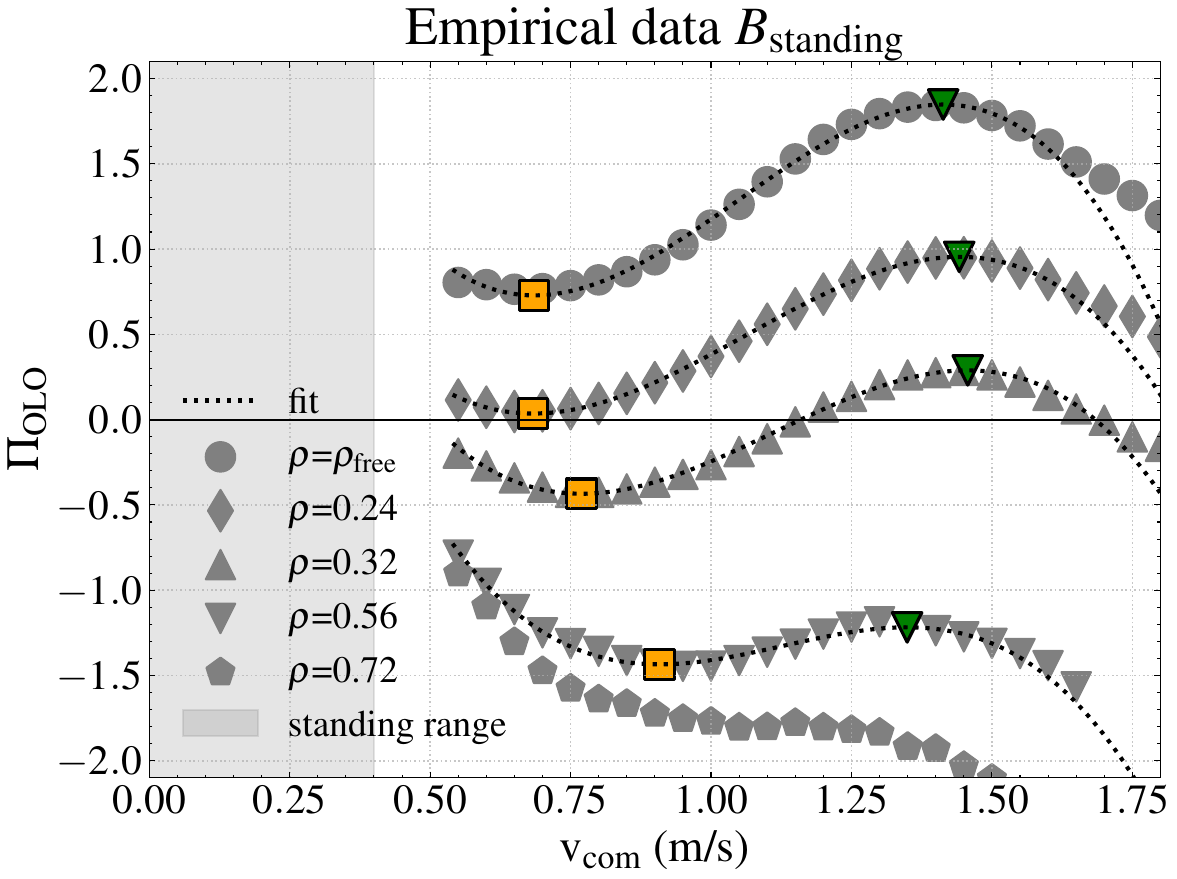}}{b}{-.2cm}{-.4cm}
		\figinnerlab{
		\includegraphics[width=0.35\linewidth]{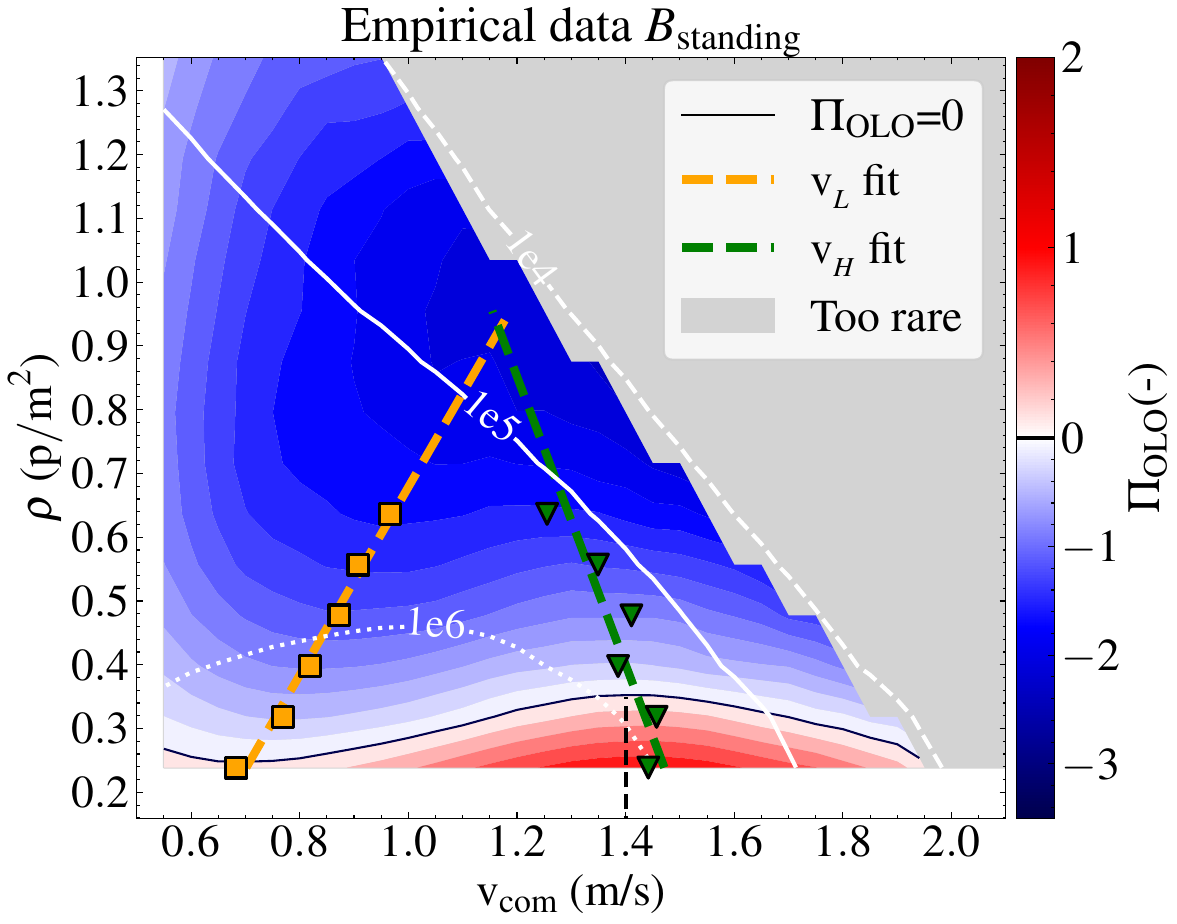}}{c}{-.2cm}{-.4cm}
		\figinnerlab{
		\includegraphics[width=0.35\linewidth]{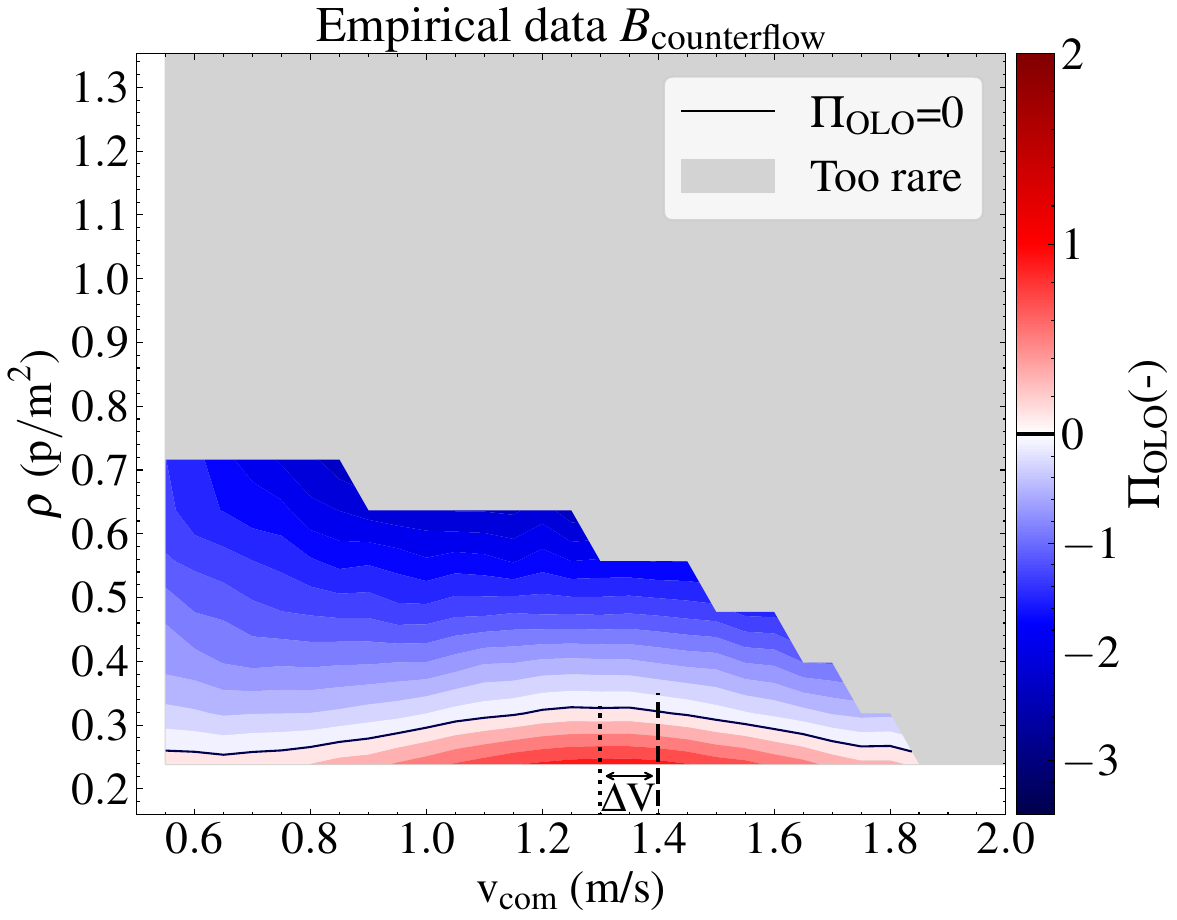}}{d}{-.2cm}{-.4cm}
		\caption{
			$\lgr$ as a function of dyad speed $\comspeed$ for (a) a dyad in free-flow conditions $\binfree$, (b,c) a dyad traversing a standing crowd $\binstanding$
			and (d) a dyad going against a crowd $\binagainst$.
			The dyad configuration preference changes with dyad speed $\comspeed$ and density $\density$.
			(a) In $\binfree$ conditions, $\lgr$ varies non-monotonically with $\comspeed$,
			showing a minimum at $\vcomlow$ and a maximum at $\vcomhigh$, with $\lgr > 0$ throughout
			(i.e., $\Prob(\stateslr) > \Prob(\statesud)$, abreast more likely).
			(b) In $\binstanding$ conditions, the overall shape of $\lgr$ is preserved and systematically decreases as $\density$ increases:
			higher densities promote in-file configurations ($\lgr < 0$).
			(c) Contour plot of the standing-crowd condition across all recorded densities,
			showing that velocities $\vcomlow$ and $\vcomhigh$ approach one another, meeting at finite density (linear fits).
			The black dashed line marks the speed at which the $\lgr = 0$ contour reaches its maximum $\density$.
			(d) Contour plot for the opposing-crowd condition, $\binagainst$, exhibiting similar behavior to (c). The black dotted line marks the speed at which the $\lgr = 0$ contour reaches its maximum $\density$, shifted by $\Delta V = 0.1\,\mathrm{m/s}$ to lower speeds. Note that the counter-flow regime $\lgr$ trend can be obtained from the standing flow case with good accuracy via a rescaling and translation transformation (see \refsec{sec:modelgeneric}).
			In (c-d), the black contour line marks the $\lgr = 0$ abreast-in-file transition ($\Prob(\stateslr) = \Prob(\statesud)$); gray regions denote insufficient data (less than $10^4$ data points - in panel (c), contours outlining the amount of data in different regions are in white).
		}\label{fig8}
\end{figure}

By our definition of the free flow regime, $\binfree$ (\refeqq{eq:binningdef}), there is no crowd near the dyad; hence, the variables $\density$ and $\vprox$ have no effect on the dyad formation.
Consequently, $\lgr$ depends only on the dyad speed $\comspeed$, i.e., in formulas
\begin{equation}
	\lgr(\param \, |\, B = \binfree) = \lgr^{\text{free}}(\comspeed)
\end{equation}
In \reffig{fig8}a, we report the measured values $\lgr^{\text{free}} = \lgr^{\text{free}}(\comspeed)$.
We observe that $\lgr^{\text{free}}$ is strictly positive and, in particular,
\begin{equation}
	\lgr^{\text{free}}(\comspeed) \geq
	0.65 > 0 \qquad \forall\, \comspeed \in [0.5,1.75]\,m/s.
\end{equation}
Hence,  in line
with previous findings~\cite{DyadDistanceMoussaid2010, zanlungo_potential_2014}, the abreast configuration is always more likely than the in-file configuration - indeed by a factor $ \approx 2^{0.65} \approx 1.57$ or more. Moreover, $\lgr^{\text{free}}$ depends on speed in a non-monotonic way, and we can isolate two critical velocity values:
\begin{align}
	 & \vcomlow \approx 0.70\, m/s \quad \lgr^{\text{free}}(\comspeed) \approx 0.65\quad \comspeed \leq \vcomlow &  & \text{\textit{minimum}}           \\
	 & \vcomhigh \approx 1.35\, m/s \quad \lgr^{\text{free}}(\vcomhigh) \approx 1.9                              &  & \text{\textit{relative maximum}}.
\end{align}
Note that $\vcomlow$ and $\vcomhigh$ are, respectively, smaller and larger than the free flow modal speed ($\mu_{dyad} = 1.23\,m/s$, \reffig{fig2}b); moreover, at $\comspeed = \vcomhigh$, abreast configurations are almost four times more likely than in-file cases.

We conjecture that these formations reflect the dyad's certainty regarding its movement goal and the urgency to reach it:
\begin{itemize} %
	\item at low speeds ($\comspeed \approx \vcomlow$), uncertainty about the destination may result in less structured configurations;
	\item  at moderate speeds ($\vcomlow < \comspeed < \vcomhigh$), increased speed may indicate greater goal clarity, shifting the focus toward communication and comfort~\cite{zanlungo_potential_2014}, which is facilitated by abreast formations;
	\item at high speeds ($\comspeed > \vcomhigh$), as the dyad enters the running regime, the movement is likely driven by urgency, and efficient movement becomes the priority. This reduces the likelihood of abreast configurations due to potential speed mismatches between members.
\end{itemize}
The non-monotonic trend of $\lgr$ is a strong feature that extends beyond the free flow regime, and it is the central element in our model.
In particular, the simplest model featuring a relative minimum and maximum satisfies
\begin{equation}
	\frac{\partial \lgr^{\text{free}}(\comspeed)}{\partial \comspeed} \propto -(\comspeed - \vcomlow)(\comspeed - \vcomhigh),
	\label{eq:part_deriv_no_density}
\end{equation}
which, up to a scaling factor $\scalefactor > 0$ and an offset $\addfactor$, gives (cf. dotted line in \reffig{fig8}a)
	\begin{equation}
		\label{eq:ModelNoDens}
		\lgr^{\text{free}}(\comspeed) = -\scalefactor \left(\frac{\comspeed^3}{3} - \frac{\comspeed^2}{2} (\vcomlow + \vcomhigh) + \comspeed \, \vcomlow \, \vcomhigh\right) + \addfactor. %
	\end{equation}
This simple polynomial model matches the data accurately between $0.5\,\mathrm{m/s}$
and $1.6\,\mathrm{m/s}$, while it fails at  velocities higher than $1.6\,\mathrm{m/s}$ (very rare), since the measured $\lgr^{\text{free}}(\comspeed)$ has a slower decay and likely no zero-crossing (i.e., we expect that at high velocity dyads have no reason to switch to a predominantly in-file configuration).

\subsection{$\lgr$ in static crowd conditions, $\binstanding$}\label{subsec:OLOstanding}
In comparison with the free flow regime, the case of a dyad walking through a standing crowd adds the additional crowd density parameter, $\density$, to the system.
In formulas, we now consider the relation
\begin{equation}
	\lgr(\param\, |\, B = \binstanding)= \lgr^{\text{standing}}(\density, \comspeed),
\end{equation}
which we report for five selected density levels in \reffig{fig8}b and in terms of contours in \reffig{fig8}c.
The phenomenology is quite rich, and has the following features:
\begin{itemize}
	\item for density levels
	      \begin{equation}
		      \density < 1.05\,\densityunit =\colon \density_T,
	      \end{equation}
	      $\lgr$ decreases with $\density$: higher density yields a general preference towards in-file.

	      Moreover, regardless of the speed $\comspeed$, at density
	      \begin{equation}\label{eq:density-for-infile-only}
		      \rho > 0.35\,\densityunit =\colon \density_C
	      \end{equation}
	      the in-file configuration is always preferred. This holds even up to a ten-fold factor around the state $(\comspeed = 1.2\,\textrm{m/s},\rho = \density_T)$.

		We hypothesize that this trend arises because in-file configurations facilitate easier navigation through a crowd.

	      Conversely, when $\density > \density_T$, $\lgr^{\text{standing}}$ increases slightly with density, though  remaining negative (in-file dominates).
	      We hypothesize that at high densities, the dyad's agency diminishes,
	      and configuration is increasingly dictated by crowd conditions rather than individual preference;

	\item at densities $\density \gtrsim 0.6\,\densityunit$, $\lgr^{\text{standing}}$ is systematically closest to zero at the lowest velocity levels considered ($\comspeed \approx 0.5\,\mathrm{m/s}$). In other terms, at lower velocities, there is higher uncertainty about the preferred configuration of a dyad, e.g., due to uncertainty in destination, as mentioned in the free-flow case;

	\item increasing $\density$ from the free-flow regime, when
	      $\density = 0.32\,\densityunit=\densityzerocross$, $\lgr^{\text{standing}}$ crosses zero for the first time: the abreast configurations become as likely as the in-file. At even higher density levels, abreast states become increasingly less likely than in-file configurations. Moreover, at $\density = \densityzerocross$ when a dyad walks at  velocity $\comspeed \approx 0.75\,\mathrm{m/s} (=\vcomlow)$ the in-file configuration is about $1.4$ times more likely than the abreast. Such a likelihood  completely reverses at $\comspeed \approx 1.4\,\mathrm{m/s} (=\vcomhigh)$;

	\item the non-monotonic dependency of $\lgr$ on $\comspeed$ observed in the free-flow case extends to the standing case, up to $\density \approx 0.7\,\densityunit$. We have two critical velocities,  $\vcomlow = \vcomlow(\density)$ and $\vcomhigh = \vcomhigh(\density)$, at which the likelihood of abreast configurations remains, respectively, lowest and highest (marked, respectively, with green squares and green triangles in \reffig{fig8}a-c).

	      The distance $|\vcomhigh - \vcomlow|$ diminishes as the density increases. Effectively, at $\density \approx 0.7\,\densityunit$, $\lgr^{\text{standing}}$ has a saddle-node-like bifurcation and its critical points disappear, acquiring a monotonically decreasing behavior;

	\item in the $(\comspeed,\density)$ plane, not all the states are possible. High density conditions for which $\comspeed$ is also high are unlikely. This corresponds to the gray-shaded region in \reffig{fig8}c, which marks cases that are  physically impossible or very unlikely (fewer than $\approx 10.000$ measurements).
	      The boundary of these impossible/unlikely conditions satisfies $\partial \rho/\partial \comspeed <0$, signaling, as expected, that the maximum feasible speed through a crowd diminishes with density. This is likely due to increased physical constraints and reduced maneuverability.

\end{itemize}

We model $\lgr^{\text{standing}}$ by generalizing $\lgr^{\text{free}}$ (\refeqq{eq:ModelNoDens}). Primarily, we allow $\vcomlow$,  $\vcomhigh$ to depend linearly on the density, say
\begin{align}
	\vcomhigh(\density) = \alpha_{h} \density + \beta_{h} \qquad\qquad
	\vcomlow(\density)  = \alpha_{l} \density + \beta_{l}, %
\end{align}
with parameters identified through linear fits of the fundamental diagrams   $\vcomlow = \vcomlow(\density)$ and $\vcomhigh = \vcomhigh(\density)$. Note that $\partial \lgr^{\text{standing}} / \partial \comspeed$ scales as the product of two fundamental diagrams. After integration, this gives
	\begin{equation}
		\lgr^{\text{standing}}(\comspeed, \density) = -\scalefactor \left(\frac{\comspeed^3}{3} - \frac{\comspeed^2}{2} (\vcomlow(\density)+\vcomhigh(\density)) + \comspeed \vcomlow(\density) \vcomhigh(\density)\right) + \addfactor,
		\label{eq:PImodelStanding}
	\end{equation}
Note that a more accurate fit could be obtained by allowing also linear dependence of the multiplying factor $\scalefactor$ on the density
\begin{equation}
	\scalefactor = \scalefactor(\density) = \alpha_\scalefactor \density + \beta_\scalefactor;
\end{equation}
this increased complexity nevertheless contributes only a $\approx 5\%$ variation
(cf. values in \reftab{tab:fit-params-with-dens}).
\begin{table}[h!]
	\centering
	\caption{Parameters of our $\lgr^{\text{standing}}(\comspeed, \density)$ model (\refeqq{eq:PImodelStanding}).}\label{tab:fit-params-with-dens}
	\begin{minipage}{0.48\linewidth}
		\centering
		\begin{tabular}{|l|l|l|}
			\hline
			Symbol                  & Value                     & Units                                              \\
			\hline
			$\alpha_{l}$            & \vcomlowslopefitWithDens  & $\mathrm{m^3\,s^{-1}\,p^{-1}}$                     \\\hline
			$\beta_{l}$             & \vcomlowfitWithDens       & $\mathrm{m\,s^{-1}}$                               \\\hline
			$\alpha_{h}$            & \vcomhighslopefitWithDens & $\mathrm{m^3\,s^{-1}\,p^{-1}}$                     \\\hline
			$\beta_{h}$             & \vcomhighfitWithDens      & $\mathrm{m\,s^{-1}}$                               \\\hline
			$\alpha_{\scalefactor}$ & \fitScaleslopefitWithDens & $\mathrm{m}^{-1}\,\mathrm{s}^{3}\,\mathrm{p}^{-1}$ \\\hline
			$\beta_{\scalefactor}$  & \fitScalefitWithDens      & $\mathrm{m^{-3}\,s^{3}}$                           \\\hline
			$\addfactor$            & \DfitWithDens             & -                                                  \\
			\hline
		\end{tabular}
	\end{minipage}%
\end{table}

\begin{figure}[h]
		\centering
				\figinnerlab{
		\includegraphics[width=0.35\linewidth]{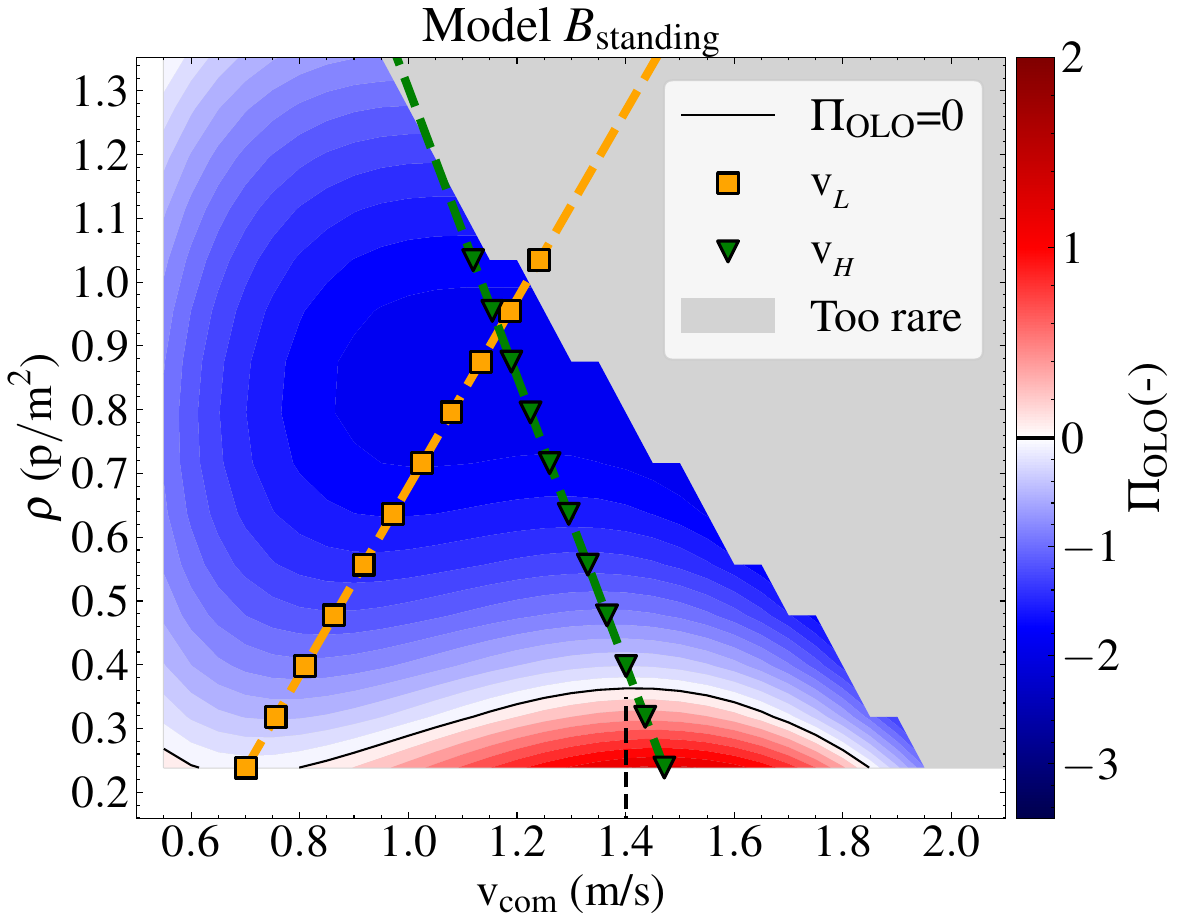}}{a}{-.2cm}{-.4cm}
		\figinnerlab{
		\includegraphics[width=0.35\linewidth]{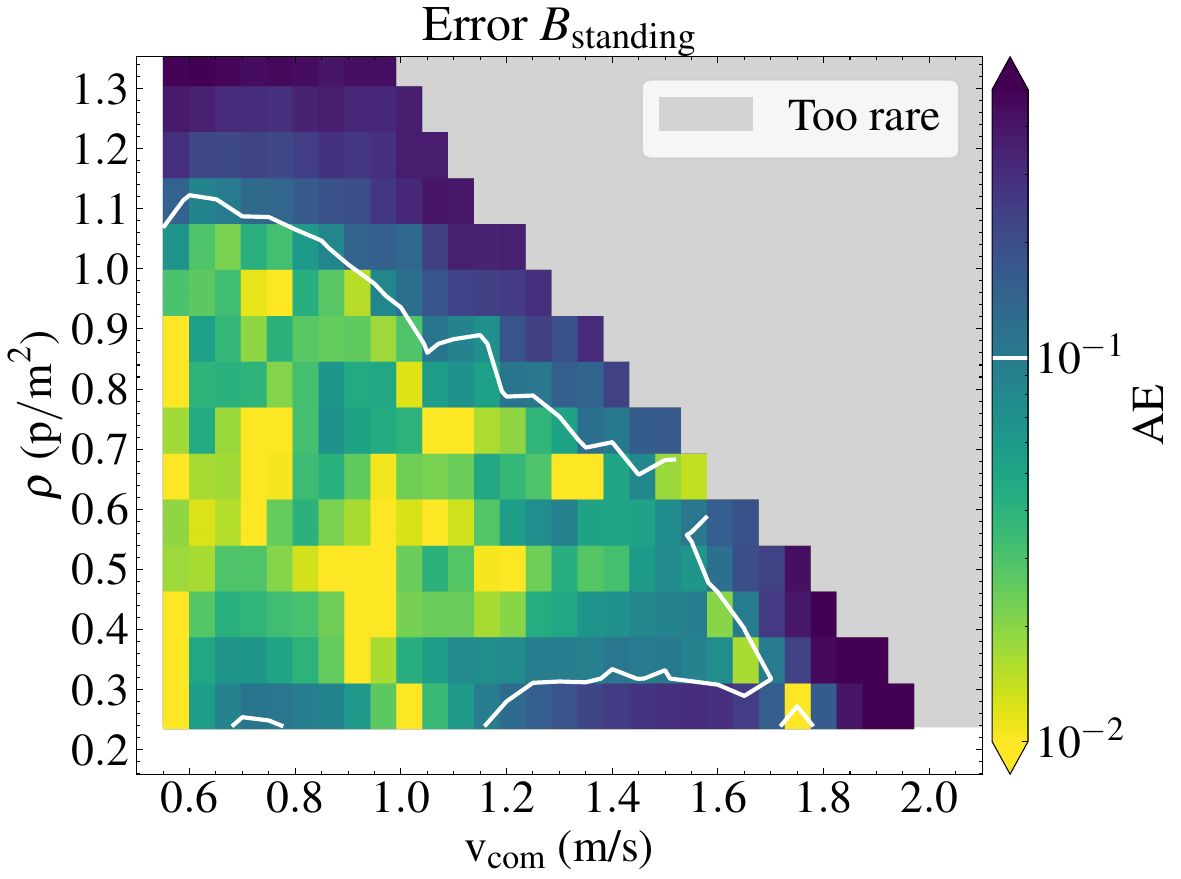}}{b}{-.2cm}{-.4cm}
		\figinnerlab{
		\includegraphics[width=0.35\linewidth]{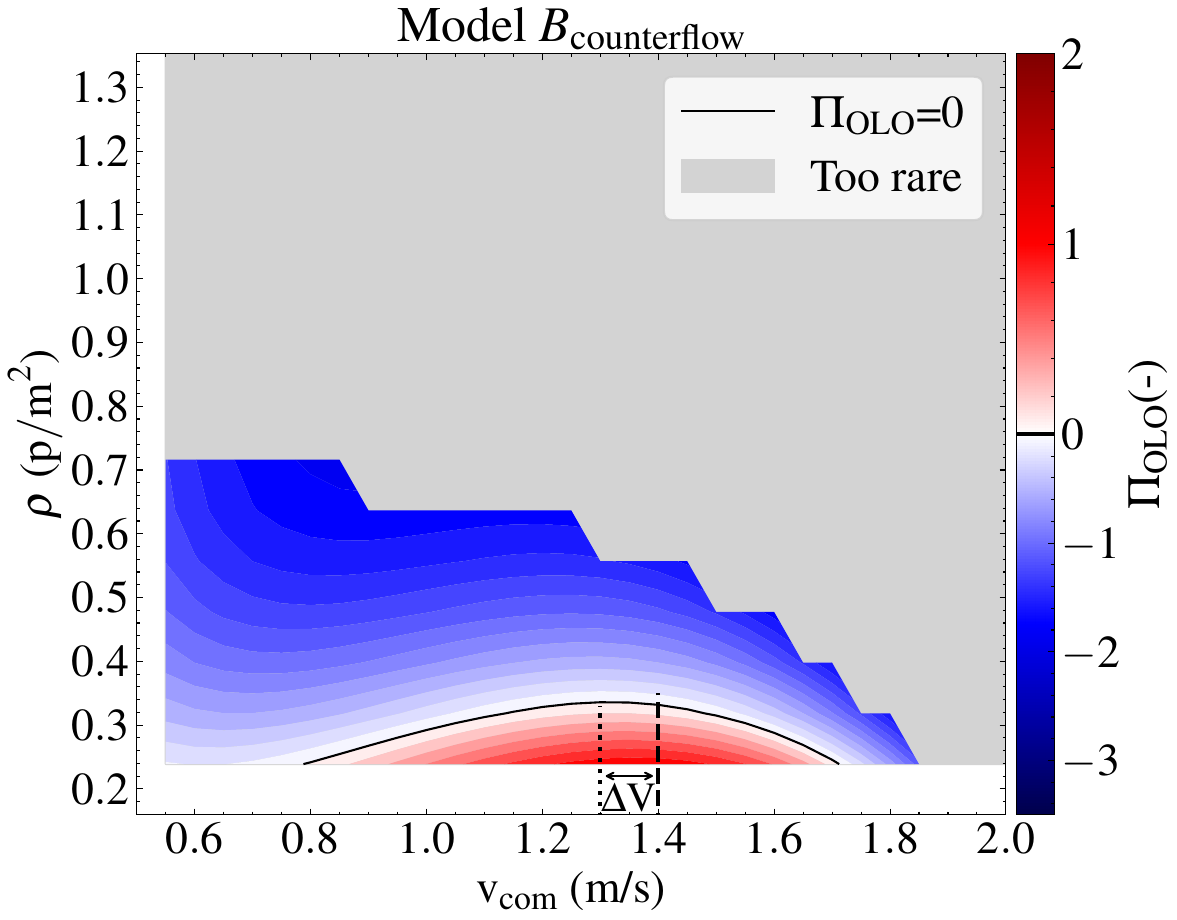}}{c}{-.2cm}{-.4cm}
		\figinnerlab{
		\includegraphics[width=0.35\linewidth]{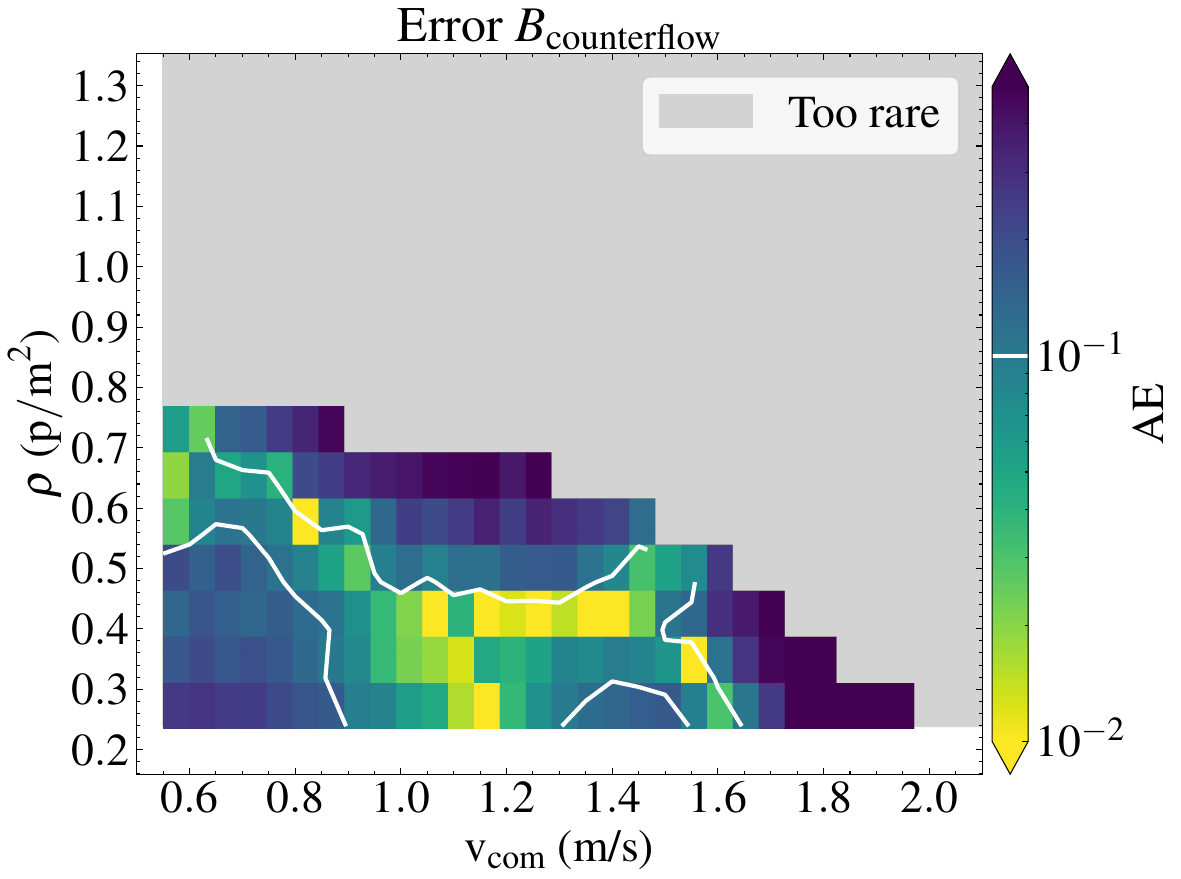}}{d}{-.2cm}{-.4cm}
		\caption{
			Contour plots of $\lgr$ (\refeqq{eq:big_pi}) models for standing $\binstanding$ (\refeqq{eq:PImodelStanding}) and counter-flow $\binagainst$ (\refeqq{eq:pimodelagainst}) conditions.
			(a) Standing crowd model showing polynomial-like structure with characteristic velocities $\vcomlow(\density)$ and $\vcomhigh(\density)$ that converge as density increases toward $\density_T = 1.05\,\densityunit$.
			(b) Absolute error (AE) between model (a) and empirical data (\reffig{fig8}c), demonstrating good predictive accuracy for $\comspeed < 1.7$~m/s and $\density < 1.1\,\densityunit$.
			(c) Counter-flow model obtained by applying velocity offset $\Delta V = 0.1$~m/s and density compression $Z = 1.08$ to the standing case (\refeqq{eq:pimodelagainst}), exhibiting similar functional form but shifted to lower velocities and compressed density range.
			In (a, c), the black contour line marks $\lgr = 0$, i.e., $\Prob(\stateslr) = \Prob(\statesud)$, and gray regions indicate insufficient data (less than $10^4$ measurements).
			(d) Absolute error between model (c) and empirical data (\reffig{fig8}d), showing accurate predictions for $\comspeed < 1.6$~m/s.
			Deviations in both models occur primarily at domain edges and in rare high-velocity conditions.
		}\label{fig9}
\end{figure}

This simple model of the $\lgr$ behavior in the standing regime shows very good agreement with the measurements: in \reffig{fig9}a, we report the model's contour plot, which recovers all the qualitative features observed above.
For a quantitative assessment, in \reffig{fig9}b we report the absolute error (AE) between predicted and empirical $\lgr$ values ($\text{AE}(\comspeed,\rho) =  \abs{(\lgr^{model}(\comspeed,\rho) - \lgr^{data}(\comspeed,\rho))} $), demonstrating that the model is accurate up to $\comspeed \approx 1.7$\,m/s and densities up to $1.1\,\densityunit$. 
\refSIsec{ap:valid_model_region} includes an empirical equation (\refeqq{eq:validmodelstanding}) for the region in the $(\comspeed,\density)$ plane in which our model is defined.

\subsection{$\lgr$ in generic flow conditions}\label{sec:modelgeneric}
\begin{figure}[h]
		\centering
				\includegraphics[width=0.42\linewidth]{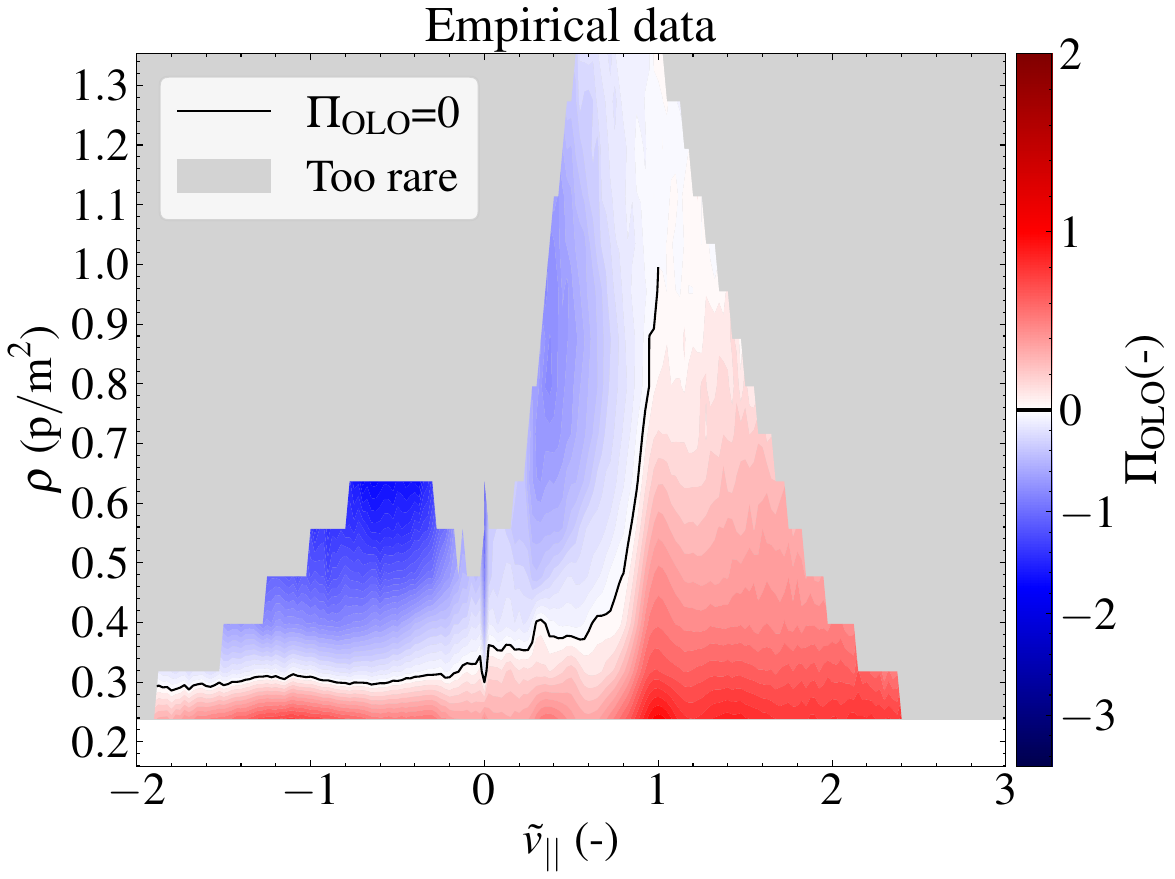}
		\caption{
			$\hatlgr$ as a function of local density $\density$ and relative parallel velocity $\vrelpar$ (cf.~\refeqq{eq:velpar}), showing configuration preferences for $\binagainst$ ($\vrelpar < 0$) and $\binwith$ ($\vrelpar > 0$).
			Three distinct behavioral regimes are evident:
			(i) Counter-flow conditions ($\vrelpar < 0$) show density-dependent transitions with minimal velocity dependence, similar to the standing crowd case;
			(ii) Co-flow with crowd overtaking dyad ($\vrelpar > 1$) shows strong abreast preference that decreases with density;
			(iii) Co-flow with dyad overtaking crowd ($0 < \vrelpar < 1$) exhibits a sharp transition from counter-flow-like behavior to abreast preference.
			The black contour line marks $\hatlgr = 0$ ($\Prob(\stateslr) = \Prob(\statesud)$).
			Gray regions denote insufficient data.
		}\label{fig10}
\end{figure}

Here, we relax the previous regime constraints and examine $\lgr$ under generic co-flow and counter-flow conditions.
The formation behavior in co- and counter-flow conditions changes quite dramatically. To showcase it, we proxy first the flow regime using the relative parallel velocity variable, $\vrelpar$ (\refeqq{eq:velpar}). In \reffig{fig10}, we report the contours of $\lgr$ in dependency of the relative parallel velocity and density, i.e., of the relation
\begin{equation}\label{eq:lgrhat}
	\lgr(\param \mid  \forall \comspeed, \forall B) =  \hatlgr(\density, \vrelpar).
\end{equation}
\noindent We can observe substantially different behaviors:
\begin{itemize}
	\item  in counter-flow conditions ($\vrelpar < 0$),
	      the density $\density \approx \density_C$ (\refeqq{eq:density-for-infile-only}) marks the $\hatlgr = 0$ threshold line, with practically no dependency on the relative velocity $\vrelpar$. In other words, the transition density, already observed in the standing case, extends to the counter-flow case. 
		  In the following \refsec{subsec:OLOagainst}, we will examine the counter-flow case in greater detail and show how our $\lgr$ model for standing crowd extends to this regime up to a constant shift.

	\item the co-flow case ($\vrelpar > 0$) features two sub-cases, sharply marked by the almost asymptotic behavior of the threshold line $\hatlgr = 0$ around the speed-matching conditions $\vrelpar \approx 1$ (i.e., $\frac{\partial \density}{\partial \vrelpar}|_{\hatlgr = 0, \vrelpar = 1} \gg 1$).
	      A dyad co-flowing and being overtaken by the surrounding crowd shows a preference for abreast configurations:
	      \begin{equation}
		      \hatlgr\left(\, \vrelpar > 1\,,\, \density < 1\,\densityunit\,\right) > 0.
	      \end{equation}
	      Such a preference decreases with density and, as we approach $\density \approx 1\,\densityunit$, uncertainty in formation ($\hatlgr\approx 0$) is highest. 
		  We interpret this as follows: when the dyad is being overtaken by the crowd, it is likely in a leisure walking condition, and the abreast configuration is favored, e.g., to allow for interactions. 
		  Such a ``relaxed'' abreast state becomes, however, harder and harder to maintain with density.
	      Moreover, cases with high density and high relative velocity are clearly unlikely to happen, as represented by the  gray-shaded areas, which mark insufficient measurements;
	\item for dyads overtaking the crowd ($0 < \vrelpar < 1$), $\hatlgr$ shows two regimes separated by a sharp boundary: a counter-flow-like regime in which in-file configurations dominate above a critical density, and an abreast-dominated regime reached once the speed-matching condition is exceeded.
	      Note that, as commented after \refeqq{eq:velpar}, $\vrelpar\approx 0^+$ is a case only possible when the dyad walks much faster than the surrounding crowd, which might effectively appear to the dyad as standing.
\end{itemize}

\subsubsection{$\lgr$ in counter-flow crowd conditions, $\binagainst$}\label{subsec:OLOagainst}
We conclude by analysing and modeling the final regime, $\binagainst$, in which the dyad is moving against a crowd. As shown in \reffig{fig10}, we can neglect the role of $\vrelpar$ (as \reffig{fig10} shows that $\hatlgr$ is nearly independent of $\vrelpar$ for $\vrelpar < 0$), and consider the dependency
\begin{equation}
	\lgr(\param \mid B = \binagainst, \forall \vrelpar) = \lgr^{\text{counterflow}}(\comspeed,\density),
\end{equation}
whose contours we report in \reffig{fig8}d.
Notably, $\lgr^{\text{counterflow}}$  exhibits a similar functional form to $\lgr^{\text{standing}}$ (\reffig{fig8}c), i.e.,  dyads exhibit similar behavior, when crossing a standing or a counter-flowing crowd, yet with minor differences:
\begin{itemize} %
	\item  all previously observed features from the $\binstanding$ regime (see \refsec{subsec:OLOstanding}) are slightly compressed in the $\density$ direction by a factor $Z = 1.08$ and occur at slightly lower velocities, offset by $\Delta V = 0.1\,\mathrm{m/s}$.

		This is likely due to increased navigational complexity, which reduces the overall traversal speed of dyads and narrows the velocity range in which the previous behaviors emerge.
	      A compression along the density axis suggests that dyads perceive counter-flowing crowds as effectively denser than stationary ones;
	\item the region with insufficient data is notably larger, with no observations for $\density > 0.8\,\densityunit$.
	      This likely indicates that dyads tend to avoid moving against a crowd beyond this density threshold.
\end{itemize}
Due to the structural similarities, we prioritize interpretable, low-parameter models and base our model for $\lgr^{\text{counterflow}}$ on the standing-crowd case $\lgr^{\text{standing}}$ (\refeqq{eq:PImodelStanding}):
\begin{equation}
	\lgr^{\text{counterflow}}(\comspeed, \density) = \lgr^{\text{standing}}(\comspeed + \Delta V, Z\,\density),
	\label{eq:pimodelagainst}
\end{equation}
where $\Delta V$ is the effective velocity offset and $Z$ is the effective density compression factor.
The contour plot resulting from the model is depicted in \reffig{fig9}c.
As shown in \reffig{fig9}d, the model achieves low AE and closely matches the observed transition lines and main trends.
Deviations are confined to higher velocities ($>1.6\,\mathrm{m/s}$) and higher densities ($>0.6\,\densityunit$).
\refSIsec{ap:valid_model_region} includes an empirical equation (\refeqq{eq:validmodelagainst}) for the region in the $(\comspeed,\density)$ plane in which our model is defined.

\section{Discussion}\label{sec:discussion}

In this paper, we presented a large-scale probabilistic analysis of the dynamics of dyads, i.e., groups of two pedestrians, walking in real-world crowded environments.
We leveraged a multi-year anonymous pedestrian trajectory dataset from Eindhoven Central Station (NL), in which we  automatically identified  over \amountdyads~dyads.
Given the anonymity constraint, our dataset included only position features and tracking (i.e., no orientation or any further kinematic or personal features).
To automate the dyad identification process, we relied on proxemics heuristics (distance consistency) and leveraged a previously proposed graph-based approach.
The dyad dataset that we have established is several orders of magnitude larger than those currently used,
unlocking robust conditional analyses of both average behavior and fluctuations in dyad dynamics.

At the core of our work is a collection of probabilistic phenomenological relations that connect the dynamics of dyads to those of the surrounding crowds through kinematic observables.
Such observables are the dyad center-of-mass speed, its spatial formation, the crowd density in the neighborhood of the dyad, and the crowd relative velocity.
On these bases, we employed a large set of quantities that can be deterministically derived from these observables.
With data anonymity in mind, we argue that these provide a minimal, yet practically exhaustive, family of relevant observables.

Our probabilistic analysis provides a thorough phenomenological description of dyads, supplying a collection of fundamental diagram-like notions.
Throughout our work, we analysed finer and finer aspects of the dynamics of dyads, starting indeed from traditional fundamental diagrams, i.e., density-speed relations.
Regardless of the specific flow condition, dyads move  $\approx 10\%$ slower than a generic pedestrian in a crowd for comparable macroscopic density.
Also, regardless of the specific flow condition, we observe that dyads in the abreast configuration move faster than in-file at low density, and this trend reverses at high density.
The specific crossover density is flow regime-dependent.
As one could expect, the transition occurs at a lower density ($\rho \approx 0.5\,\densityunit$) for counter-flowing and standing crowds, compared to co-flowing crowds (crossover at $\rho \approx 0.8\,\densityunit$).
Dyads in counter-flow also move substantially slower than dyads in co-flow by at least $10\%$.

The relevant  phenomenology of dyads extends, however, beyond the velocity of the center-of-mass, since the degrees of freedom include a ``rotational'' component and a relative distance component.
At low density, dyads are most likely observed in abreast configuration. This holds robustly until density levels of $0.3-0.4\,\densityunit$.
The rotational degree of freedom is enabled by increasing the dyad velocity and/or the surrounding density, with this abreast-to-in-file transition being enhanced when the dyad is in counter-flow or when it walks through a standing crowd (in opposition to the co-flow case). The distance between the dyad members also has a dependency on the density.
Yet, we observe that the distance in abreast configuration is almost density independent (modal abreast distance $\approx 0.6\,\mathrm{m}$).
On the contrary, in in-file conditions, the modal distance decreases significantly from $\approx 1.3\,\mathrm{m}$ in free-flow to $\approx 0.9\,\mathrm{m}$ at $\rho\approx 1.3\,\densityunit$.
These different trends showcase the importance of treating the abreast and in-file conditions separately.

To quantitatively investigate how the formation changes with respect to the dyad and crowd state, we introduced  a new scalar variable mapping the log-likelihoods of dyad formations:
$\lgr$. $\lgr$ provides a probabilistic measure of the relative preference (indeed, the observational likelihood) for abreast versus in-file formations depending on the crowd state.
Conceptually, $\lgr$ can also be interpreted as the energy difference between the abreast and in-file states for a given crowding condition under a Boltzmann-like assumption. Through $\lgr$, we accurately traced the boundary between abreast and in-file states ($\lgr = 0$) as the surrounding crowd state changes.

The most interesting regimes for the formation and $\lgr$ are the counter-flow cases, including their limits: free-flow (the surrounding crowd has $0$ density) and standing crowd  (the surrounding crowd has $0$ velocity).
Up to $\density \approx 0.7\,\densityunit$, $\lgr$ exhibits a non-monotonic dependence on dyad velocity.
In addition, there are two density-dependent critical velocity values at which $\lgr$ is minimum and maximum.
These are the velocities with, respectively, the lowest and highest likelihoods of being in the abreast state.
Additionally, the velocities are also lower and higher than the average dyad velocity for the same density.
We conjecture that these velocities separate regimes of directional uncertainty, which also reverberate on the dyad formation and on running regimes, in which abreast is preferred as much as possible.
These two critical velocities merge at around $\density \approx 0.7\,\densityunit$ and $\lgr$ becomes monotonically decreasing with the dyad velocity.
Notably, transitioning from a standing crowd to counter-flow has little influence on $\lgr$.
A counter-flowing crowd marginally reduces the density and velocity levels at which a given $\lgr$ value is observed.
In other terms, a counter-flowing crowd acts on the dyad formation as a standing crowd of slightly higher density would, while additionally shifting the dyad's effective velocity range downward, together compressing the density dynamic range.

We proposed a compact and accurate phenomenological model for $\lgr$ hinging on the aforementioned density-dependent critical velocity values - which we fit as linear velocity-density relations.
Effectively, $\lgr$ (its derivative with respect to the dyad velocity) behaves as the product of two traditional fundamental diagrams.
As such, it can be modeled with a few parameters and, overall, a cubic polynomial. 
Finally, in co-flow states, $\lgr$ typically indicates an abreast preference that weakens with increasing density, approaching $\lgr \approx 0$ at high densities. 

Three limitations of the present study merit consideration.
\noindent First, we willingly excluded an analysis of ``sideways'' crossings between crowd and dyads - limiting ourselves to the better defined counter- and co-flow states and limits thereof. Such a sideways crossing condition is rich with corner cases, whose complexity is left to forthcoming studies.

\noindent Second, the proximity region used here for density estimation is isotropic (a circle of radius $R$); an anisotropic variant -- e.g.\ a half-disc or ellipse aligned with $\vcom$ -- could potentially better capture the forward-biased nature of pedestrian perception, something that is not done here because we can not directly infer gaze direction.

\noindent Finally, as a single-site study, the results reported here may reflect conditions specific to Eindhoven Central Station; the phenomenological structure of the relations -- and in particular the $\lgr$ framework -- is, however, built on purely kinematic observables and is expected to transfer to comparable environments. Hence, validation across different sites is a natural direction for future work.

In general, our analysis opens the possibility of detailed microscopic modeling of dyad dynamics across various flow regimes. 
Indeed, given the prevalence of dyads in pedestrian crowds, this is instrumental for accurate and generalizable models of group dynamics and, by extension, crowd dynamics as a whole, as well as for fundamental studies of dyad behavior.
Note that  scaling the dyad detection and analysis to millions of %
trajectories
was essential to achieve the statistical resolution necessary to quantify configuration transitions systematically - something inaccessible to analyses hinged on smaller datasets.
The $\lgr$ framework provides a compact, yet expressive scalar measure of dyad configuration,
enabling potential integration of empirically-grounded group dynamics into both microscopic crowd-simulation models and macroscopic active-matter theories.

\section*{Acknowledgments}
The authors wish to acknowledge Tom Harmsen, who contributed initial implementations of the analysis as part of his Bachelor's final project at TU/Eindhoven.

\bibliography{bib}
\newpage

\begin{appendices}

  {\noindent\Large\textbf{S1 Appendix}}\\[0.5em]
  {\noindent\large Supplementary appendices for ``Probabilistic dynamics of small groups in crowd flows'' by
	Chiel van der Laan and Alessandro Corbetta.}\\[0.8em]
  \noindent DOI: \href{https://doi.org/10.48550/arXiv.2511.13181}{10.48550/arXiv.2511.13181}

\bigskip
	\section{Definition of the standing threshold $\vstandingthresh$}\label{ap:whystanding}
	\begin{figure}[h]
		\centering
		\includegraphics[width=0.5\linewidth]{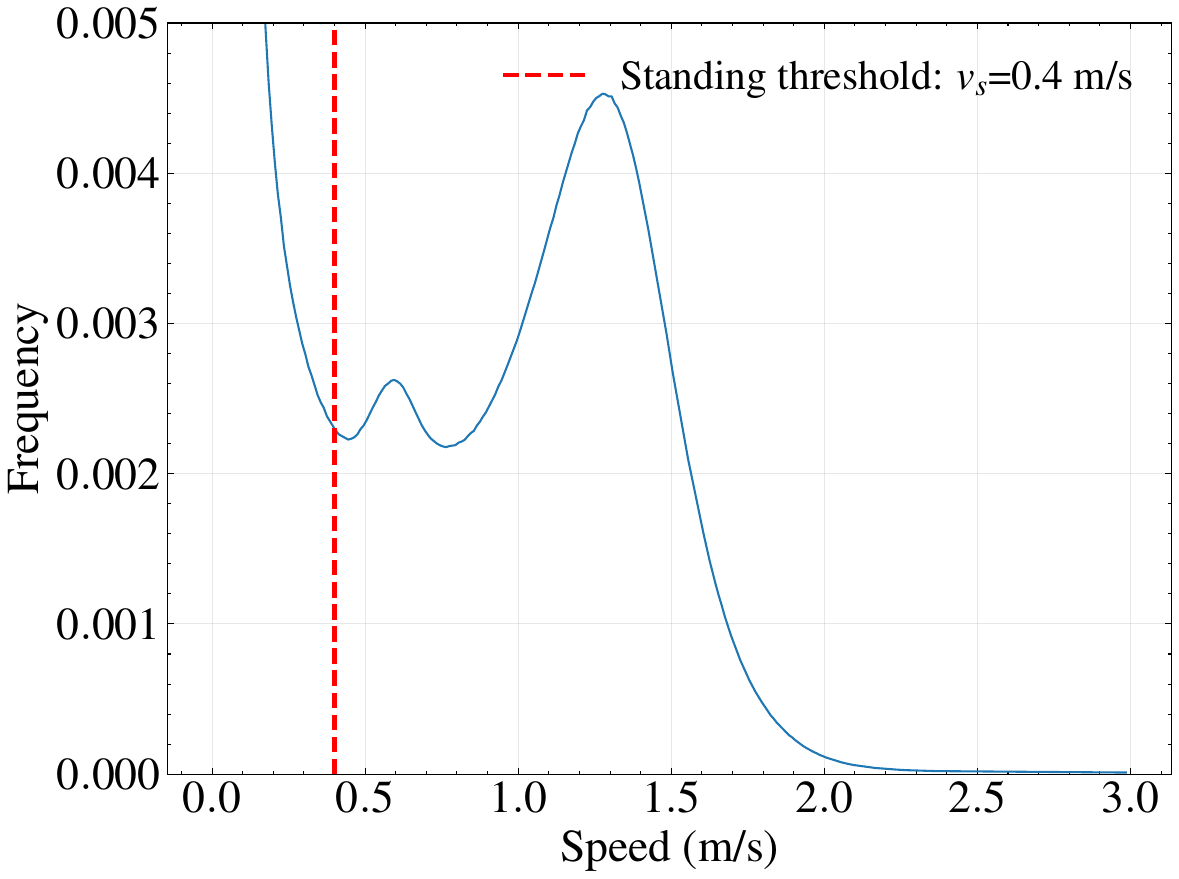}
		\caption{Probability density function of pedestrian speeds from twenty randomly selected days in our dataset.
			The red dashed line marks the standing threshold $\vstandingthresh = 0.4\,\mathrm{m/s}$, which is set between the stationary peak at $0.0\,\mathrm{m/s}$ and the slow-walking peak at $0.6\,\mathrm{m/s}$.
		}\label{fig:APPstanding}
	\end{figure}

	\noindent To distinguish between standing and moving pedestrians, we analyze the speed distribution of pedestrians in our dataset (\reffig{fig:APPstanding}). The distribution exhibits three distinct peaks: stationary pedestrians at $\approx0.0\,\mathrm{m/s}$, slow walkers at $\approx0.6\,\mathrm{m/s}$, and normal walkers at $\approx1.3\,\mathrm{m/s}$. We set the standing threshold at $\vstandingthresh = 0.4\,\mathrm{m/s}$, positioned between the stationary and slow-walking peaks.

	\section{Dyad detection algorithm -- technical details}\label{app:dyad-detection}

\begin{figure}[t]
		\centering
		\includegraphics[width=.4\linewidth]{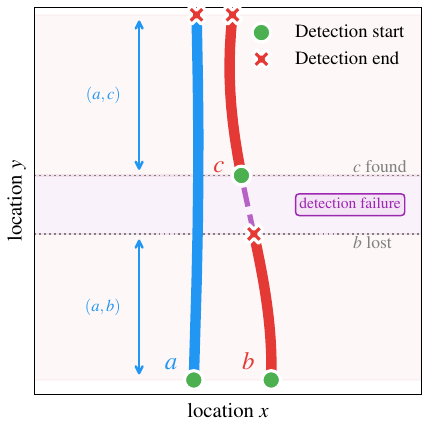}
		\caption{%
		Hardening the dyad detection algorithm in case of temporary tracking failures: example case.
		We sketch the trajectories, in blue and red, of a genuine dyad. 
		The ``blue'' trajectory is correctly estimated and attributed identifier $a$. On the contrary, the tracker temporarily loses the ``red'' trajectory, resulting in distinct identifiers, $b$ and $c$, appearing in disjoint time intervals. The idealized procedure in \refsec{subsec:dyad-detection} would unnecessarily disregard similar cases. The technical workaround in \refeqq{eq:smart-connect} correctly retains these measurements.
		}\label{fig:dyad-detection}
	\end{figure}

	The procedure described in \refsec{subsec:dyad-detection} outlines the idealized detection scheme.
	In practice, to ensure only physically meaningful trajectory segments are analyzed, two imperfections in the raw dataset need to be addressed:
	\begin{itemize}
		\item \textbf{Detection failures.} A pedestrian may temporarily go undetected and reappear as a new trajectory, incorrectly causing connected components of size $>2$ (see \reffig{fig:dyad-detection}).
		\item \textbf{Boundary artifacts.} Velocities are unreliable near trajectory endpoints, requiring trimming of each co-observation interval.
	\end{itemize}
	These entail minor technical modifications of the dyad detection approach that we hereby describe.

	\paragraph{Detection failures.}
	Detection failures require special attention, because the procedure in \refsec{subsec:dyad-detection} might unnecessarily reject valid dyads.
	For example, in the scenario of \reffig{fig:dyad-detection}, pedestrian $b$ is temporarily lost by the tracker and later re-identified as a new trajectory $c$.
	The procedure introduced in \refsec{subsec:dyad-detection}
	would identify $a$, $b$, and $c$ as part of a connected component of size~$3$, incorrectly discarding both valid pairs $(a,b)$ and $(a,c)$. 
	
	To prevent such false rejections, we resolve dyad membership at each time step $t$ rather than over the full trajectory.
	Concretely, we define the ``active edge set'' $\mathcal{E}_0(t)$ as
	\begin{equation}
		\mathcal{E}_0(t) = \left\{ (\traj_i(t), \traj_j(t)) \;\middle|\; (\traj_i, \traj_j) \in \mathcal{E}_0 \wedge t \in \trajtime_i \cap \trajtime_j \right\}.
	\end{equation}
	This gives a snapshot of all co-present pairs at each time instant $t$.
	A genuine dyad member can be paired with at most one partner at any instant.
	We therefore mark a pedestrian as ``ambiguous'' at time $t$ ($\mathcal{A}(t)$) if they are a member of more than one active candidate edge in $\mathcal{E}_0(t)$, i.e., their connected component is strictly larger than $2$ or, in formulas,  
	\begin{equation}
		\mathcal{A}(t) = \left\{ k \;\middle|\; \bigl|\{ (\traj_i(t), \traj_j(t)) \in \mathcal{E}_0(t) : i = k \text{ or } j = k \}\bigr| > 1 \right\}.
	\end{equation}
This allows us to replace the connected component pruning with a logic that retains only pairs from $\mathcal{C}$ in which neither partner is at any time ``ambiguous''.
	In equations, it holds:
	\begin{equation}
		\mathcal{C}_\mathcal{A} = \left\{ (\traj_i, \traj_j) \in \mathcal{C} \;\middle|\; \forall t:\, i \notin \mathcal{A}(t) \text{ and } j \notin \mathcal{A}(t) \right\}.\label{eq:smart-connect}
	\end{equation}
	With this approach, both ($a,b$) and ($a,c$) from the case of \reffig{fig:dyad-detection} are retained as valid dyads.

	\paragraph{Boundary artifacts.}
	Velocity estimates derived exclusively from positional data become unreliable near the start and end of each trajectory.
	To exclude these unreliable segments, we trim the trajectory by the length of our Savitzky-Golay filter window $\Delta t_{\rm trim}$.
	In equations, this trimmed co-observation interval $\trajtime_{ij}^{\,\Delta}$ reads:
	\begin{equation}
		\trajtime_{ij}^{\,\Delta} = \bigl[\min(\trajtime_i \cap \trajtime_j) +
		\Delta t_{\rm trim},\; \max(\trajtime_i \cap \trajtime_j) - \Delta t_{\rm trim}\bigr],
	\end{equation}	
	which, equivalently to \refeqq{eq:walkingdomain}, results in the ``trimmed walking time domain'' $\tilde{\trajtime}_{ij}^{\,\Delta}$:
	\begin{equation}
		\tilde{\trajtime}_{ij}^{\,\Delta}  = \bigl\{\, t \in \trajtime_{ij}^{\,\Delta}\;\big|\;
		\|\mathbf{v}_i(t)\| > \vstandingthresh,\
		\|\mathbf{v}_j(t)\| > \vstandingthresh \,\bigr\}.
	\end{equation}
	The trimming necessitates a modification to $\mathcal{D}$ (cf.~\refeqq{eq:dyadset}) that now has to satisfy three criteria:
	(i) to discard excessively short trajectories, as in~\refeqq{eq:dyadset}, the co-observation interval must be longer than $t_{\min}$;
	(ii) to ensure we did not trim away only parts where the dyad was walking, the trimmed trajectories must at least be walking for $\tilde{t}_{\min}^{\,\Delta}$; and
	(iii) to check for proximity in this new domain, 
	the mean inter-pedestrian distance during $\tilde{\trajtime}_{ij}^{\,\Delta}$ must be smaller than $\distanceThreshold$.
	The modified equation thus reads
	\begin{equation}\label{eq:improved-dyad}
		\mathcal{D} = \left\{ (\traj_i, \traj_j) \in \mathcal{C}_\mathcal{A} \;\middle|\;
		\begin{aligned}
			&|\trajtime_{ij}| > t_{\min} \\
			&|\tilde{\trajtime}_{ij}^{\,\Delta}| > t_{\min}^{\,\Delta} \\
			&\langle d_{ij} \rangle_{\tilde{\trajtime}_{ij}^{\,\Delta}} < \distanceThreshold
		\end{aligned}
		\right\},
		\qquad
		\begin{aligned}
		&t_{\min}            = 8.0~\mathrm{s}\\
		&t_{\min}^{\,\Delta} = 4.0~\mathrm{s}\\
		&\distanceThreshold  = 1.5~\mathrm{m} \\
		&\Delta t_{\rm trim} = 2.2~\mathrm{s}.
		\end{aligned}
	\end{equation}

	\section{Sensitivity analyses}\label{ap:sensitivity}

	\begin{figure}[h]
		\centering
		\figinnerlab{\includegraphics[width=0.47\linewidth]{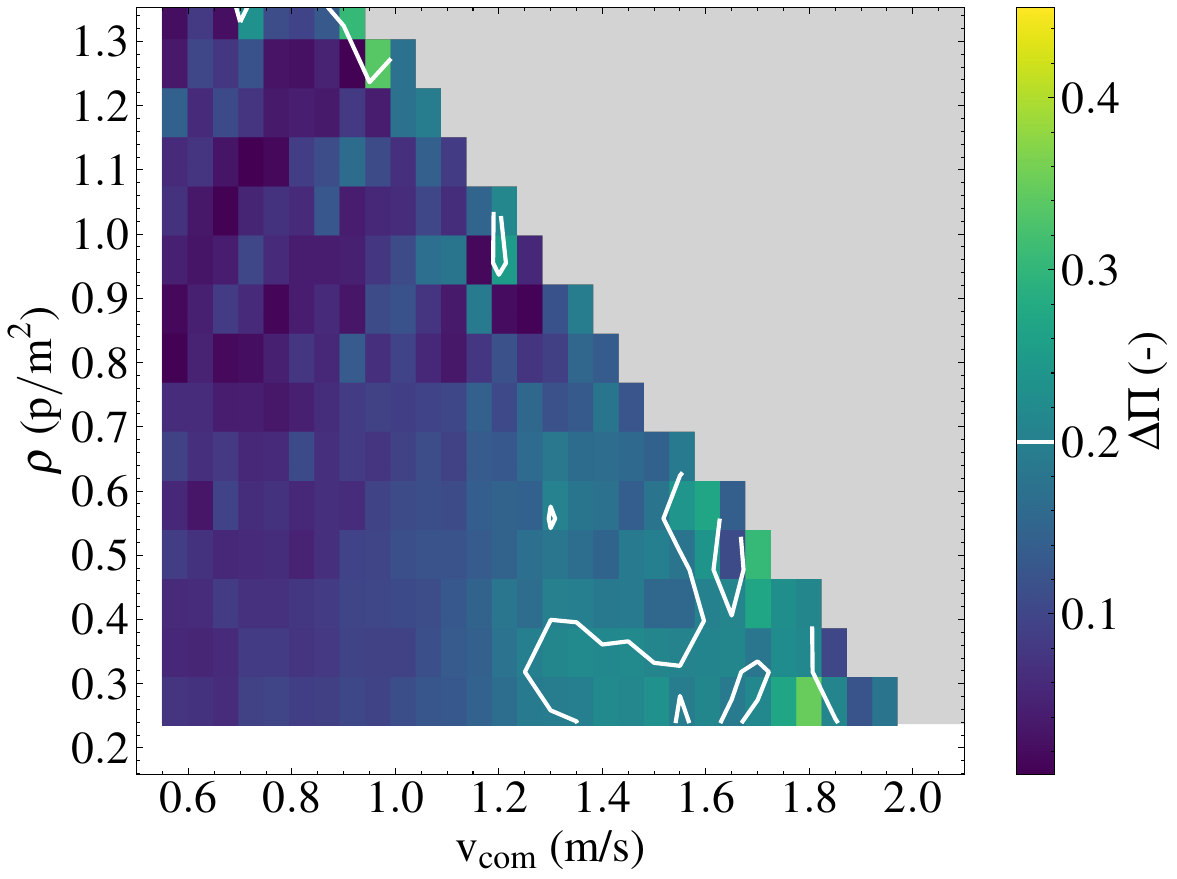}}{a}{-.2cm}{-.4cm}
		\figinnerlab{\includegraphics[width=0.47\linewidth]{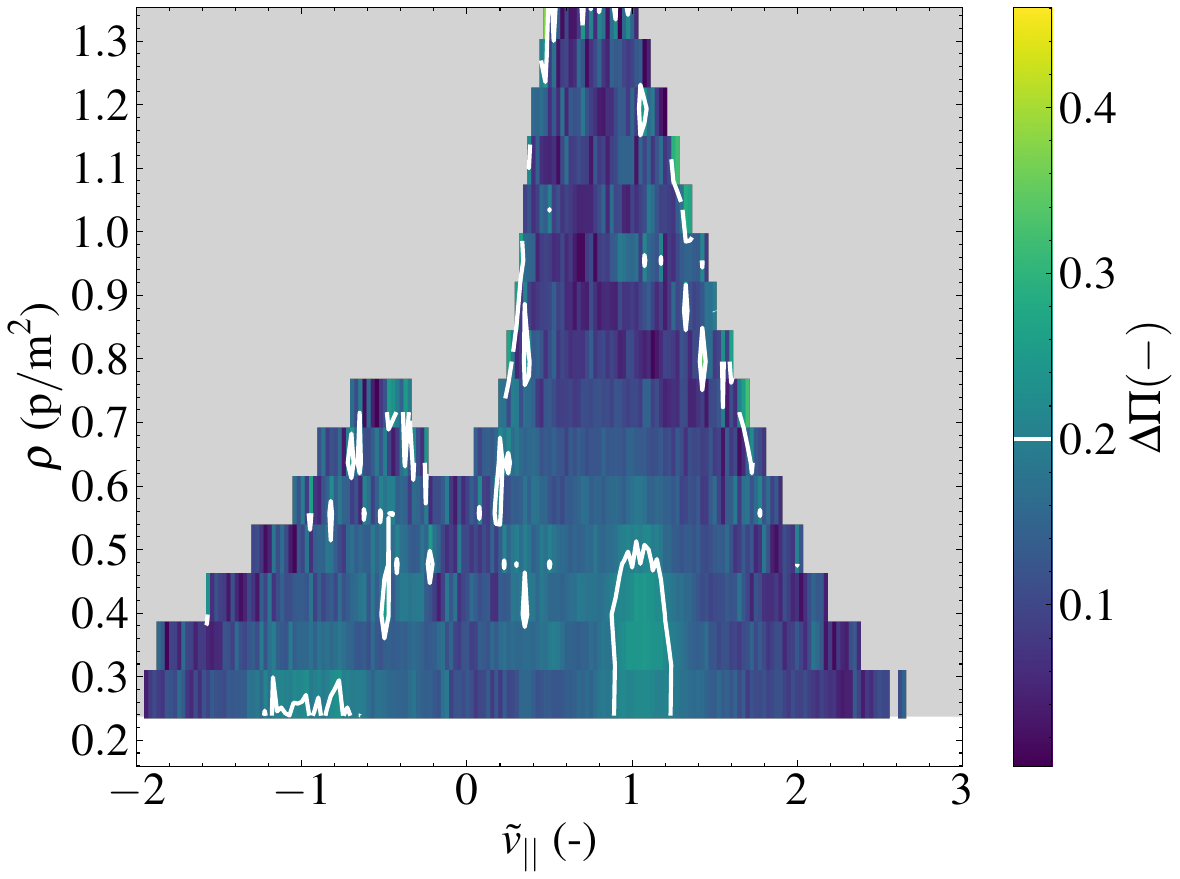}}{b}{-.2cm}{-.4cm}
		\caption{%
		Sensitivity of the empirical formation indicator $\lgr$ to a joint $\pm10\%$ perturbation of the dyad detection parameters in \refeqq{eq:improved-dyad}, relative to the original parameter set though $\Delta \Pi$ (cf.~\refeqq{eq:deltapi}).
		(a) $\Delta \Pi$ for $\lgr^{\text{standing}}(\comspeed,\density)$ (cf.~\reffig{fig8}c), which remains consistently lower than $0.2$ (with $\lgr$ typically having order $1$).
		(b) $\Delta \Pi$ for $\hatlgr(\vrelpar,\density)$ (cf.~\reffig{fig10}) 
		which also remains typically well below $0.2$.%
		}\label{fig:APPsens_params}
	\end{figure}

	\noindent We assess the robustness of our dyad detection algorithm and our overall findings with respect to $\pm10\%$ perturbations of key parameters:
	\begin{itemize}
		\item the detection thresholds $t_{\min}$, $t_{\min}^{\,\Delta}, \distanceThreshold$ that determine which pairs are retained as dyads. 
		\item the radius $\closeradius$ that governs local density estimation.
	\end{itemize}
	We show that our results are practically insensitive to these perturbations.

	\paragraph{Robustness of the detected dyad set.}
	We verify that the detected dyad set is insensitive to the precise choice of the three detection thresholds $t_{\min}$, $t_{\min}^{\,\Delta}$, and $\distanceThreshold$ (cf.~\refeqq{eq:connections}, ~\refeqq{eq:improved-dyad}) by applying a joint $\pm 10\%$ perturbation, yielding the parameter and related dyad sets:
\begin{itemize}
  \item $-10\%$ perturbation (dyad set $\mathcal{D}_-$): $t_{\max} = 7.0\,\mathrm{s}$, $\tilde{t}_{\max} = 3.3\,\mathrm{s}$, and $d_{\max} = 1.4\,\mathrm{m}$;
  \item original parameters (dyad set $\mathcal{D}$): $t_{\max} = 8.0\,\mathrm{s}$, $\tilde{t}_{\max} = 4.0\,\mathrm{s}$, and $d_{\max} = 1.5\,\mathrm{m}$; 
  \item $+10\%$ perturbation (dyad set $\mathcal{D}_+$): $t_{\max} = 8.8\,\mathrm{s}$, $\tilde{t}_{\max} = 4.4\,\mathrm{s}$, and $d_{\max} = 1.6\,\mathrm{m}$.
\end{itemize}
As reported in \reftab{tab:dyad_overlap}, the dyad set $\mathcal{D}$ has more than $83\%$ overlap (in terms of data samples, indicated by the operator $|\cdot|$) with the perturbed sets $\mathcal{D}_+$ and $\mathcal{D}_-$. Most importantly, these differences in dyad selection account to no practical distinction when it comes to quantify the statistical properties of the dyads and our models thereof. 

	\begin{table}[h!]
		\centering
		\caption{Pairwise relative overlap between dyad sets detected under the original parameter set ($\mathcal{D}$) and under the $\pm10\%$ perturbed parameter sets ($\mathcal{D}_-$ and $\mathcal{D}_+$). 
		}\label{tab:dyad_overlap}
\begin{tabular}{|c|c|c|}\hline
  \multicolumn{2}{|c|}{Relative overlap (\%)}  & Jaccard (\%)\\\hline
$|\mathcal{D} \cap \mathcal{D}_-|$/$|\mathcal{D}| = 86\%$ & $|\mathcal{D} \cap \mathcal{D}_-|$/$|\mathcal{D}_-| = 87\%$ & $|\mathcal{D} \cap \mathcal{D}_-|$/$|\mathcal{D} \cup \mathcal{D}_-| = 80\%$ \\\hline
$|\mathcal{D} \cap \mathcal{D}_+|$/$|\mathcal{D}| = 89\%$ & $|\mathcal{D} \cap \mathcal{D}_+|$/$|\mathcal{D}_+| = 83\%$ & $|\mathcal{D} \cap \mathcal{D}_+|$/$|\mathcal{D} \cup \mathcal{D}_+| = 80\%$ \\\hline
$|\mathcal{D}_- \cap \mathcal{D}_+|$/$|\mathcal{D}_-| = 87\%$ & $|\mathcal{D}_- \cap \mathcal{D}_+|$/$|\mathcal{D}_+| = 79\%$ & $|\mathcal{D}_- \cap \mathcal{D}_+|$/$|\mathcal{D}_- \cup \mathcal{D}_+| = 73\%$ \\ \hline
\end{tabular}
	\end{table}

	To confirm that this robustness translates to the main results, \reffig{fig:APPsens_params} reports the pointwise absolute differences with respect to the original values of $\lgr^{\text{standing}}(\comspeed,\density)$ and $\hatlgr(\vrelpar,\density)$ under all three parameter sets.
	Specifically, we consider the indicator 
	\begin{equation}\label{eq:deltapi}
		\Delta \Pi(\vcom,\density ) = \max \left\{ |\Pi_\mathcal{D} - \Pi_{\mathcal{D}_-}| , |\Pi_\mathcal{D} - \Pi_{\mathcal{D}_+}| \right\}.
	\end{equation}
	The empirical values have with maximum pointwise deviation from the original below $0.2$ across large parts of the domain.

	\paragraph{Sensitivity of density-dependent analyses to $\closeradius$.}
	The radius $\closeradius$ (cf.~\refeqq{eq:density}) determines the local density $\density$ used in all analyses.
	By construction, $R$ does not affect the dyad identification procedure. 
	We demonstrate that $\pm 10\%$ perturbations in $R$ reflect into predictable shifts in $
	\density$.

\begin{figure}[t]
		\centering
		\figinnerlab{\includegraphics[width=0.47\linewidth]{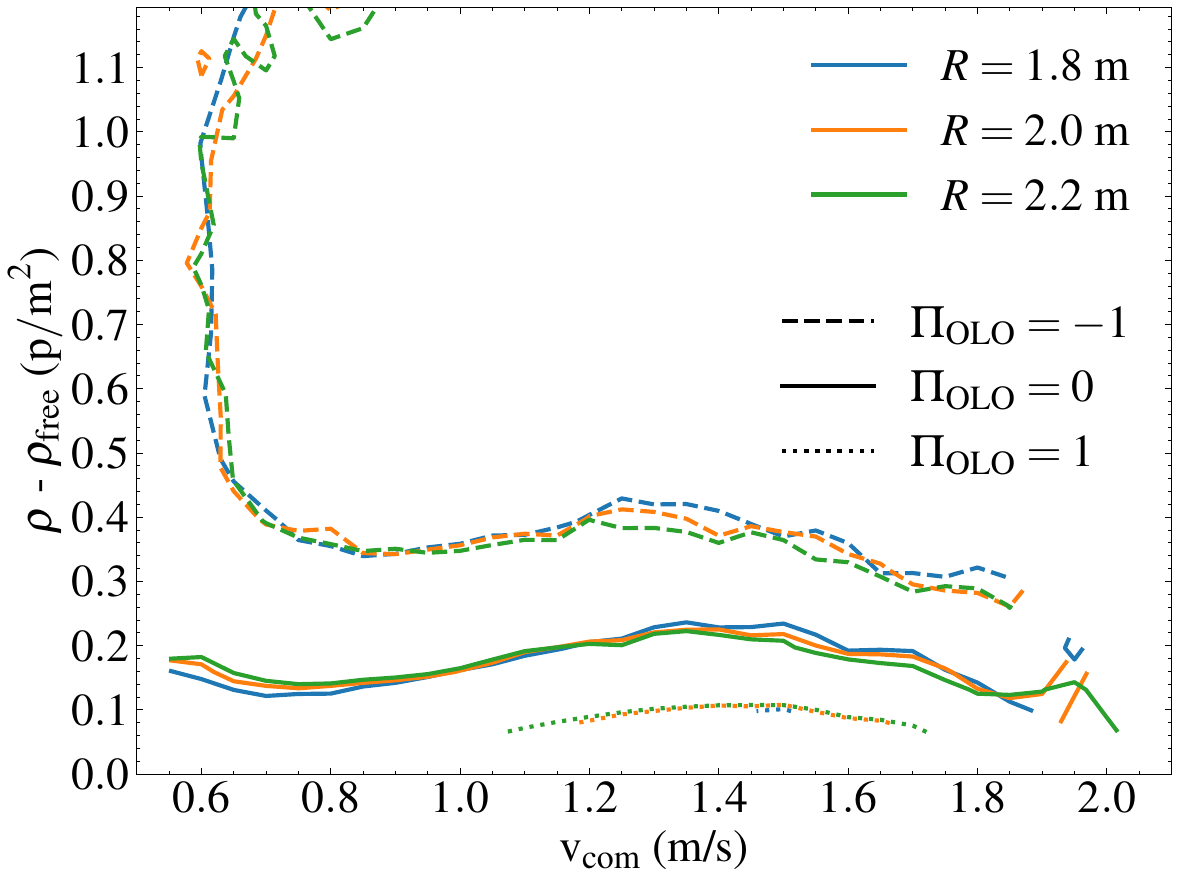}}{a}{-.2cm}{-.4cm}
		\figinnerlab{\includegraphics[width=0.47\linewidth]{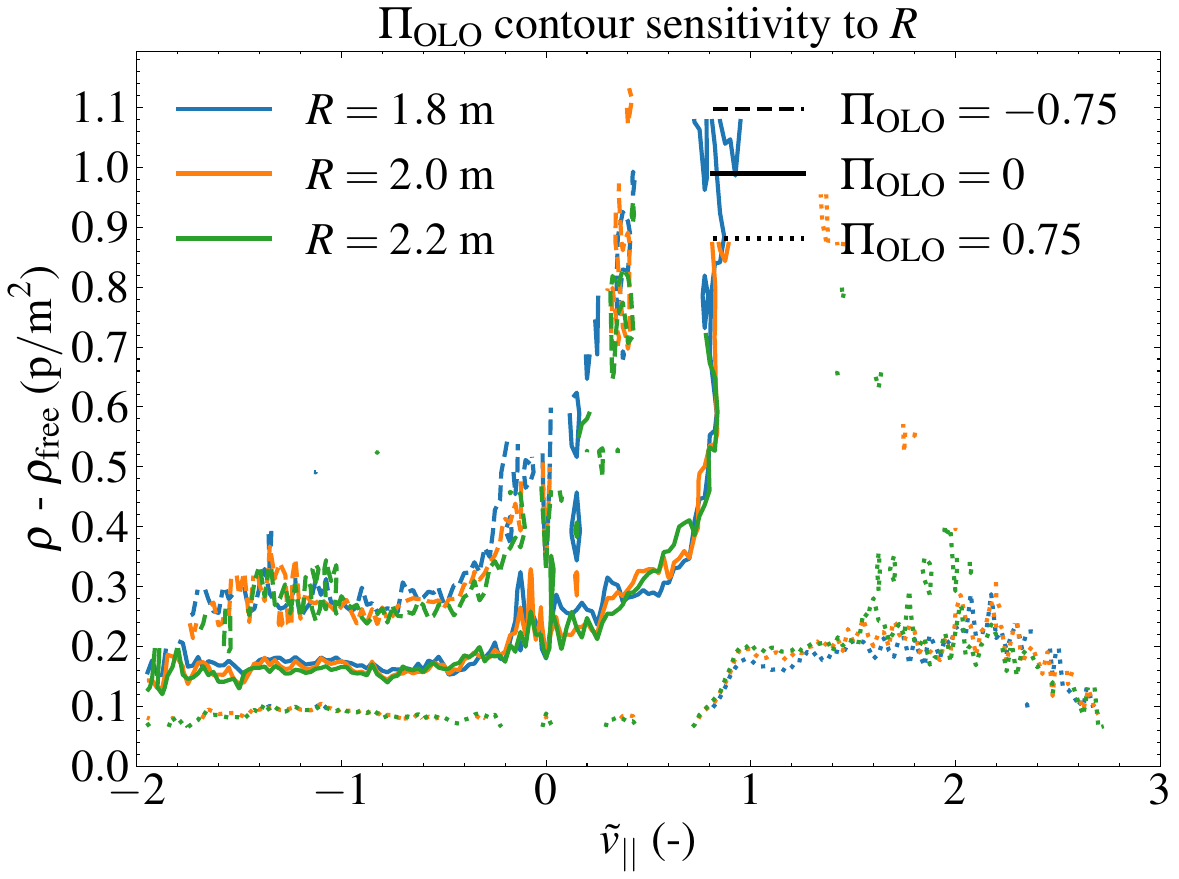}}{b}{-.2cm}{-.4cm}
		\caption{%
			Sensitivity of empirical $\lgr$ values to a $\pm10\%$ variation of the density radius $\closeradius$ in \refeqq{eq:density}. 
			Contours computed with $R=1.8\,\mathrm{m}$, $R=2.0\,\mathrm{m}$ (as in \refsec{subsec:KeyAndCoord}), and $R=2.2\,\mathrm{m}$ are shown in each figure as blue lines, orange lines, and green lines, respectively. For each radius, three contour levels are displayed using solid, dashed, and dotted line styles. 
			(a) Three $\lgr^{\text{standing}}(\comspeed,\density_{\mathrm{crowd}})$ contours for the three radii, showing substantial overlap.
			(b) Three $\hatlgr(\vrelpar,\density_{\mathrm{crowd}})$ contours, also exhibiting strong overlap.
			These figures indicate that the observed differences are primarily due to a predictable shift along the density axis, as in \refeqq{eq:density_shift}.
		}\label{fig:APPsens_R}
	\end{figure}

	\noindent Our density definition (cf.~\refeqq{eq:density}) can be decomposed into a crowd and a dyad contribution:
\begin{equation}
  \density = \frac{\closecount}{\pi \closeradius^2} = \frac{\closecount - 2}{\pi \closeradius^2} + \frac{2}{\pi \closeradius^2} = \density_\mathrm{crowd}(R) + \densityfreeflow(R),
\end{equation}
where, under the $\pm 10\%$ variations, $\density_\mathrm{crowd} = \density - \densityfreeflow$ is 
effectively insensitive to changes in $\closeradius$.
In equations, for any perturbed radius $R'$ and original radius $R$, it holds:
\begin{equation}
  \density(R') - \densityfreeflow(R') \approx \density(R) - \densityfreeflow(R).
\end{equation}
Therefore, the aforementioned predictable shift for a perturbed radius $R'$ with respect to our original choice $R$ is:
\begin{equation}\label{eq:density_shift}
  \density(R') \approx \density(R) - \densityfreeflow(R) + \densityfreeflow(R') 
  = \frac{\closecount}{\pi \closeradius^2} - \frac{2}{\pi \closeradius^2} + \frac{2}{\pi \closeradius'^2}.
\end{equation}
	Figure~\ref{fig:APPsens_R} confirms the validity of this shift for both $\lgr^{\text{standing}}(\comspeed,\density)$ and $\hatlgr(\vrelpar,\density)$ by plotting three contours as a function of $\density_{\mathrm{crowd}}$ instead of $\density$. 
	As one can observe, there is near perfect overlap, reinforcing the weak sensitivity on $R$.

	\section{Modal distance definition}\label{ap:modal}
	\begin{figure}[h]
		\centering
		\figinnerlab{
			\includegraphics[width=0.45\linewidth]{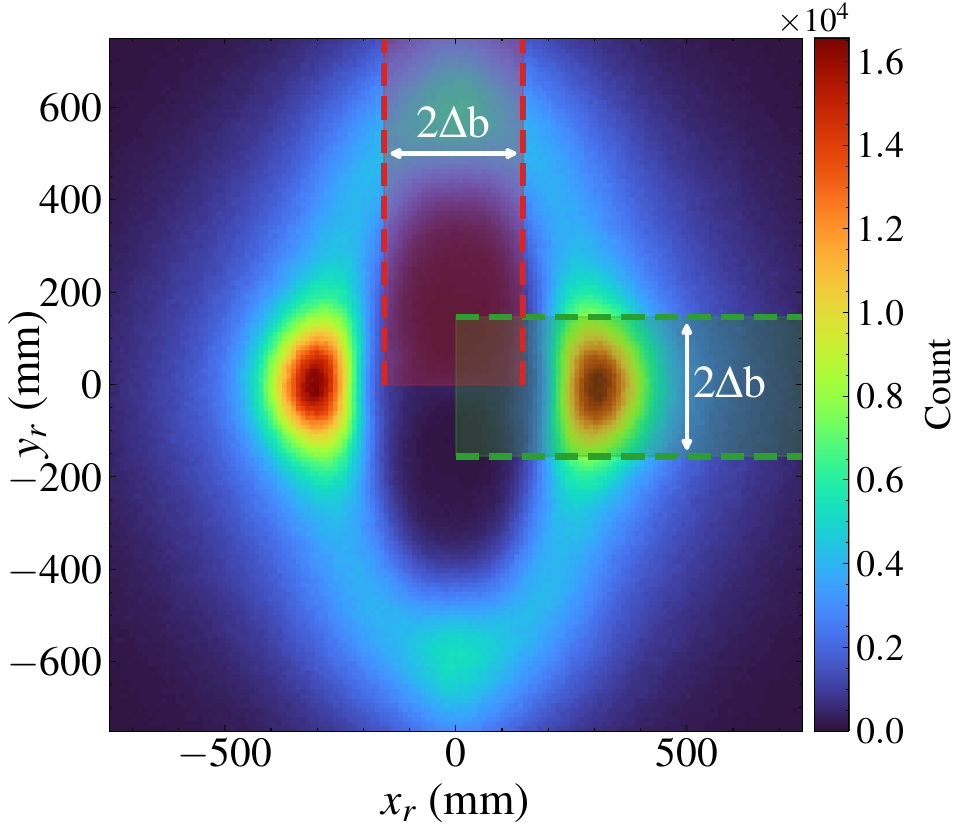}}{a}{-.2cm}{-.4cm}
		\figinnerlab{
			\includegraphics[width=0.45\linewidth]{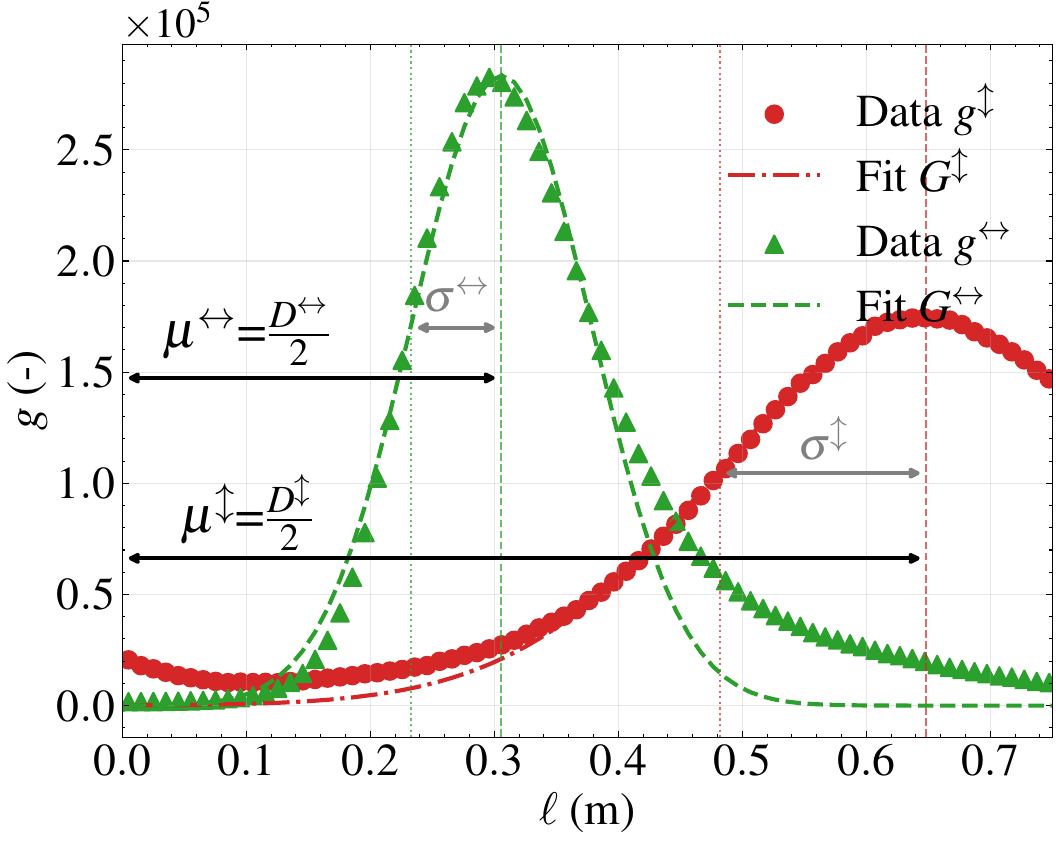}}{b}{-.1cm}{-.4cm}

		\caption{(a) Probability distribution $\Prob(\xr, \yr)$, the bands in green and red indicate the integration domains for $\modalletter^{\leftrightarrow}(\xr)$ and $\modalletter^{\updownarrow}(\yr)$ respectively (\refeqq{eq:integrationdomains}).
			(b) The data points retrieved from the integrations for $\modalletter^{\leftrightarrow}$ in green triangles and $\modalletter^{\updownarrow}$ in red circles with their corresponding $G$ fits.
		}\label{fig:appendixModalDistance}
	\end{figure}

	To extract characteristic interpersonal distances for each configuration type (abreast and in-file), we integrate the probability distributions $\Prob(\xr, \yr)$ over narrow bands around each principal axis (\reffig{fig:appendixModalDistance}a depicts the integration domains). In formulas, let
	\begin{equation}
		\modalbandlr(\xr) = \int_{-\deltaband}^{\deltaband} \Prob(\xr, \yr) \, d\yr, \quad
		\modalbandud(\yr) = \int_{-\deltaband}^{\deltaband} \Prob(\xr, \yr) \, d\xr
		\label{eq:integrationdomains}
	\end{equation}
	where $\deltaband = \dBandMeter\,\mathrm{m}$ defines the bandwidth. As illustrated in \reffig{fig:appendixModalDistance}b, we fit each marginal distribution with a Gaussian:
	\begin{equation}
		\Gausfit(\ell) \propto \exp\left(-\frac{(\ell-\mu)^2}{2\sigma^2}\right),
		\label{eq:gaussian_fit}
	\end{equation}
	where $\mu$ is the modal position and $\sigma$ is the standard deviation. Finally, we define the interpersonal distance as twice the value of $\mu$, as  in \refeqq{eq:interpersonal_distances} (cf.   \reffig{fig:appendixModalDistance}b).

	\section{Region of validity for the $\lgr$ model}\label{ap:valid_model_region}
\begin{figure}[h]
		\centering
		\figinnerlab{
			\includegraphics[width=0.45\linewidth]{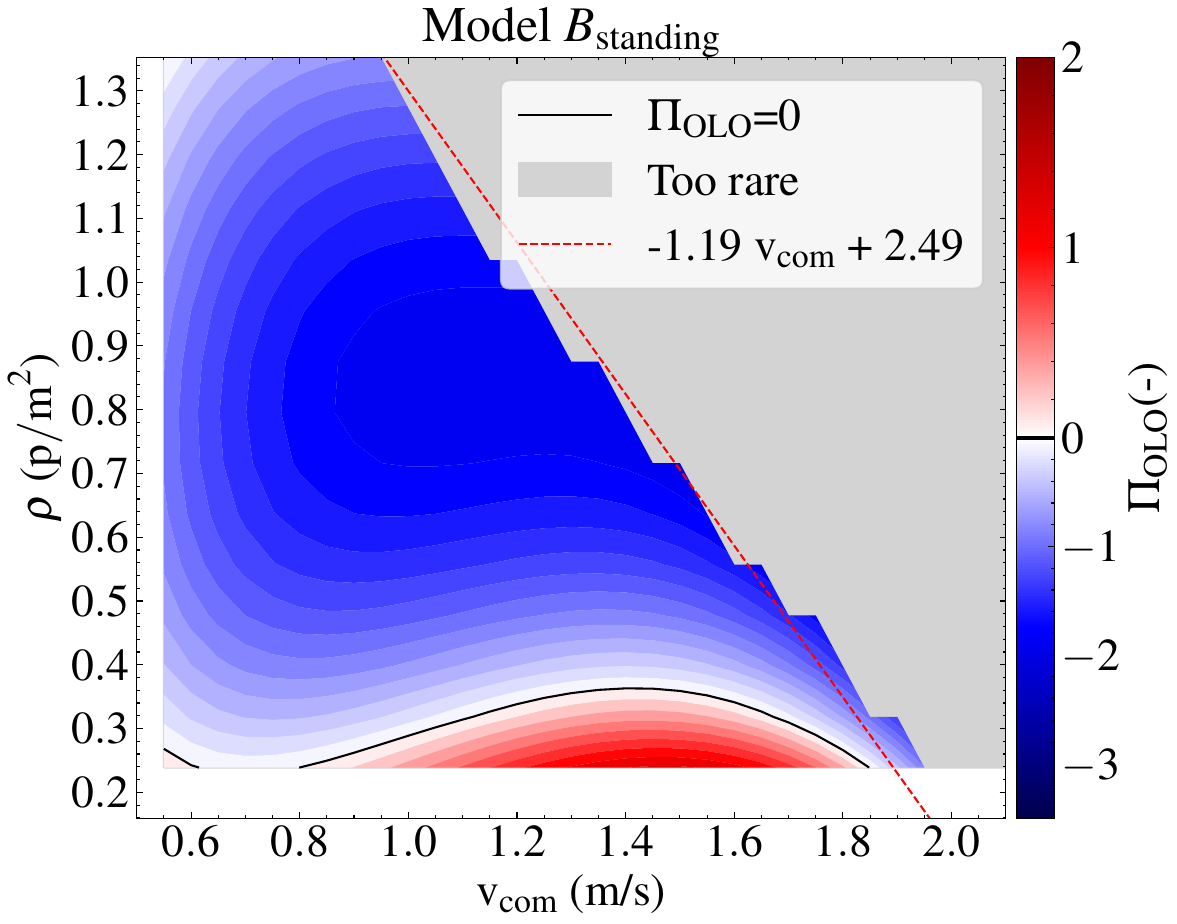}}{a}{-.2cm}{-.4cm}
		\figinnerlab{
			\includegraphics[width=0.45\linewidth]{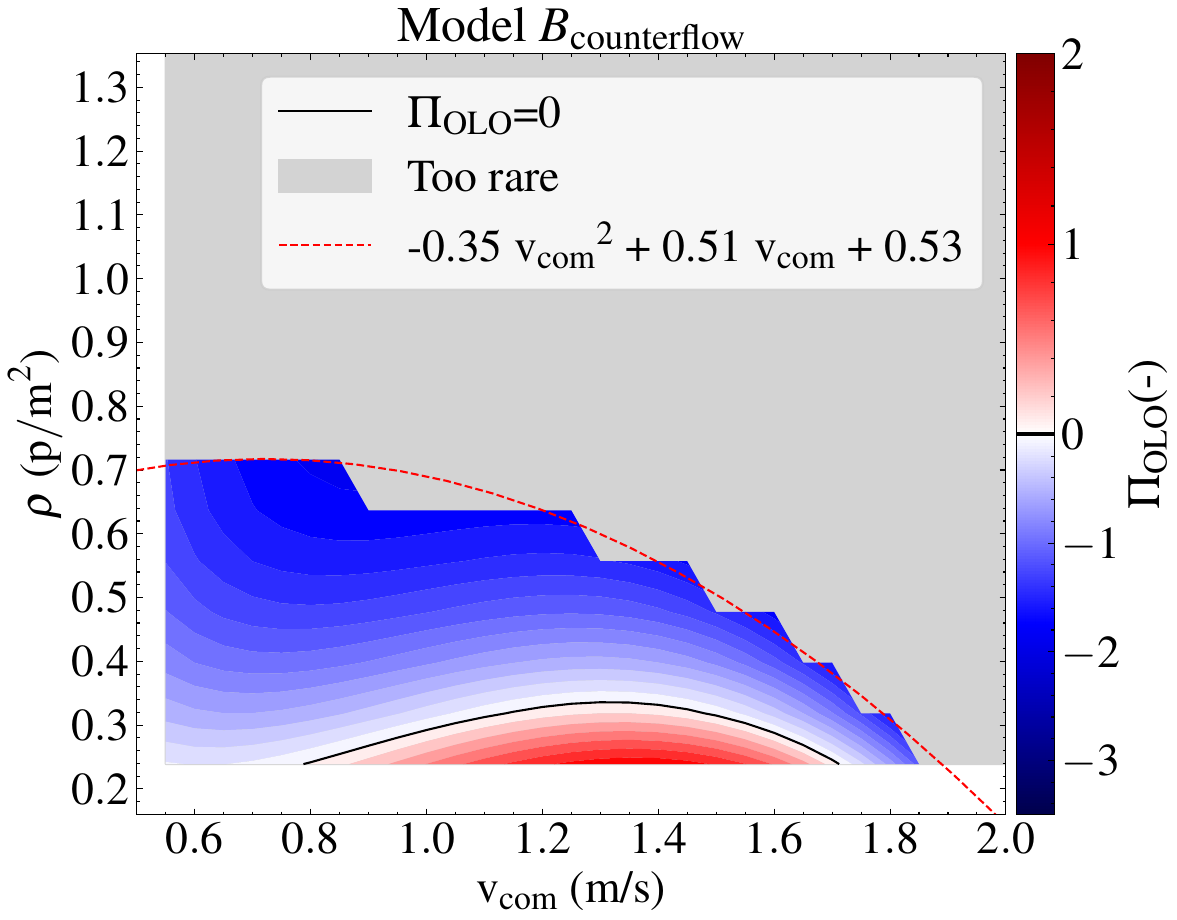}}{b}{-.2cm}{-.4cm}
		\caption{Models for $\lgr^{\text{standing}}(\comspeed, \density)$ (\refeqq{eq:PImodelStanding}) and $\lgr^{\text{counterflow}}(\comspeed, \density)$ (\refeqq{eq:pimodelagainst}), with their respective validity regions.
			(a) Standing crowd model (\refeqq{eq:PImodelStanding}) with validity boundary (dashed red line, \refeqq{eq:validmodelstanding}).
			(b) Counter-flow model (\refeqq{eq:pimodelagainst}) with validity boundary (dashed red line, \refeqq{eq:validmodelagainst}).
			Gray regions indicate insufficient data for reliable model predictions.
		}\label{fig:APPmodellimits}
	\end{figure}

	In \refsec{subsec:OLOstanding} and \refsec{subsec:OLOagainst}, we provided phenomenological models for $\lgr$ in standing crowd and counter-flow conditions, respectively. These models are defined only in regions where sufficient data is available.
	Specifically, the model for $\lgr^{\text{standing}}(\comspeed, \density)$ (\refeqq{eq:PImodelStanding}) has validity in the region
	\begin{equation}
		\{(\comspeed, \density) \colon \densityfreeflow = 0.16\,\densityunit < \density < A\,\comspeed + B ,  \quad \comspeed > \vstandingthresh\}, \label{eq:validmodelstanding}
	\end{equation}
	where $A = -1.19\, \mathrm{ps}/\mathrm{m}^3$, $B = 2.49\,\densityunit$.
	Whereas the  model for $\lgr^{\text{counterflow}}(\comspeed, \density)$ (\refeqq{eq:pimodelagainst}) has validity in
	\begin{equation}
		\{(\comspeed, \density) \colon \densityfreeflow < \density < C\,\comspeed^2 + D\,\comspeed + E, \quad \comspeed > \vstandingthresh\},
		\label{eq:validmodelagainst}
	\end{equation}
	where $C=-0.35\,\mathrm{ps}^2/\mathrm{m}^4$, $D=0.51\, \mathrm{ps}/\mathrm{m}^3$, and $E=0.53\,\densityunit$.
	In \reffig{fig:APPmodellimits}, we illustrate these valid regions.

\newpage
\section{Robustness of the $\lgr$ model parameters}\label{ap:robustness}

\begin{figure}[h]
    \centering
    \includegraphics[width=0.49\linewidth]{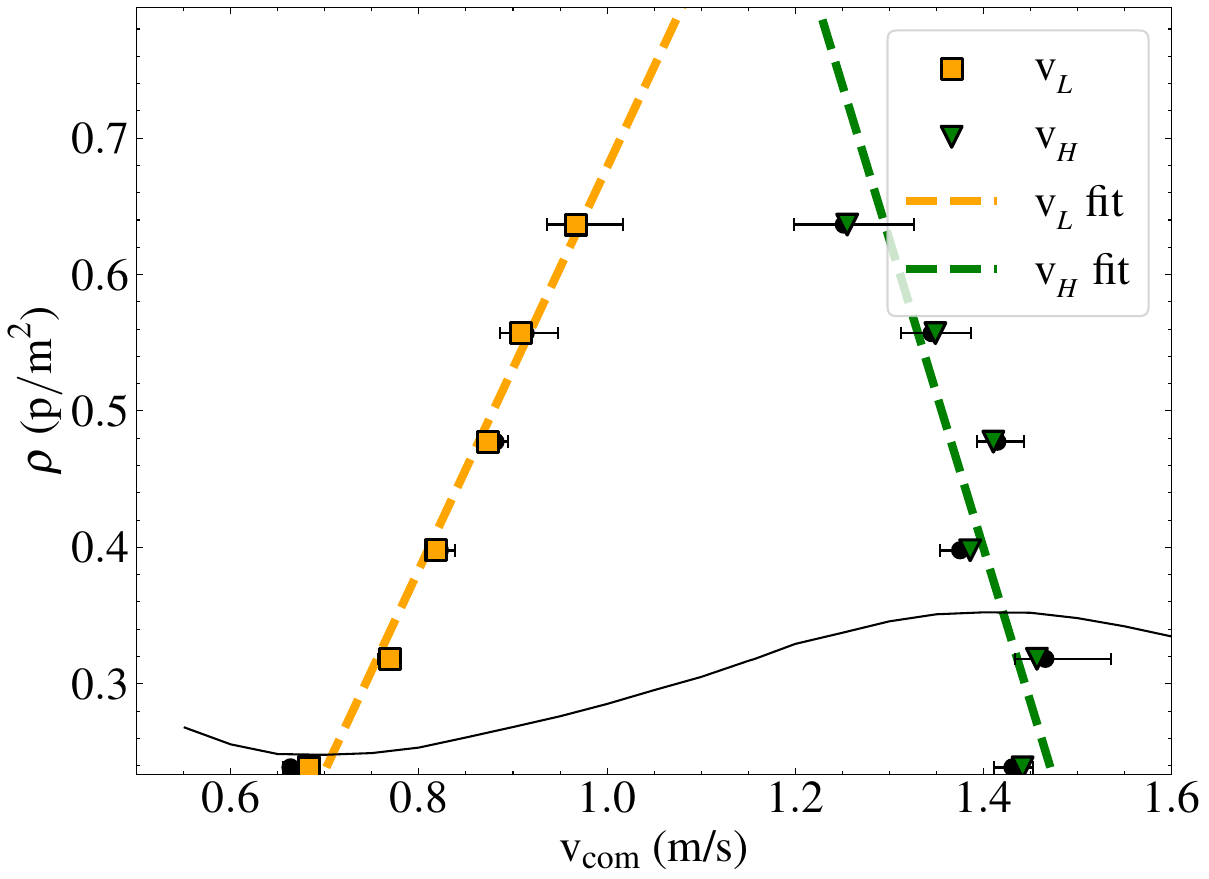}
    \caption{
		Split-sample validation of the parameters $\vcomlow(\density)$ and $\vcomhigh(\density)$ for the standing crowd model (cf. \refeqq{eq:PImodelStanding}).
The empirical $\vcomlow$ and $\vcomhigh$ (respectively, yellow squares and green triangles) are shown along with their corresponding fits (dashed lines).
In black the empirical split-sample mean (circles) and error bars (cf.~\refeqq{eq:split_error_bars}) are overlaid, with a black contour line that marks $\lgr = 0$ from \reffig{fig8}c as a reference.
    }\label{fig:split_2d}
\end{figure}

To assess the robustness of the fitted parameters $\vcomlow(\density)$ and $\vcomhigh(\density)$ (cf.\ \refsec{subsec:OLOstanding} and \reftab{tab:fit-params-with-dens}), 
we perform a 5-fold split-sample validation at the trajectory level.
The 900-day dataset is partitioned randomly into five non-overlapping subsets of 180 days each. For each of these folds, $s = 1,\ldots,5$, the complete pipeline (dyad detection, binning, and model fitting) is executed independently.
The spread across folds is shown as asymmetric error bars, with the lower and upper extents given by
\begin{equation}\label{eq:split_error_bars}
    \sigma^-_{\vcomlow}(\density_k) = \bar{v}_L(\density_k) - \min_s \vcomlow^{(s)}(\density_k),
    \qquad
    \sigma^+_{\vcomlow}(\density_k) = \max_s \vcomlow^{(s)}(\density_k) - \bar{v}_L(\density_k),
\end{equation}
where $\bar{v}_L(\density_k) = \frac{1}{5}\sum_s \vcomlow^{(s)}(\density_k)$ is the mean across folds, and analogously for $\vcomhigh$.
\reffig{fig:split_2d} shows the mean and error bars for the empirical $\vcomlow$ and $\vcomhigh$ values across folds, together with the fitted curves from the full dataset.
The narrow error bars for $\vcomlow$ indicate that these parameters are robustly identified across folds.
For $\vcomhigh$, the error bars are wider, likely reflecting greater data variability in these regimes.

\end{appendices}

\end{document}